\documentclass[pra,aps,floats,floatfix,twocolumn,showpacs,longbibliography,superscriptaddress,notitlepage,preprintnumbers,10pt]{revtex4-1}

\usepackage[USenglish]{babel}

\usepackage{amsmath,amssymb,amsfonts}

\usepackage{graphicx}
\usepackage{dcolumn}
\usepackage{bm}
\usepackage[normalem]{ulem} 

\usepackage[colorlinks=true,linkcolor=blue,citecolor=red]{hyperref}

\usepackage[T1]{fontenc} 
\usepackage{lmodern}

\usepackage[usenames,dvipsnames]{color}
\newcommand{\rozmiarjeden}{0.485\textwidth}

\newcommand{\COMPA}{COM$_1$}
\newcommand{\COMPB}{COM$_2$}
\newcommand{\COMNA}{AFM}
\newcommand{\COIPA}{COI$_1$}
\newcommand{\COIPB}{COI$_2$}
\newcommand{\COIPC}{COI$_3$}
\newcommand{\COINA}{AFI$_1$}
\newcommand{\COINB}{AFI$_2$}
\newcommand{\FLI}{NOM}
\newcommand{\MI}{NOI}
\newcommand{\MIP}{NOI$_{\textrm{A}}$}
\newcommand{\MIN}{NOI$_{\textrm{B}}$}

\begin{document}
\preprint{Submitted to: PHYSICAL REVIEW B}
\title{Extended Falicov-Kimball model: Exact solution for finite temperatures}
\author{Konrad Jerzy Kapcia}
\email[Corresponding author; e-mail: ]{konrad.kapcia@ifj.edu.pl}
\affiliation{Institute of Nuclear Physics, Polish Academy of Sciences, ul. W. E. Radzikowskiego 152, PL-31342 Krak\'{o}w, Poland}
\author{Romuald Lema\'nski}
\email[e-mail: ]{r.lemanski@intibs.pl}
\affiliation{Institute of Low Temperature and Structure Research, Polish Academy of Sciences, ul. Ok\'olna 2, PL-50422 Wroc\l{}aw, Poland}
\author{\fbox{Stanis\l{}aw Robaszkiewicz}}
\email[Deceased on 7 June 2017]{\ }
\affiliation{Faculty of Physics, Adam Mickiewicz University in Pozna\'n, ul. Umultowska 85, PL-61614 Pozna\'n, Poland}
\date{\today}
\begin{abstract}
The extended Falicov-Kimball model is analyzed exactly for finite temperatures in the limit of large dimensions.
The onsite, as well as the intersite density-density interactions represented by the coupling constants $U$ and $V$, respectively, are included in the model.
Using the dynamical mean field theory formalism on the Bethe lattice we find rigorously the temperature dependent density of states (DOS) at half-filling.
At zero temperature ($T=0$) the system  is ordered to form the checkerboard pattern and the DOS has the gap 
$\Delta(\varepsilon_F) > 0$ at the Fermi level, if only $U\neq 0$ or $V\neq 0$. 
With an increase of $T$ the DOS evolves in various ways that depend both on $U$ and $V$.  
If  $U <0$ or $U > 2V$, two additional subbands develop inside the principal energy gap. 
They become wider with increasing $T$ and at a certain $U$- and $V$-dependent temperature $T_{MI}$ they join with each other at $\varepsilon_F$. 
Since above $T_{MI}$ the DOS is positive at $\varepsilon_F$, we interpret $T_{MI}$ as the transformation temperature from insulator to metal. 
It appears, that $T_{MI}$ approaches the order-disorder phase transition temperature $T_{OD}$ for $|U|=2$ and $0<U\lesssim 2V$, but otherwise $T_{MI}$ is substantially lower than $T_{OD}$.
Moreover, we show that if $V\lesssim 0.54$ then $T_{MI}=0$ at two quasi-quantum critical points $U_{cr}^{\pm}$ (one positive and the other negative), whereas for $V\gtrsim 0.54$ there is only one negative $U^-_{cr}$.
Having calculated the temperature dependent DOS we study thermodynamic properties of the system starting from its free energy $F$ and then we construct the phase diagrams in the variables $T$ and $U$ for a few values of $V$. 
Our calculations give that inclusion of the intersite coupling $V$ causes the finite temperature phase diagrams to become asymmetric with respect to a change of sign of $U$.
On these phase diagrams we detected stability regions of eight different kinds of ordered phases, where both charge-order and antiferromagnetism coexists (five of them are insulating and three are conducting) and three different nonordered phases (two of them are insulating and one is conducting).
Moreover, both continuous and discontinuous transitions between various phases were found.
\end{abstract}




\maketitle


\section{Introduction}

Correlated electron systems exhibit many diverse and interesting properties, as e.g. charge and magnetic ordering, superconductivity, mixed valence, metal-insulator phase transition, etc.~\cite{MicnasRMP1990,ImadaRMP1998,YoshimiPRL2012,FrandsenNatCom2014,CominScience2015,%
NetoScience2015,CaiNatPhys2016,HsuNatCom2016,PelcNatCom2016,ParkPRL2017,NovelloPRL2017}.
Unfortunately, it is very difficult to find complete solutions even for simple models that describe these systems, therefore many issues are not yet explained satisfactorily~\cite{NagaokaPRB1966,LiebPRL1968,LiebPhysA1986,%
GeorgesRMP1996,FreericksRMP2003}.

One of a few methods of reliable studying of strongly correlated fermion systems is the dynamic mean field theory (DMFT) \cite{DongenPRL1990,GeorgesRMP1996,FreericksRMP2003,FreericksBook2006,KotliarRmp2006}, which is the exact theory in the limit of high dimensions ($D\rightarrow+\infty$) or, equivalently, of large coordination number.
And one of a few models for which this method can be used to achieve accurate results in the thermodynamic limit is the Falicov-Kimball  model (FKM) \cite{FalicovPRL1969,BrandtMielsch1989,BrandtMielsch1990,BrandtMielsch1991,BrandtUrbanek1992,%
FreericksPRB2000,ZlaticFreericksLemanskiCzycholl2001,FreericksRMP2003,ChenPRB2003,HassanPRB2007,Lemanski2014,Lemanski2016}, sometimes referred to as the simplified Hubbard model. 
Recently, the FKM has been also investigated by a cluster extension of the DMFT~\cite{haldar.laad.16,haldar.laad.17a,haldar.laad.17b}.
The simplest version of the FKM describes spinless electrons interacting with localized ions via only the local (on-site) Coulomb coupling $U$.

So far, the FKM has been used to describe various effects, such as crystal formation, mixed valence, metal-insulator phase transition, and so on, e.g., Refs.~\cite{FalicovPRL1969,GeorgesRMP1996,ImadaRMP1998,FreericksRMP2003}. 
In particular, in Refs.~\cite{Lemanski2014,Lemanski2016} there were analyzed exactly properties of this model (in $D\rightarrow \infty$) related to the order-disorder phase transition caused by a rise of temperature $T$ and the associated to it insulator-conductor transition.
The results reported  in \cite{Lemanski2014} show that the rising $T$ causes an evolution of the density of states (DOS) consisting, e.g., in formation of additional bands within the main energy gap.
It turns out that above a certain $U$-dependent temperature $T_{MI}(U)$, $0 \leq T_{MI}(U) \leq T_{OD}(U)$, where $T_{OD}(U)$ is the order-disorder transition temperature, the DOS at the Fermi level $\rho (\epsilon_F)$ becomes positive (in the ordered phase).
Moreover, it was fix there the value $U_{cr}$, which points out the quasi-quantum critical point, for which $T_{MI}(U_{cr}) = 0$, what means that if $U=U_{cr}$, then $\rho (\epsilon_F)> 0$ for any $T> 0$ (the details are also discussed in the subsequent sections of this work).

The results obtained for the simplest FKM have prompted us to investigate the extended FKM (EFKM), which, in addition to the local interactions, also includes nonlocal couplings represented by the Coulomb's repulsion force $V$ between electrons located on neighboring sites of the crystal lattice. 
As it is well known, in some systems this effect can be quite significant and sometimes it can lead even to a change in nature of the metal-insulator phase transition \cite{LemanskiPRB2017,AmaricciPRB2010,KapciaPRB2017}.

The effects of Coulomb's interactions between electrons located on neighboring lattice sites have already been intensively investigated for the extended Hubbard model, e.g., Refs.~\cite{AmaricciPRB2010,AyralPRL2012,AyralPRB2013,HuangPRB2014,GiovannettiPRB2015,KapciaPRB2017,%
TerletskaPRB2017,AyralPRB2017,KapciePRE2017,TerletskaPRB2018} and references therein, while for the EFKM only very few results have been reported so far \cite{DongenPRB1992,DongenAP1997,GajekJMMM2004,Farkasovsky2008,HamadaJPhysSocJap2017,LemanskiPRB2017}. 
As far as we know, exact results for the EFKM were only reported in Refs. \cite{DongenPRB1992,DongenAP1997,LemanskiPRB2017} for $D \rightarrow \infty$,
but Refs.~\cite{DongenPRB1992,DongenAP1997} only refer to the limiting cases of $U \rightarrow 0$ and $U \rightarrow \infty$ and they did not attain the lowest temperatures.

In our previous work \cite{LemanskiPRB2017} we obtained the exact solution of the EFKM, but only in the ground state and discussed a few properties at infinitesimally small temperatures.
In this work we report our exact results also for the EFKM, but those obtained at finite temperatures.
This paper can be also viewed as an extension of the work \cite{Lemanski2014}, where the simplest version of the FKM was investigated at finite temperatures. 
Our findings end up with the resulting phase diagram of the EFKM at finite temperatures for a wide range of interaction parameters $U$ and $V$.

Here, we need to emphasize that the inclusion of repulsion between electrons located on neighboring sites significantly increases the level of difficulty in studying the system. 
This is because in the ordered phase all calculated physical quantities are then expressed not only explicitly by order parameter $d$, as it is in the case of the simplest version of the FKM, but also by additional 
$d$-dependent parameter $d_1(d)$, which for given $d$ has to be determined from the self-consistent equation (all details and the precise definitions of  parameters $d$ and $d_1$ are given in the next paragraphs).
In the ground state the task is relatively simple because then $d=1$, and for given values of $U$ and $V$ parameter $d_1$ has a certain fixed value. 
But at finite temperature it becomes challenging, as then both $d$ and $d_1$ depend on $T$.

The rest of the paper is organized as follows. 
In Sec.~\ref{sec:modelandmethod} the model considered is introduced (Sec.~\ref{sec:sub:EFKM}), and the
equations for Green's functions are determined with dynamical mean field theory (Sec.~\ref{sec:sub:DMFT}). 
Section~\ref{sec:results} is devoted to a discussion of analytical (Sec.~\ref{sec:sub:exactformulas}) and numerical solutions at $T>0$, including phase diagrams of the model (Sec.~\ref{sec:sub:phasediagrams}). 
Finally, in Sec.~\ref{sec:conclusions} the results of this work are summarized and the conclusions are provided.

\section{Model and method}
\label{sec:modelandmethod}

\subsection{Extended Falicov-Kimball model}
\label{sec:sub:EFKM}

Here, we study the same Hamiltonian $\hat{H}$ as was used in Ref. \cite{DongenPRB1992} and then in Ref. \cite{Lemanski2014} on the Bethe lattice. 
It includes electrons' kinetic energy term $\hat{H}_t$, Coulomb interaction terms (onsite $\hat{H}_U$ and intersite $\hat{H}_V$) and also $\hat{H}_{\mu}$ term representing an influence of the chemical potential.
Thus, the considered Hamiltonian has the following form:
\begin{eqnarray}
\label{eq:ham}
\hat{H} = \hat{H}_t + \hat{H}_U + \hat{H}_V + \hat{H}_{\mu},
\end{eqnarray}
where
\begin{eqnarray*}
\hat{H}_t &=& \frac{ t}{\sqrt{Z}}\sum\limits_{\left\langle i,j \right\rangle}\left(\hat{c}^+_{i\downarrow} \hat{c}^{\ }_{j\downarrow}+\hat{c}^+_{j\downarrow} \hat{c}^{\ }_{i\downarrow}\right),
\quad \hat{H}_U = U\sum\limits_{i}\hat{n}_{i\uparrow}\hat{n}_{i\downarrow},  \\
\hat{H}_V&=&\frac{2V}{Z}\sum\limits_{\left\langle i,j \right\rangle,\sigma,\sigma '}\hat{n}_{i\sigma}\hat{n}_{j\sigma '},
\quad \quad \quad \hat{H}_{\mu}=-\sum\limits_{i,\sigma}\mu_{\sigma}\hat{n}_{i\sigma},
\end{eqnarray*}
with $Z$ being the coordination number.
$\hat{n}_{i\sigma}=\hat{c}^+_{i\sigma}\hat{c}^{\ }_{i\sigma}$ is the occupation number and $\hat{c}^+_{i\sigma}$ ($\hat{c}^{\ }_{i\sigma}$) denotes the creation (annihilation) operator of an electron with spin $\sigma=\uparrow,\downarrow$.
Electrons with spin $\sigma=\uparrow$ ($\downarrow$) are localized (itinerant), that is why here we call them ions (electrons), respectively.
The prefactors in $\hat{H}_t$ and $\hat{H}_V$ have been chosen such that they yield a finite and non-vanishing contribution to the free energy per site in the limit $Z\rightarrow\infty$.
$\left\langle i,j \right\rangle$ denotes the sum over nearest-neighbor pairs.
At half-filling, i.e., for $n=1$ ($n=  \tfrac{1}{L} \sum_{i,\sigma} \left\langle \hat{n}_{i\sigma} \right\rangle $, $L$ is the number of lattice sites) the chemical potential $\mu$ for the both types of electron is given by $\mu \equiv \mu_\sigma = \tfrac{1}{2}U+2V$ (and $n_\sigma=1/2$ for both $\sigma=\uparrow,\downarrow$, $n_\sigma=  \tfrac{1}{L} \sum_{i} \left\langle \hat{n}_{i\sigma} \right\rangle $) \cite{DongenPRB1992}.

Note that in this work the model is analyzed on the Bethe lattice, which is alternate one, i.e., it can be divided into two equivalent sublattices.
We take $t$ as an energy unit, i.e., $t=1$, and basically  interaction couplings $U$ and $V$, temperature $k_BT$ ($k_B$ denotes the Boltzmann constant), gap at the Fermi energy $\Delta(\varepsilon_F)$, and energies $\varepsilon$ are given in units of $t$.
Nevertheless, for a clarity, $t$ will be explicitly given in some expressions (similarly like in Ref.~\cite{LemanskiPRB2017}).  
We also assume that $V$ interaction is repulsive, i.e., $V>0$.

\subsection{Dynamical mean field theory}
\label{sec:sub:DMFT}

The dynamical mean field theory (DMFT) enables exact studies of the correlated electron systems, including EFKM,
in the high-dimension limit \cite{DongenPRL1990,GeorgesRMP1996,FreericksRMP2003,FreericksBook2006,KotliarRmp2006,Lemanski2014,LemanskiPRB2017}.
Moreover, it was proven that the nonlocal interaction term $V$ can be treated at the Hartree level because the exchange (Fock) and the correlation energy due to the intersite term are negligible in that limit \cite{MullerHartmannZPB1989,mullerhartmann1989IJMPB,MetznerVollhardtPRL1989,MetznerZPhysB1989}.

The basic quantity calculated within the DMFT is the retarded Green's function $G(z)$, which is defined
for the complex $z$ with $\textrm{Im}(z)>0$.
Due to the fact that we are dealing with the system composed of two sublattices, we need to determine two Green's functions $G^+$ and $G^-$ separately for ``$+$'' and ``$-$'' sublattice.
Here, we use the Green's functions derived by van Dongen for the EFKM on the Bethe lattice in the limit of large dimension \cite{DongenPRB1992}.
The formulas  [for $t=1$ and at half-filling (i.e., for $\mu=U/2+2V$)] have the following forms:
\begin{eqnarray}
       G^+(z)=\frac{z+v+\frac{1}{2}Ud -G^-(z)}{[z+v+\tfrac{1}{2}U -G^-(z)]
       [z+v-\tfrac{1}{2}U -G^-(z)]} \nonumber \\
\label{eq2B2:greenfuntions}	\\
       G^-(z)=\frac{z-v-\frac{1}{2}Ud -G^+(z)}{[z-v+\tfrac{1}{2}U -G^+(z)]
       [z-v-\tfrac{1}{2}U -G^+(z)]}, \nonumber
\end{eqnarray}
where $v=V(d+d_1)$.
In such an approach we have two parameters $d$ and $d_1$, which need to be determined self-consistently.
$d$ stands for the order parameter (because it is obtained by minimalization of the free energy), which is equal to the difference in mean occupancies of the localized electrons on sublattices '$+$' and '$-$':
\begin{equation}
\label{eq:defpard}
 d = n_{\uparrow}^{+} - n_{\uparrow}^{-},
\end{equation}
whereas $d_1$ is the difference of the mean occupancies of the itinerant electrons on the both sublattices:
\begin{equation}
\label{eq:defpard1}  
 d_1 = n_{\downarrow}^{+} - n_{\downarrow}^{-}.
\end{equation}
In the above definitions $ n_{\sigma}^{\alpha} := \langle \hat{n}_{i\sigma}\rangle $ for any $i\in \alpha$, where $\alpha=+,-$ denotes the sublattice index.
In fact, $d$ and $d_1$ are not independent quantities, as for a given temperature $T$ and for a given parameter $d$ the value of $d_1$ can be determined unambiguously (excluding the case of coexistence of two phases at the first-order transition points, as it is discussed further). 
However, $d$ needs to be found from the condition for a minimum of the free energy.

Notice that due to the equivalence of two sublattices in the Bethe lattice the solution with parameters $d$ and $d_1$ is equivalent to the solution in which the parameters have the opposite signs (i.e., in which they are equal to $-d$ and $-d_1$, respectively).
As a consequence, in the rest of the paper, we consider the solutions with $d\geq0$ only. 
With such a choice, as it will be shown further, solutions with both signs of $d_1$ can be found depending on values of the interactions and the temperature. 

The mentioned parameters can be associated to charge polarization $\Delta_Q$ and staggered magnetization $m_Q$ by the following relations:  

\begin{equation}
\label{eq:defDeltaQmQ}
 \Delta_Q = \tfrac{1}{2}\left(d+d_1\right) \quad \textrm{and} \quad m_Q=\tfrac{1}{2}\left(d-d_1\right),
\end{equation}%
which create a different, but equivalent, set of parameters.
For $d>0$ one gets that $\Delta_Q>0$ and $m_Q>0$ because $d>|d_1|$ for any finite value of $U$ or $V$ (for $d=0$ one always gets $d_1=0$, hence also $\Delta_Q=0$ and $m_Q=0$). 
For calculations presented in this work it is not relevant, which combination [$(d,d_1)$ or $(\Delta_Q,m_Q)$] is used.
Note also that any of $d$, $d_1$, $\Delta_Q$, and $m_Q$ reflects breaking of the system in the ordered phases mentioned below.
Additionally, a use of $d$ and $d_1$ ensures a simple correspondence to previous works on the FKM \cite{DongenPRL1990,FreericksPRB1999,ChenPRB2003,MaskaPRB2006,HassanPRB2007,Lemanski2014} and the EFKM \cite{DongenPRB1992,LemanskiPRB2017}.

At $T=0$, i.e., for $d=1$, the solutions of the set of equations (\ref{eq2B2:greenfuntions}) can be written in a simple analytical form, because then the system reduces to the quadratic equation for $G^{+}(z)$ or $G^{-}(z)$. 
The formulas for $G^\pm(z)$ at the ground state have the following form:
\begin{equation}\label{eq:GSgreenfunk}
G^{\pm}(z) = \frac{4z^2-A^2-\sqrt{(4z^2-A^2)(4z^2-A^2-16)}}{4(2z \pm A)},
\end{equation}
where $A=2V(1+d_1)-U$. 
Their analyses are reported in Ref. \cite{LemanskiPRB2017}. 
However,  finding solutions of (\ref{eq2B2:greenfuntions}) at arbitrary temperature $T>0$ is equivalent to finding of roots of the polynomial: of the third rank in the disordered phase (when $d=0$ and $d_1=0$, $G^{+}=G^{-}$, and it is independent on $V$) \cite{Lemanski2014,HubbardPRSLA1964,VelickyPR1968,DongenAP1997} and of the fifth rank in a general case, when $0<d<1$. 
So, even though we have no simple analytical formulas for $G^{\pm}(z)$ when $T>0$, we are able to get very precise numerical values on $G^+(z)$ and $G^-(z)$ for any $T$ and any values of $U$ and $V$.

For $V=0$, the formulas on  $G^+$ or $G^-$ and their analysis in the whole temperature region are provided in Ref.~\cite{Lemanski2014}.
In a general case of $V \geq 0$ one needs to solve the following fifth rank polynomial equation on $G^{+}$
[if we know $G^{+}$ then we can find $G^{-}$ from Eq.~(\ref{eq2B2:greenfuntions})]:
\begin{eqnarray}
W(G^+) & := & a_0 + a_{1}G^{+} + a_{2}(G^{+})^2 + a_{3}(G^{+})^3 \nonumber \\
& + & a_{4}(G^{+})^4 + a_{5}(G^{+})^5 = 0
\label{eq2B3:polynomialgreen}
\end{eqnarray}
Coefficients $a_0$, $a_1$, $a_2$, $a_3$, $a_4$, and $a_5$ are functions of $z$, $U$, $V$, $d$ and $d_1$.
Since the explicit expressions for these coefficients are rather lengthy, we present them into Appendix~\ref{sect:a}.

As can be seen from formula (\ref{eq2B2:greenfuntions}), the entire temperature dependence of the Green's functions, and thus also some other characteristics of the system that are expressed by them, such as, e.g., the DOS and the energy gap at the Fermi level, comes merely from the temperature dependence of parameters $d(T)$ and $d_1(T)$. 
Of course, all other thermodynamic characteristics of the system also depend on these functions, in addition to the explicit dependence on $T$.  
Therefore, our primary task is to determine $d(T)$ and $d_1(T)$.

The procedure of finding $d(T)$ and $d_1(T)$ is as follows. 
First, having determined $G^+$and $G^-$ we calculate the DOS functions $\rho^+$ and $\rho^-$ from the standard formulas:
 \begin{equation}
       \rho^{\pm} (U,V,d,d_1;\varepsilon)=-\frac{1}{\pi} \textrm{Im}\left[G^{\pm}(U,V,d,d_1;\varepsilon +i0)\right].
       \label{eq2B3:dos}
       \end{equation}
Then, for given parameters $U$, $V$, and $T$ we solve the self-consistent equation for $d_1$ [Eq.~(\ref{eq:defpard1})], from which we get the dependence of $d_1$ as a function of $d$.
Concentrations $ n_{\downarrow}^{\pm}$ appearing in Eq.~(\ref{eq:defpard1}) are calculated from the expression
\begin{equation}
 n_{\downarrow}^{\pm}=\int_{-\infty}^{\varepsilon_F}\frac{\rho^{\pm}(U,V,d,d_1;\varepsilon)}{1+\exp\left[\varepsilon / \left(k_{B}T\right)\right]}d\varepsilon
\label{2B6}
\end{equation}
(in our case the Fermi level is located at $\varepsilon_F=0$).

\begin{figure}
	\includegraphics[width=\rozmiarjeden]{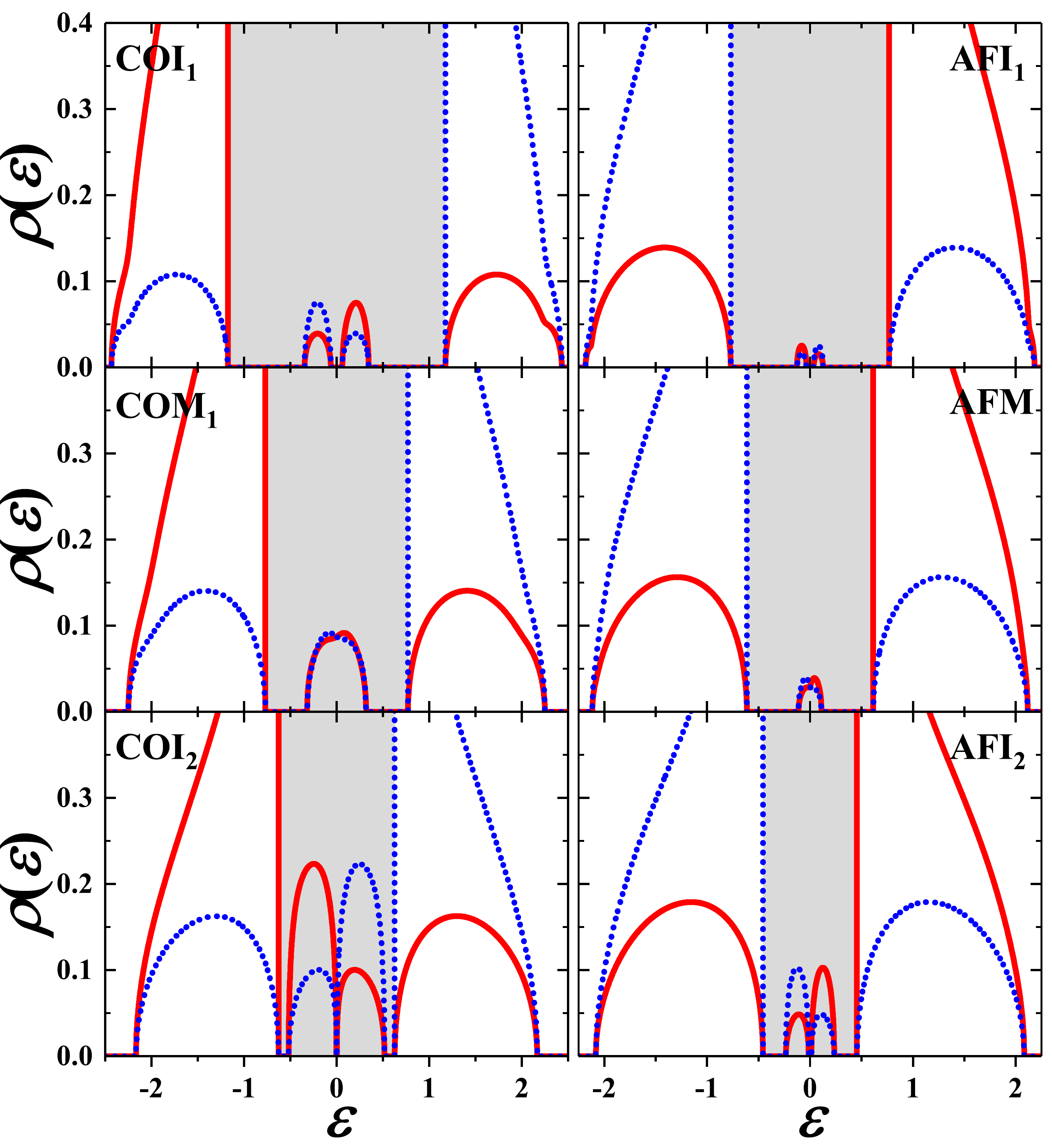}
	\caption{The itinerant electron densities of states $\rho^{+}$ (solid line) and $\rho^{-}$ (dotted line) in each sublattice 
	(i) left column: for $T=0.15$ and $V=0.1$ in phases with $d_1>0$: 
	{\COIPA} ($U=-2.0$, $d=0.949$, $d_1=0.801$), 
	{\COMPA} ($U=-1.5$, $d=0.698$, $d_1=-0.630$), and 
	{\COIPB} ($U=-1.0$, $d=0.724$, $d_1=0.502$), 
	from the top, respectively;
	(ii) right column: for $T=0.05$ and $V=0.1$ in phases with $d_1<0$: 
	{\COINA} ($U=1.6$, $d=0.994$, $d_1=-0.701$),	 
	{\COMNA} ($U=1.3$, $d=0.987$, $d_1=-0.630$), and 
	{\COINB} ($U=1.0$, $d=0.947$, $d_1=-0.524$), 
	from the top, respectively.
	The Fermi level is located at $\varepsilon=\varepsilon_F=0$.
	The gray shadows indicate schematically the principal gap between the main (lower and upper) bands, where the subbands appear at $T>0$. 
	In all insulators $\rho^{\pm}(\varepsilon_F)=0$ and $\Delta(\varepsilon_F)>0$.
	\label{fig:DOSfull1}%
		}
\end{figure}

Next, we insert function $d_1(T,d)$ into the expression for free energy $F$ given by Eqs. (\ref{eq2B1}), (\ref{eq2B1a}), and (\ref{eq2B1b}), and finally, we minimize $F$ over $d$, from where we get $d(T)$ and then also $d_1(T)$ from Eqs. (\ref{eq:defpard1}) and (\ref{2B6}).

\section{Results}
\label{sec:results}

The model predicts the existence of a variety of ordered and nonordered (NO) phases.
The full phase diagram of the model is quite complex.
In particular, the diagram exhibits five distinguishable regions in which an  ordered insulator occur ({\COIPA}, {\COIPB}, and {\COIPC} with $d_1>0$;  {\COINA} and {\COINB} with $d_1<0$). 
Three distinguishable regions of ordered metal are also present ({\COMPA} and {\COMPB} with $d_1>0$; {\COMNA} with $d_1<0$). 
Note that in these ordered phases both a charge order and an antiferromagnetic order  exist simultaneously, i.e., $\Delta_Q \neq 0$ and $m_Q \neq 0$ [excluding $U\rightarrow-\infty$ and $V\rightarrow+\infty$ limits ($U<2V$), where $m_Q\rightarrow0$ as well as $U\rightarrow+\infty$, where $\Delta_Q\rightarrow0$].
Moreover, the nonordered metal ({\FLI}) is found to be stable is some range of the model parameters. 
This {\FLI} is not the Fermi liquid phase due to a huge scattering rate 
that is maintained even when approaching 
the ground state (for details for the FKM see Refs. \cite{SiPRB1992,DongenAP1997,FreericksRMP2003}).
Finally, two separated ranges of the model parameters are determined, where the nonordered insulator (denoted here as {\MI}) occurs.
In the {\FLI} and the {\MI} phases $d=0$ and $d_1=0$ (or $\Delta_Q=0$ and $m_Q=0$, equivalently).
One should be also aware of the fact that the {\MI} of the FKM has a quite different nature than the Mott insulator described by the HM \cite{FreericksRMP2003,PhilippEPJB2017}.

In this paper all phases with $d_1>0$, i.e., the phases with dominant charge order ($\Delta_Q>m_Q$), are  
named as charge-ordered (CO) ones
and all phases with $d_1<0$, i.e., the phases where antiferromagnetism dominates ($m_Q>\Delta_Q$), are 
named as antiferromagnetic (AF) ones (note that we assumed that $d\geq0$, cf. Sec.~\ref{sec:sub:DMFT}).
The lower index (which can be $1$, $2$, and $3$) labels the phases with the same properties (details are given further in the text).
We also introduced the denotations {\MIP} and {\MIN} to distinguish two region of the {\MI} occurrence for $U<-2$ and $U>2$, respectively. 
In the following sections we characterize the properties of each phase and construct the full phase diagram of the model.

\subsection{Exact formulas for finite temperatures}
\label{sec:sub:exactformulas}

\subsubsection{Density of states}
\label{sec:sub:sub:DOS}

In order to be able to perform accurate calculations, it is extremely important to precisely determine the boundaries of energy bands. 
In our case, this is possible thanks to the polynomial form of the equation for the Green function, as it is expressed in Eq. (\ref{eq2B3:polynomialgreen}). 
Indeed, the simultaneous solution of Eq. (\ref{eq2B3:polynomialgreen}) and the equation $dW(G^{+})/dG^{+}=0$, which can be written explicitly as
\begin{eqnarray}
\frac{dW(G^+)}{dG^+} & := & a_{1}+2a_{2}(G^{+}) + 3a_{3}(G^{+})^2 \nonumber \\
& + & 4a_{4}(G^{+})^3+5a_{5}(G^{+})^4=0
\end{eqnarray}
allows us (at fixed values $U$, $V$, $d$ and $d_1$)  to obtain energies at which the edges of energy bands occur.
Coefficients  $a_1$, $a_2$, $a_3$, $a_4$, and $a_5$ are defined in Appendix~\ref{sect:a}.

\begin{figure}
	\includegraphics[width=\rozmiarjeden]{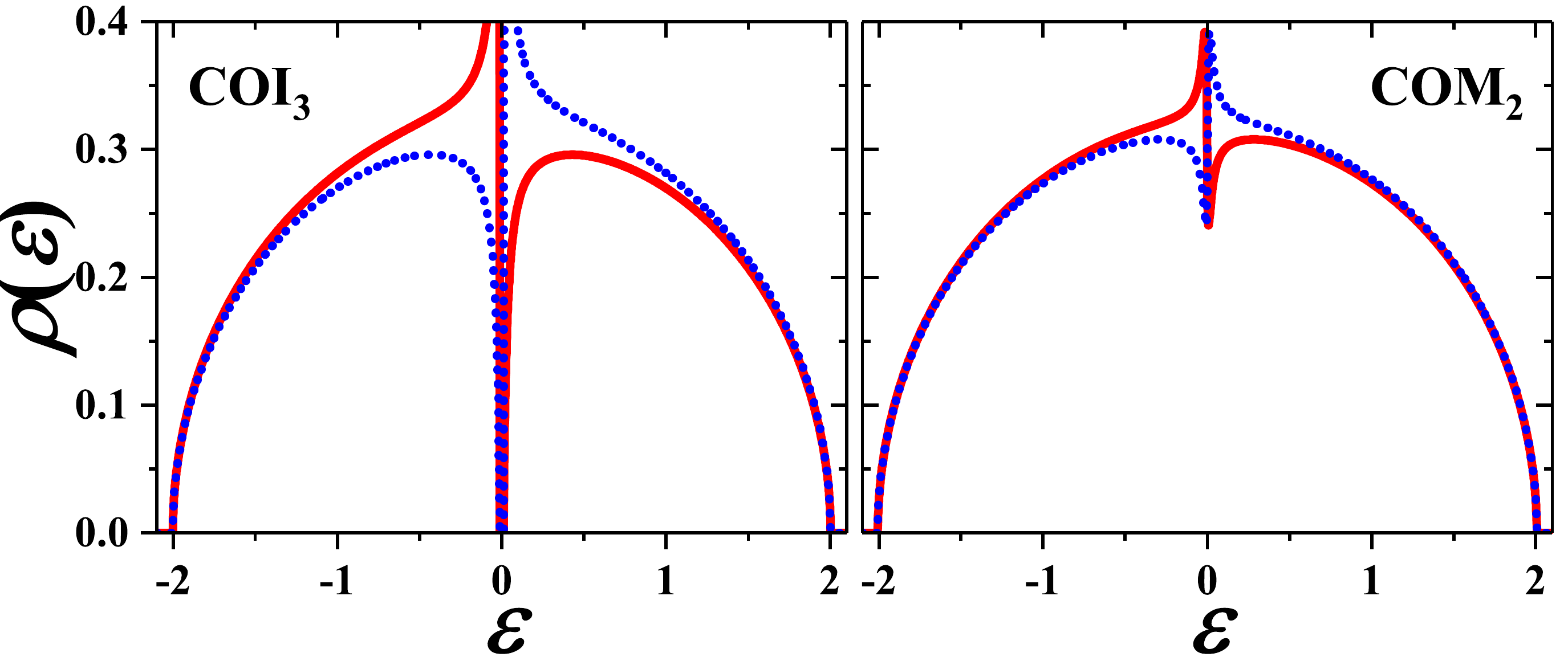}
	\caption{The itinerant electron densities of states $\rho^{+}$ (solid line) and $\rho^{-}$ (dotted line) in each sublattice for $U=0.16$ and $V=0.1$ for phases with $d_1>0$: 
	{\COIPC} ($T=0.040$, $d=0.799$, $d_1=0.051$; left panel) and {\COMPB} ($T=0.047$, $d=0.234$, $d_1=0.013$; right panel).
	The Fermi level is located at $\varepsilon=\varepsilon_F=0$.
	$\rho^{\pm}(\varepsilon_F)=0$ and $\Delta(\varepsilon_F)>0$ in the {\COIPC} .
	\label{fig:DOSfull2}%
		}
\end{figure}

The exact formula on the density of states (DOS) at the Fermi level $\rho (\varepsilon_F)=\left(\rho^+(\varepsilon_F) + \rho^-(\varepsilon_F)\right)/2$ is expressed in Eq.~(\ref{eqDOS_FermiLevel}) below through parameters $U$, $V$, $d$ and $d_1$. 
\begin{equation}
\rho (\varepsilon _F) = \left\{
\begin{array}{r}
\frac{1}{\pi}  \quad \text{if } U=0 \text{ and } (d+d_1)V=0 \\
  \\ \quad
\frac{1}{\pi}\left|\frac{\textrm{Im}\sqrt{w(U,V,d,d_1)}}{8 [(d + d_1)V]^2 -2U^2}\right|  \quad \text{otherwise},
\end{array} \right.
\label{eqDOS_FermiLevel}
\end{equation}
where
\begin{eqnarray}
& & w(U,V,d,d_1) = U^2 (U^4 - 4 U^2 +4d^2) \nonumber \\
& & +  8 (U^2-2) d U (d + d_1) V - 8 ( U^4 - 2U^2 -2 ) [ ( d + d_1 ) V ]^2  \nonumber \\
& & -  32 d U [(d + d_1)V]^3 + 16 U^2 [ ( d + d_1 ) V ]^4. 
\end{eqnarray}
It applies throughout the entire temperature range, but only through parameters $d (T)$ and $d_1(T)$.
Consequently, in the disordered phase with $d=0$ and $d_1=0$ it does not change with temperature.
Notice also that, in a general case of any value of $\varepsilon$, it is not possible to do determine  expression for $\rho(\varepsilon)$ analogous to (\ref{eqDOS_FermiLevel}).

In Figs.~\ref{fig:DOSfull1} and~\ref{fig:DOSfull2} DOS functions $\rho^{+}(\varepsilon)$ and $\rho^{-}(\varepsilon)$ [$\rho^\pm (\varepsilon ) \equiv  \rho^\pm( U,V,d, d_1; \varepsilon )$ evaluated from Eq.~(\ref{eq2B3:dos})] in all found distinguishable ordered phases are presented for several representative values of the model parameters.
The phases presented in Fig.~\ref{fig:DOSfull2} can occur only if $V>0$, whereas those shown in Fig.~\ref{fig:DOSfull1} are present on the phase diagram also for $V=0$ (cf. also Refs.~\cite{Lemanski2014,Lemanski2016}).
Note that $\rho^+(\varepsilon_F)=\rho^-(\varepsilon_F)=0$ in all insulating phases and the gap at the Fermi level $\Delta(\varepsilon_F)$ is finite (but small, and thus not clearly visible in the figures).
At the half-filling, it turns out that  $\rho^{+}(\varepsilon)=\rho^{-}(-\varepsilon)$ and the total (resultant) DOS $\rho(\varepsilon)=(1/2)\left[\rho^{+}(\varepsilon)+\rho^{-}(\varepsilon)\right]$ is symmetric, but $\rho^{+}(\varepsilon)$ and $\rho^{-}(\varepsilon)$, when considered separately, are not symmetric. 
The obvious observation is that in phases with $d_1>0$ the $\rho^+$ has the larger weight than $\rho^-$ in the main band below the Fermi level (the lower main band)  due to the fact that $n_\downarrow^+>n_\downarrow^-$.
Namely, $\rho^+$ has lower weight in its upper main band than in the lower main band. 
Likewise, $\rho^-$ has greater weight in its upper main band than in the lower main band.
In all phases with $d_1<0$ the relation between weights in main bands is reversed.

\begin{figure}
	\includegraphics[width=\rozmiarjeden]{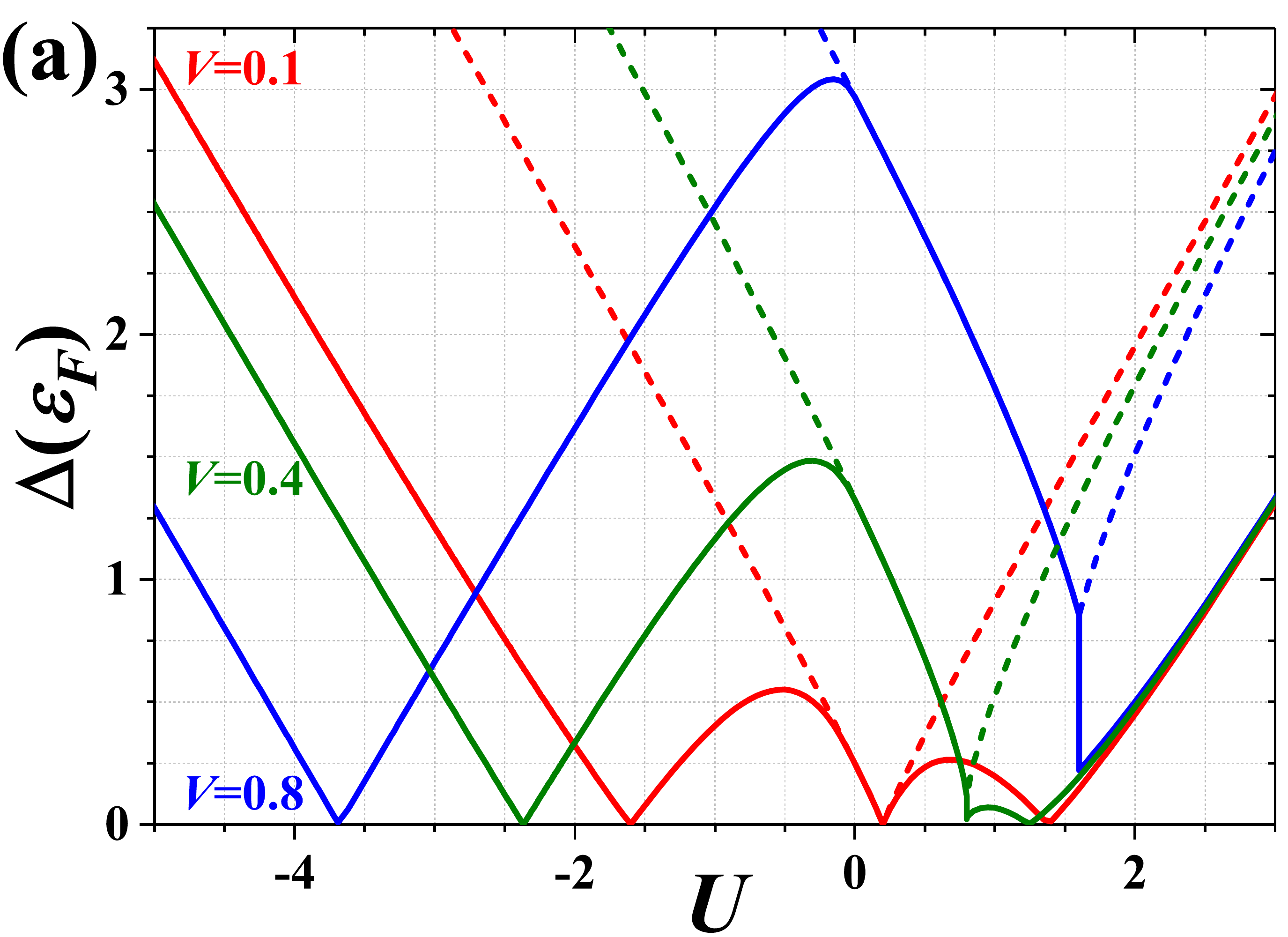}\\
	\includegraphics[width=\rozmiarjeden]{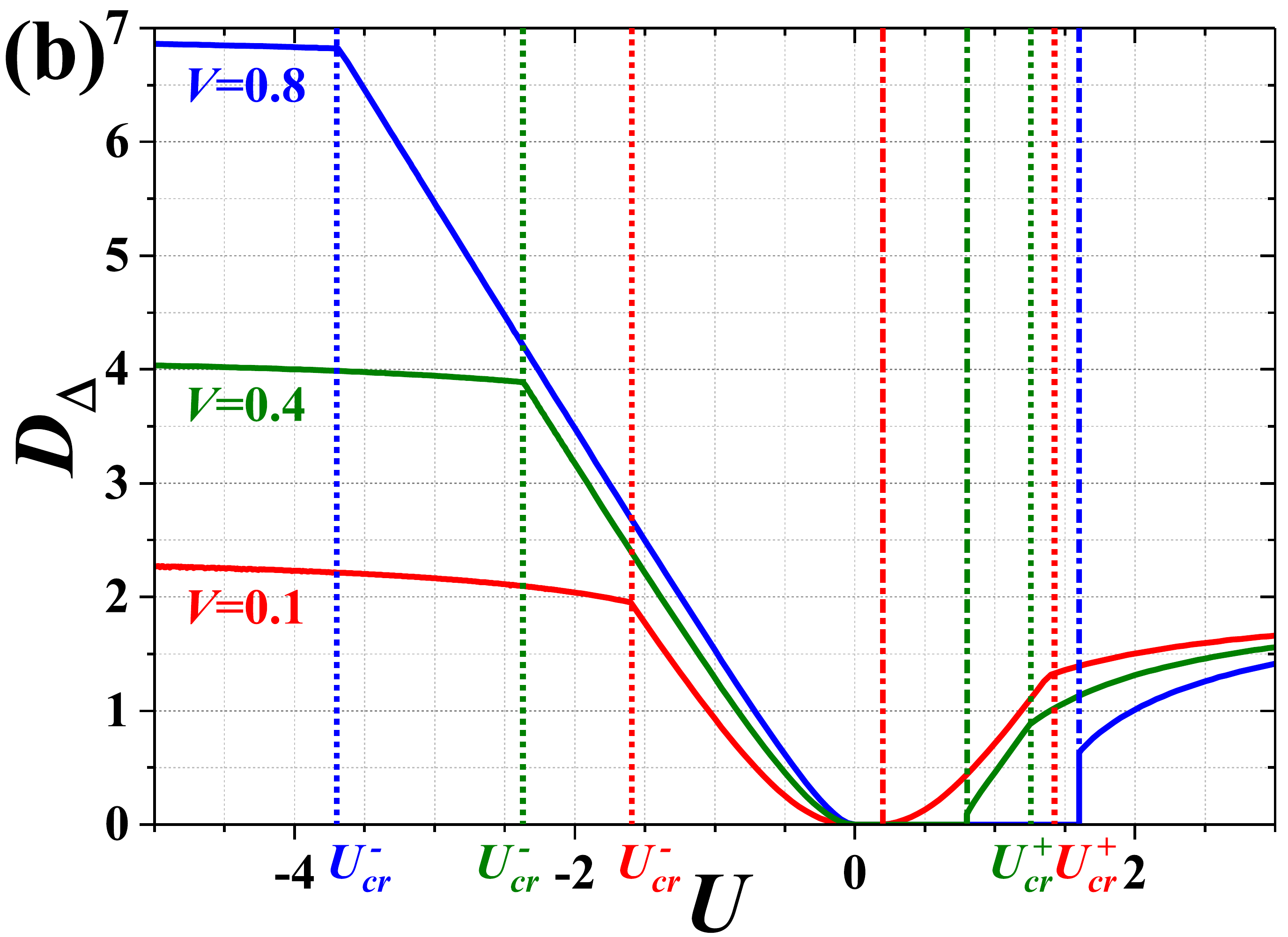}
	\caption{%
	(a) Energy gap $\Delta (\varepsilon _F)$ at the Fermi level as a function of $U$ at $T=0$ [i.e., $\Delta(\varepsilon_F)_{T=0}$, dashed lines] and in the limit $T \rightarrow 0^+$ [i.e., $\Delta(\varepsilon_F)_{T \rightarrow 0^{+} }$, solid lines] for $V=0.1$, $V=0.4$, and $V=0.8$ (as labeled).
	(b) Difference $D_{\Delta} = \Delta(\varepsilon_F)_{T \rightarrow 0^{+}}-\Delta(\varepsilon_F)_{T=0}$ for the same values of $V$ is shown.
	The vertical dotted and dashed-dotted lines correspond to location (at $T=0$) of the quasi-critical points ($U^{\pm}_{cr}$) and of the discontinuous transition for $U=2V$, respectively (for each value of $V$ from the figure, correspondingly). 
	}
\label{EnGapV01and08}
\end{figure}

The behavior of the itinerant electrons in the phases collected in Fig.~\ref{fig:DOSfull1} is associated with the specific features of the subbands, which arise inside the principal energy gap between the main bands (the region indicated by the gray shadow in Fig.~\ref{fig:DOSfull1}).
These subbands can be present only at $T>0$ (i.e., for $d<1$), whereas at $T=0$ (i.e., for $d=1$) the  DOS exhibits only two main bands separated by the principal gap at the Fermi level (cf.  Fig.~1 of Ref.~\cite{LemanskiPRB2017}).  
In particular, the characteristic feature of the  {\COIPA} which distinguishes it from the {\COIPB} is that $\rho^+$ ($\rho^-$) has lower (higher) weight in the subband lying below $\varepsilon_F$ than  in the subband located above $\varepsilon_F$.
In contrast, in the {\COIPB} $\rho^+$ ($\rho^-$) has lower (higher) weight in the subband above $\varepsilon_F$ than in the subband situated below $\varepsilon_F$.
The relation between weights in the subbands of $\rho^+$ (or $\rho_-$) in the {\COINA} is the same as in the {\COIPB}, whereas those in the {\COINB} are such as in the {\COIPA}.
One can also say that in the {\COIPA} and the {\COINA} the fillings of the main bands and of the subbands are inverted for both $\rho^+$ and $\rho^-$, i.e., the  weights in $\rho^+$ ($\rho^-$) of the subbands below and above the Fermi level are in the opposite relation than the weights in the main lower and upper bands of $\rho^+$ ($\rho^-$).

The metallic behavior of the {\COMPA} and the {\COMNA} is a result of merging of the subbands near the Fermi level, as it is visible in Fig.~\ref{fig:DOSfull1}, but the precursors of the subbands can be still visible in the DOSs of these phases (at least in the regions of their occurrence in the neighborhood of boundaries with the COIs or the AFIs, respectively, on the phase diagram). 
In these metallic phases the continuous change from a case of inverted subbands (in the {\COIPA} and the {\COINA}, in the sense discussed above)  to a case of ``noninverted'' subbands (in the {\COIPB} and the {\COINB}) occurs. 
In higher temperatures, the central subband with nonzero weight at $\varepsilon_F$ can merge with the main bands (cf. also Fig.~2 of Ref.~\cite{Lemanski2014}).

The DOSs  in the {\COIPC} and the {\COMPB} are shown in Fig.~\ref{fig:DOSfull2}.
In contrast to the DOSs discussed above, the structure of the DOS in the {\COIPC} consists only of two main bands and does not exhibit any subbands (also at $T>0$).
The metallic behavior of the {\COMPB} is associated with closing of the gap between the main bands.

To be rigorous, one should also add that for large $|U|$ in the {\COIPA} and the {\COINA} two additional subbands can appear in the DOS, one below the lower main band and one above the upper main band.
Its precursors are visible in Fig.~\ref{fig:DOSfull1} (cf. also Fig.~3 of Ref.~\cite{Lemanski2014}). 
In Fig.~1 of Ref.~\cite{Lemanski2014} the DOSs in the non-ordered phases, i.e., in the {\FLI} and in the {\MI}s, are also presented.

\subsubsection{Energy gap at the Fermi level}
\label{sec:sub:sub:energygap}

It appears, that the energy gap at the Fermi level $\Delta (\varepsilon_{F})$ is a continuous function at $T=0$ only within the interval $0 \leq U \leq 2V$, but it is discontinuous both for $U<0$ and for $2V<U$.
The continuity of $\Delta (\varepsilon_{F})$ at $T=0$ for $0 \leq U \leq 2V$ is due to the fact,
that no subbands are formed inside the principal energy gap when the temperature is raised above $T=0$ (the {\COIPC}, cf. Sec.~\ref{sec:sub:sub:DOS}). 
Indeed, at $T=0$ one has $\Delta (\varepsilon _F)_{T=0}=|2V(1+d_1)-U|$ \cite{LemanskiPRB2017},
but in the limit of $T \rightarrow 0^+$ we get the following formulas:
\begin{widetext}
\begin{equation}
\Delta (\varepsilon _F)_{T \rightarrow 0^{+}} = \left\{
\begin{array}{ll}
\left|2V(1+d_1)-U\right| & \quad \text{if } 0 \leq U < 2V  \\
&  \\
\left|\left(\sqrt{1 + 4 U^2 - 4 (1 + d_1) UV + 4 (1 + d_1)^2 U^2 V^2}-U^2-1\right)/U\right| & \quad \text{if } U < 0 \quad  \text{or}\quad 2V < U.
\end{array} \right.
\label{eq2B2}
\end{equation}
And for the special case of $U=2V$, when two ordered phases with two different energy gaps coexist at $T=0$ (a point of a discontinuous transition \cite{LemanskiPRB2017}), are two solutions:
\begin{equation}
\Delta (\varepsilon _F;U=2V)_{T \rightarrow 0^{+}} = \left\{
\begin{array}{ll}
Ud_1 & \quad \text{if } d_1 > 0  \\
&  \\
\left|\left(\sqrt{1 + 4 U^2 - 2 (1 + d_1) U^2 + (1 + d_1)^2 U^4}-U^2-1\right)/U\right| & \quad \text{if } d_1 < 0.
\end{array} \right.
\label{eq2B2a}
\end{equation}
\end{widetext}
In the above formulas [i.e., Eqs.~(\ref{eq2B2}) and (\ref{eq2B2a})] parameter $d_1$ is taken at $T=0$.
The behavior of $\Delta (\varepsilon _F)$ for a few representative values of $V$ is illustrated in Fig.~\ref{EnGapV01and08}(a). 
Figure~\ref{EnGapV01and08}(b) shows difference $D_{\Delta} = \Delta(\varepsilon_F)_{T \rightarrow 0^{+}}-\Delta(\varepsilon_F)_{T=0}$ as a function of $U$ for the same values of $V$.

It turns out that if $U$ lies outside of the region $0\leq U\leq 2V$, then $\Delta (\varepsilon _F)_{T\rightarrow 0^{+}}$ is always smaller than $\Delta (\varepsilon _F)_{T=0}$ and it attains zero at the quasi-quantum critical points $U_{cr}^{\pm}$. 
There are two such points for $V\lesssim 0.54$, one positive ($U_{cr}^{+}$) and another negative ($U_{cr}^-$), but there is only one negative $U^{-}_{cr}$ for $V\gtrsim 0.54$ (cf. also Ref.~\cite{LemanskiPRB2017}).
At the first order phase transition point at $T=0$, i.e., for $U=2V$, $\Delta (\varepsilon _F)_{T \rightarrow 0^{+}}$  exhibits discontinuous jump, but $\Delta (\varepsilon _F)_{T=0}$ is continuous.
For large positive $U$ (i.e., $U>U_{cr}^{+}$ if $V \lesssim 0.54$ and $U>2V$ if $V \gtrsim 0.54$, in the {\COINA}) or  large negative $U$ ($U<U^-_{cr}$, in the {\COIPA}), $\Delta (\varepsilon _F)_{T \rightarrow 0^{+}}$ as a function of $U$ is a monotonic function of $U$ increasing with $|U|$.
Then difference $D_\Delta$ is obviously zero for $0\leq U  \leq 2V $ (in the {\COIPC}) and it exhibits discontinuity at $U=2V$.
It is a monotonous function of $U$ for $U<0$ and $U>2V$ with a change of slope at $U_{cr}^{\pm}$ (if $U_{cr}^{+}$ exists).

In the general case of finite $T$ it is possible to precisely determine the energy gap based on the knowledge of the edges of energy bands, but before that we need to determine the values of $d(T)$ and $d_1(T)$ numerically.

\subsubsection{Free energy}
\label{sec:sub:sub:freeenergy}

Total free energy $F$ of the system per site is given by the following expression
\begin{equation}
F = F_{el} + F_{ions} + U/4 + V,
\label{eq2B1}
\end{equation}
where
\begin{eqnarray}
F_{el}& = &  \int_{-\infty}^{\infty} \left\{ \rho(U,V,d,d_1;\varepsilon ) \ln\left[1+\exp\left(-\frac{\varepsilon}{k_{B}T} \right)\right] \right\} d\varepsilon \nonumber \\ 
&- & V\left(1 - d_1^2\right)/4
\label{eq2B1a}
\end{eqnarray}
and
\begin{eqnarray}
\label{eq2B1b}
F_{ions}& = & V\left(1 - d^2\right)/4 \\
& + & k_{B} T \left[\frac{1+d}{2}\ln\left(\frac{1+d}{2}\right)+\frac{1-d}{2}\ln\left(\frac{1-d}{2}\right)\right], \nonumber
\end{eqnarray}
where $\rho( U,V,d, d_1; \varepsilon )=\left(\rho^+( \varepsilon ) + \rho^-( \varepsilon ) \right)/2$ and $\rho^\pm (\varepsilon ) \equiv  \rho^\pm( U,V,d, d_1; \varepsilon )$ is expressed by Eq.~(\ref{eq2B3:dos}).

Free energy $F$ calculated using formulas (\ref{eq2B1}), (\ref{eq2B1a}) and (\ref{eq2B1b}) has a fundamental meaning here, because only after its minimization with respect to order parameter $d$ we obtain $d(T)$ and $d_1(T)$ [from Eqs. (\ref{eq:defpard1}) and (\ref{2B6})], that enter into all physical characteristics of the system.
And on the basis of such determined free energy we constructed finite temperature phase diagrams which we present in the next section.
In these diagrams, in the phases marked as insulators (conductors) the condition $\rho_F = 0$ ($\rho_F\neq 0$) is met.
Note that the term 'insulator' is used to characterize the DOS with a gap at the Fermi level. 
It is clear that, strictly speaking, such a system would be insulating only at $T=0$.

We still need to mention that we derived formula (\ref{eq2B1a}) by generalizing the expression given for the FKM in Ref.~\cite{FreericksPRB1999}.
Then we have verified that this is equivalent to the expression presented by Van Dongen in \cite{DongenPRB1992}, but has a simpler form than that in \cite{DongenPRB1992} and allow for greater precision of calculations in the whole range of the model parameters and of temperature, which is especially important near $T=0$.
This is a consequence of the fact that in the formulation of this work one does not do summation over Matsubara frequencies.
Such a summation is difficult to perform numerically because the tails of the Green's functions are vanishing slowly with the frequencies.
Certainly, both approaches are formally equivalent, as it was shown in Ref.~\cite{ShvaikaPRB2003}, 
but in practice the method we use in this work enables us to calculate all relevant physical quantities with any precision and at any temperature, what is not the case when the summations over Matsubara frequencies are performed (cf. Refs.~\cite{HassanPRB2007,Lemanski2014} for the FKM).

\subsection{Phase diagrams}
\label{sec:sub:phasediagrams}

In this subsection we will focus on the evolution of the phase diagram of the model with increasing intersite interaction $V$.
On the diagram there are a variety of transitions, thus we discuss firstly an evolution of the order-disorder transitions (Sec.~\ref{sec:sub:sub:ODtrans}).
Next, we focus on  transitions between metallic and insulating ordered phases at lower temperatures (Secs.~\ref{sec:sub:sub:d1change} and~\ref{sec:sub:sub:nod1change}).
Finally, some dependencies of important quantities at finite temperatures are presented (Sec.~\ref{sec:sub:sub:temperaturedependences}).

At the beginning let us briefly discuss the diagram for model with $V=0$.
The detailed study of the phase diagram of the FKM is contained in Refs.~\cite{Lemanski2014,Lemanski2016}.
For $V=0$ the phase diagram is symmetric with respect to $U=0$.
However, for $U<0$ quantities $d$ and $d_1$ have the same signs (and the charge order dominates over antiferromagnetism), whereas for $U>0$ they have opposite signs (and the antiferromagnetic order is dominant; note that we assumed that $d\geq0$). 
To be precise, for $V=0$ the solution  with $d$ and $d_1$ for $U$ corresponds to the solution with $d$ and $-d_1$ occurring for $-U$.

\begin{figure}
	\includegraphics[width=\rozmiarjeden]{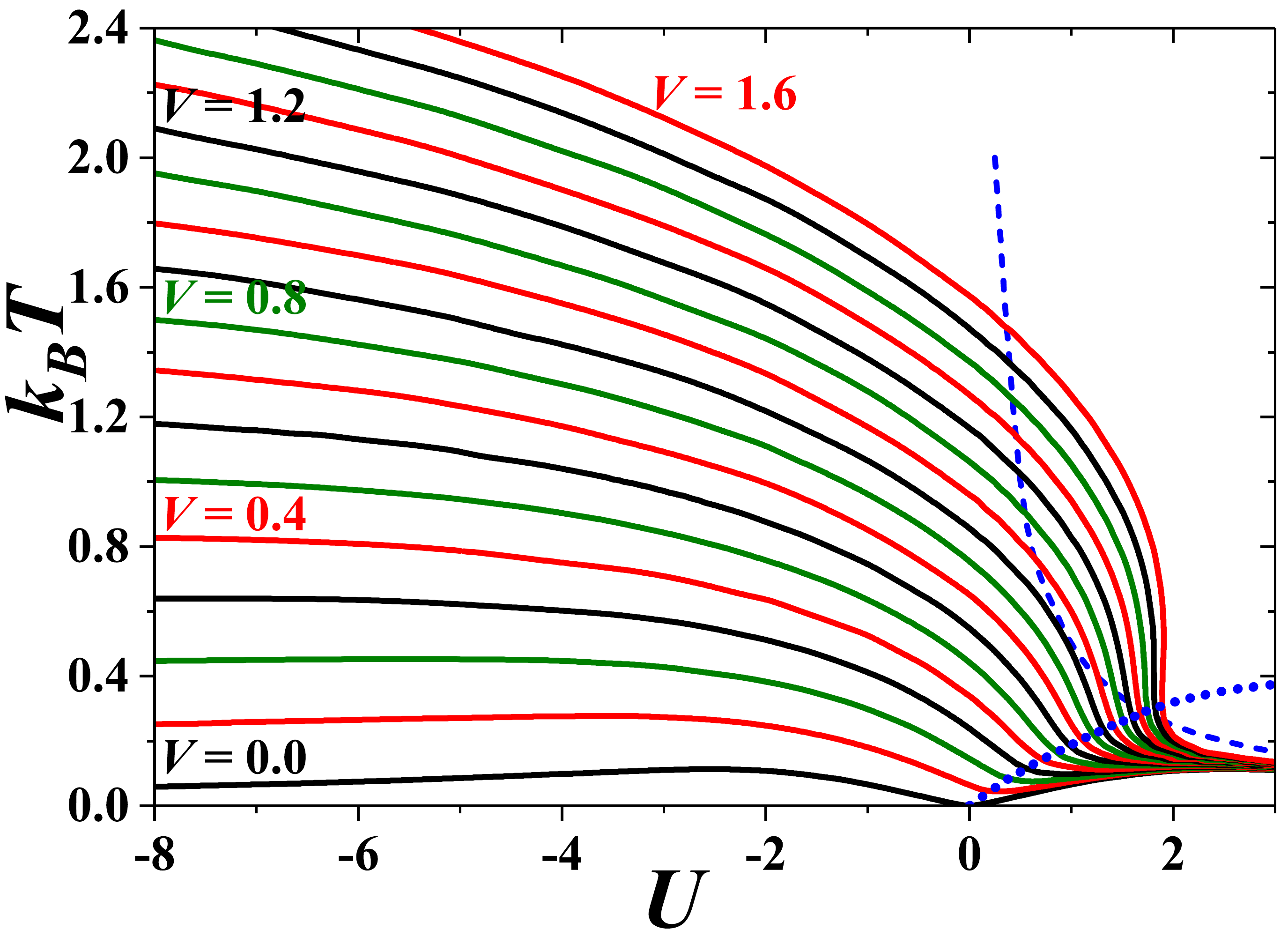}
	\caption{
	\label{fig:orderdisorder}%
		Temperature $T_{OD}$ of the order-disorder transition for different $V$ 
		(from $0.0$ to $1.6$ with a step of $0.1$).
		Only second-order lines are determined.
		The dotted line connects points at the boundaries, where parameter $d_1$ change its sign in the ordered phase below the boundary ($d_1>0$ on the left, $d_1<0$ on the right). 
		The dashed line denotes  $k_BT=1/(2U)$ dependence for large $U$. 
		Note that for large $V$ (larger that $V\approx0.7$) the order-disorder transition is first-order one for some range of $U$ (not shown in this figure).  
		}
\end{figure}

For $V=0$  a continuous order-disorder transitions  between the ordered  and the nonordered phases occurs when $T$ is raised. 
At  temperatures above the transition, two nonordered phases can exist: the {\FLI}  for $|U|<2$ and the nonordered insulator  
for $|U|>2$ ({\MIP} for $U<-2$ and {\MIN} for $U>2$).
With an increase of $|U|$, the continuous {\FLI}-{\MIP} ({\FLI}-{\MIN}) transformation occurs at $U=-2$ ($U=2$), respectively and it is not dependent on temperature.
To be precise, this is the type of metal-insulator transition predicted by Mott~(e.g., Refs.~\cite{Mott1990,Montorsi1992,Gebhard1997}), but unlike Mott's prediction that the transition would be generically discontinuous \cite{mullerhartmann1989IJMPB,DongenAP1997,FreericksRMP2003}.
The $U$-dependence of the order-disorder transition temperature  $T_{OD}$ is shown in Fig.~\ref{fig:orderdisorder} (the first solid line from the bottom is for $V=0$).
$T_{OD}$ is a nonmonotonous function of $U$.
It increases with increasing $|U|$ starting from zero at $U=0$ to reach its maximal value of $k_{B}T_{OD}\approx 0.113$ at $|U|\approx 2.61$.
With further increase of $|U|$ it decreases to zero for $U\rightarrow+\infty$.

At low temperatures (below $T_{OD}$ line) the  model exhibits six 
long-range-ordered phase, in which the orders (charge and antiferromagnetic) can coexist with the metallic or insulating behavior 
[two ordered metals: {\COMPA} and {\COMNA} (for $U<0$ and $U>0$, respectively); 
or four charge-ordered insulators: {\COIPA} and {\COIPB} (for $U<0$) as well as {\COINA} and {\COINB} (for $U>0$)]. 
The {\COMPA} ({\COMNA}) can exists only for  $0<|U|<2$ and $T>0$.
The regions of their occurrence divides the regions of the ordered insulator occurrence  into two separated areas [the {\COIPB} ({\COINB}) for $0<|U|<\sqrt{2}$ and the {\COIPA} ({\COINA}) for $|U|>\sqrt{2}$].
The {\COIPA} and the {\COIPB} for $U<0$  (the {\COINA} and the {\COINB} for $U>0$)
are differentiated by behavior of a capacity of the main lower band and a capacity of the lower subband, lying inside the main energy gap just below the Fermi level (cf. Ref.~\cite{Lemanski2016}). 
At $U^-_{cr}(V=0)=-\sqrt{2}$ ($U^+_{cr}(V=0)=\sqrt{2}$) the {\COMPA} ({\COMNA}, respectively) exist at any infinitesimally small (but finite) $T>0$.
There are so called quasi-critical points at $T=0$ \cite{LemanskiPRB2017}.

\begin{figure}
	\includegraphics[width=\rozmiarjeden]{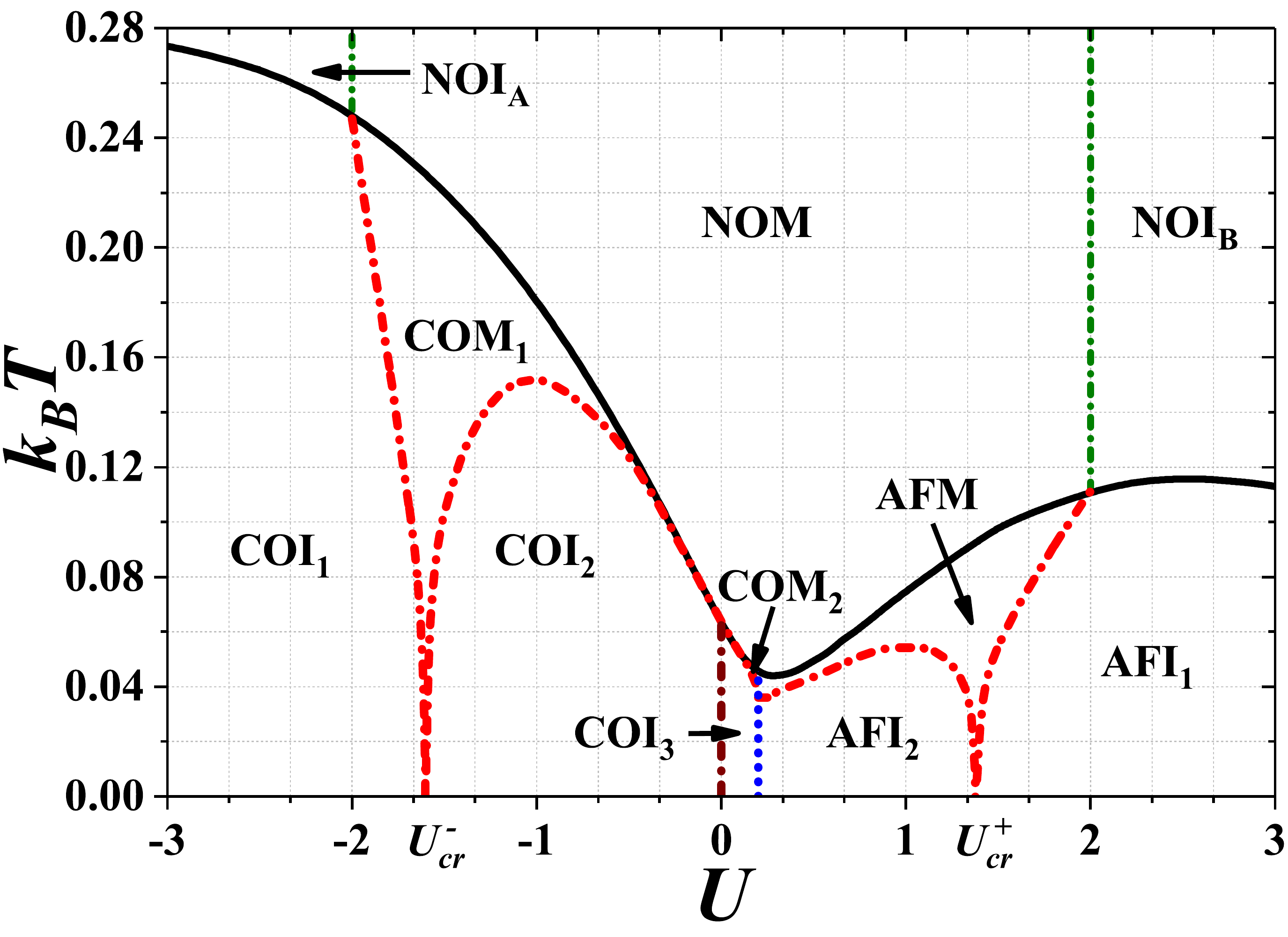}
	\caption{%
	    \label{fig:PDV01full}
		The finite temperature phase diagram for $V=0.1$.
		Solid and dashed lines denote second-order (continuous) 
		and first-order (discontinuous) transitions, respectively.		
		Each region is labeled by name of a phase, which is stable in the particular region 
		(details in text of Sec.~\ref{sec:sub:phasediagrams}).
		To determine the diagram free energies of all solutions found were compared.
		Dashed-dotted lines denotes the metal-insulator transformations, 
		which are continuous but they are not phase transitions in the usual sense 
		(details in text of Sec.~\ref{sec:sub:sub:temperaturedependences}).
		}
\end{figure}

For $V\neq0$ the phase diagram loses its symmetry with respect to $U=0$ and it needs to be discussed for negative as well as positive onsite interactions.
Nevertheless, for $V\neq0$ at temperatures above the order-disorder transition, one can distinguish three regions of the nonordered phases: two nonordered insulators: {\MIP} for $U<-2$ and {\MIN} for $U>2$, as well as the {\FLI} for $-2<U<2$. 
It turns out that the boundaries between nonordered phases do not depend either on $T$, or on $V$.
This is a consequence of the fact that $G^\pm(z)$ are dependent on $V$ only through the term $v=V(d+d_1)$ [cf. Eq.~(\ref{eq2B2:greenfuntions})] and $d=d_1=0$ in the {\FLI}, the {\MIP}, and the {\MIN}.

\begin{figure}
	\includegraphics[width=\rozmiarjeden]{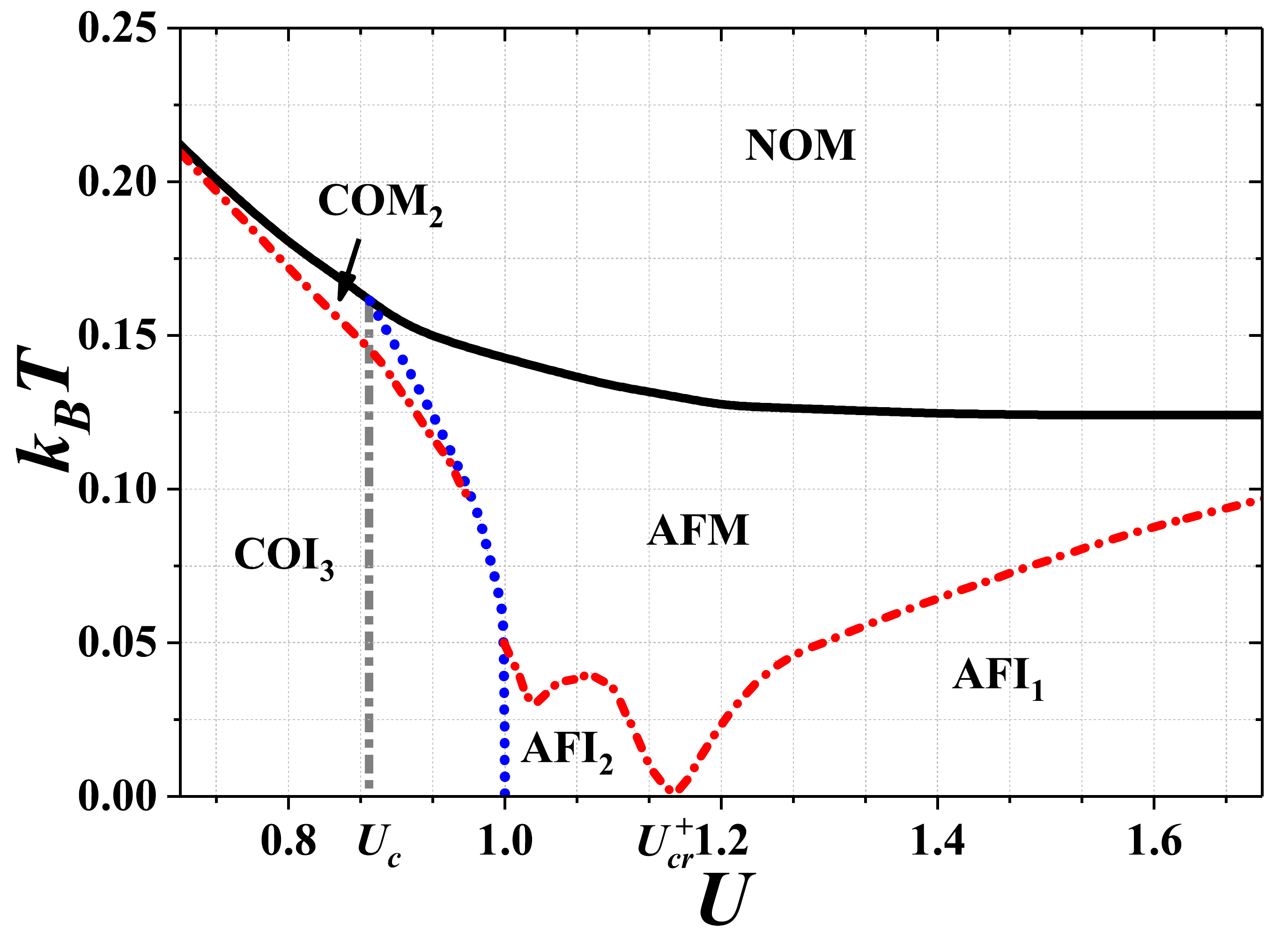}
	\caption{%
	    \label{fig:PDV05small}
		The diagram for $V=0.5$.
		The figure presents details of the phase diagram in the neighborhood of first-order transition associated to a change of sign of $d_1$ parameter.
		The vertical dashed-dotted line indicates the location of the bicritical point.
		The features of the diagram do not change in a range of $0<V<0.54$.
		Other denotations as in Fig.~\ref{fig:PDV01full}.
		}
\end{figure}

At the phase diagram for $V>0$, two new regions appear for $U\geq0$ (cf. Figs.~\ref{fig:PDV01full}, \ref{fig:PDV05small}, and \ref{fig:PDV06smallUpositive}).
One is that of the {\COIPC} extending from the ground state (in the range $0 \leq U<2V$ at $T=0$).
Another one is a region of the {\COMPB}, which appears for $0<U<U_c$ at finite temperatures between regions of the {\COIPA} and {\FLI} and is separated from that of the {\COMPA}.  
$U_c$ is the value of $U$ interaction for the bicritical point, where two second-order lines: {\COMPB}-{\FLI} and {\COMNA}-{\FLI} and one first-order line: {\COMPB}-{\COMNA} merge together (it is explicitly denoted in Figs. \ref{fig:PDV05small}, and \ref{fig:PDV06smallUpositive}).
Notice that $U_c<2V$.
The bicritical point is found for intersite repulsion smaller than $V\approx 0.7$.
The detailed discussion of the evolution of the phase diagram with increasing $V$ is contained below.

\subsubsection{Order-disorder transition}
\label{sec:sub:sub:ODtrans}

Figure~\ref{fig:orderdisorder} presents temperatures $T_{OD}$  of  the order-disorder transition for several values of $V$ obtained within an assumption that the transition is continuous. 
The diagram is determined by comparing energies of the  
ordered phases (with $d=0.04$ and both signs of $d_1$ determined self-consistently) and the NO phase.
Decreasing $T$ or $U$ (starting from the region of nonordered phase) with a small step the first point, when the free energy of the ordered phase is smaller, determines the continuous boundary.
In Fig.~\ref{fig:orderdisorder} the line of points is also indicated, where parameter $d_1$ in the ordered phase changes its sign ($d_1>0$ on the left and $d_1<0$ on the right).  
As we will show below for large $V$ (larger than $V\approx0.7$), the order-disorder transition is also first-order one in some range of intermediate values of $U$ (temperatures of such a transition are not shown in Fig.~\ref{fig:orderdisorder}). 
Notice that for $V\gtrsim 1.5$ temperature $T_{OD}(U)$ for the continuous transition exhibits reentrant behavior (e.g., the boundary for $V=1.6$ is $Z$-shaped near $U \approx 1.9$, it is slightly visible in Fig.~\ref{fig:orderdisorder}, cf. also Fig.~\ref{fig:PDV16smallUpositive}).

\begin{figure}
	\includegraphics[width=\rozmiarjeden]{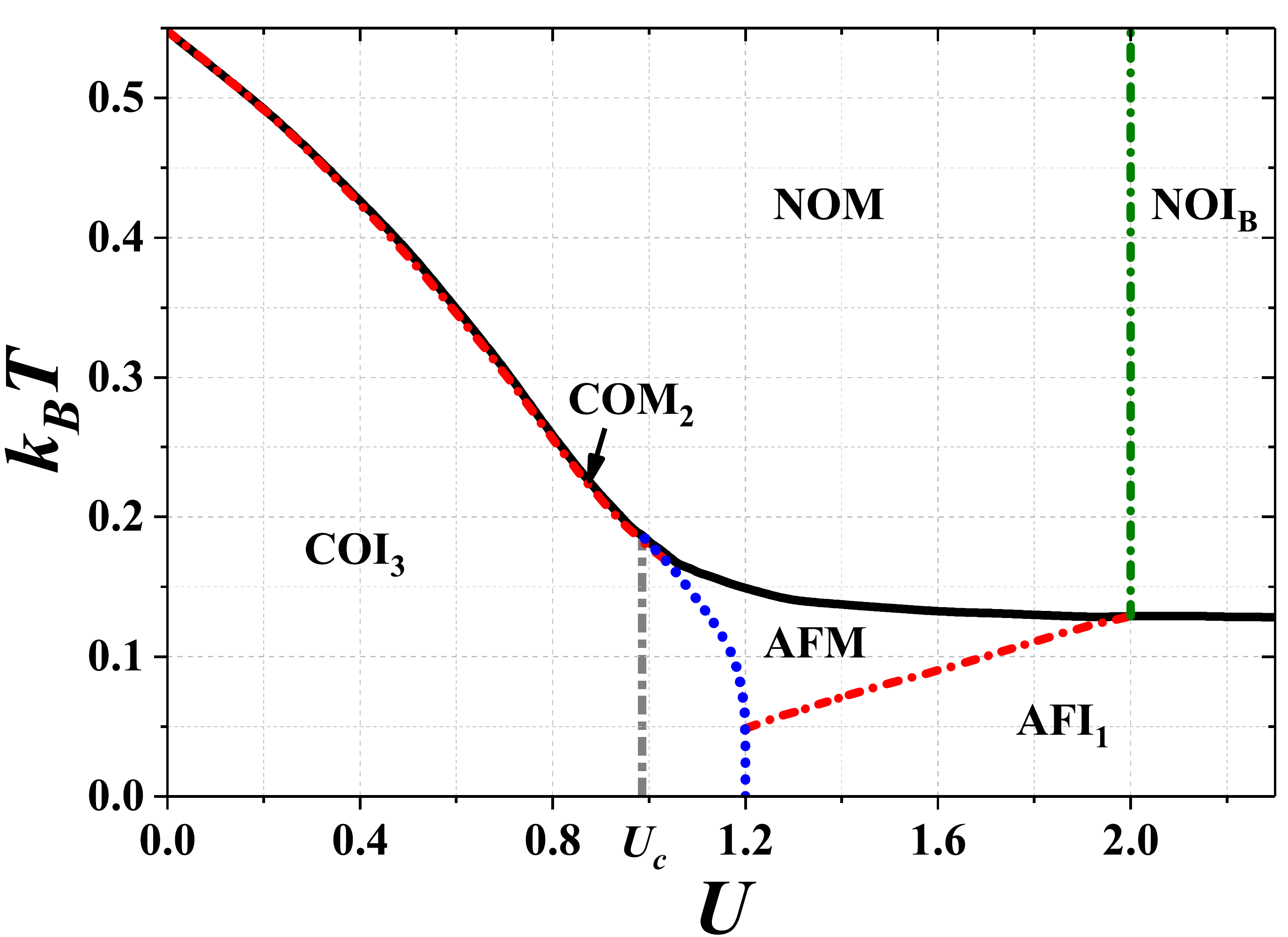}
	\caption{%
	    \label{fig:PDV06smallUpositive}
		The diagram for $V=0.6$ and $U>0$ (cf. Fig.~\ref{fig:PDV06smallUnegative}).
		The region of the {\COINB} disappeared.
		The region of the {\COMPB} is very thin.
		The vertical dashed-dotted line indicates the location of the bicritical point.
		Other denotations as in Fig.~\ref{fig:PDV01full}.		
		}
\end{figure}

For small $V \neq 0$ the order-disorder transition is indeed continuous one for any $U$, but the boundary line is no longer symmetric with respect to $U=0$, however, it retains its two maxima for both signs of $U$ (Figs.~\ref{fig:orderdisorder} and \ref{fig:PDV01full}).
The local minimum still exists between these two maxima, but it is located at $U>2V$. 
In the range $0<|U|<2$ the transition occurs between two metallic phases: 
for $-2<U<0$ between the {\COMPA} and the {\FLI},  for $0<U<U_{c}$ between the {\COMPB} and the {\FLI}, and  for $U_{c}<U<2$ between the {\COMNA} and the {\FLI} (Figs.~\ref{fig:PDV01full} and \ref{fig:PDV05small}).
For larger $|U|$, i.e., for $|U|>2$,  the transition line separates the regions of insulating phases:
for $U<-2$ --- the {\COIPA} and the {\MIP}, whereas for $U>2$ --- the {\COINA} and the {\MIN} (e.g., Fig.~\ref{fig:PDV01full})).
At the single point $U=0$ the transition is directly from the {\COIPC} to the {\FLI} 
(in the neighborhood of the transition point the {\COMPA} and the {\COMPB} are also stable).

\begin{figure}
	\includegraphics[width=\rozmiarjeden]{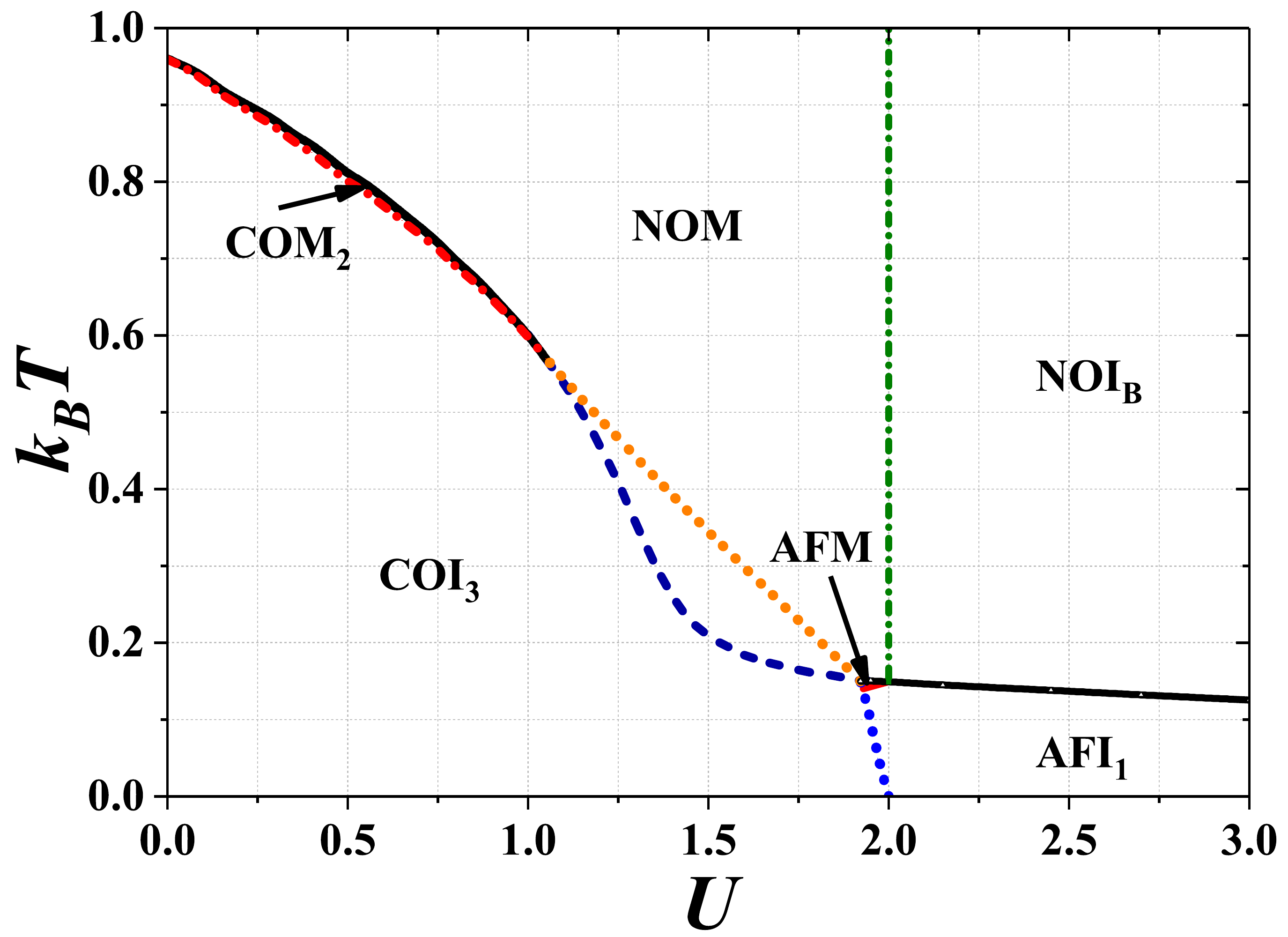}
	\caption{%
	    \label{fig:PDV10smallUpositive}
		The diagram for  $V=1.0$.
		The order-disorder {\COIPC}-{\FLI} transition (omitting the metallic phase) is discontinuous.
		The dashed line denotes the second-order boundary from Fig.~\ref{fig:orderdisorder}, which is not a boundary between the phases with the lowest free energies.
		Other denotations as in Fig.~\ref{fig:PDV01full}. 
		}
\end{figure}

For larger values of $V$,  the maximum of transition temperature $T_{OD}(U)$ for $U>0$ disappears, but the other one for $U<0$ is present for any $V$ (Figs.~\ref{fig:orderdisorder} and \ref{fig:PDV06smallUpositive}).
The disappearance of the maximum of $T_{OD}(U)$ for repulsive $U$ occurs at $V\approx0.6$.
In addition, the transition in some range of $U$ changes its order into first one (for the intersite repulsion larger than $V\approx 0.7$, Fig.~\ref{fig:PDV10smallUpositive}) with the transition temperature higher than $T_{OD}$ found with an assumption of the continuous transition.
This first-order transition is directly from the {\COIPC} to the {\FLI}, without passing through the region of the {\COMPB}.
With further increase of $V$ (for $V$ larger than $V\approx 1.05$) also discontinuous {\COIPC}--{\MIN} transition appears with simultaneous disappearance of the region of the {\COMNA} (Fig.~\ref{fig:PDV16smallUpositive}). 
When $V\rightarrow+\infty$, the results approaches  to the CO-NO transition line for the atomic limit of the model ($n=1$, $D\rightarrow+\infty$) with an occurrence of the first-order transition for $4/3\ln(2)<U/V<2$ and $2/3>k_{B}T/V>0$ (the region of the {\COINA} vanishes in this limit) \cite{MicnasPRB1984,KapciaPhysA2016,KapciaAPPA2012}.
For any finite $V$, the second-order transition for $U>0$ with increasing $T$ from the ordered phases with $d_1>0$ is only from the {\COMPB} and can be to the {\FLI} (as shown for $V=1.6$) or even to the {\MIN} (for larger $V$, not shown in the figures).
Note also that the reentrant behavior of $T_{OD}$ for the second-order transition has been also found in the atomic limit of the model \cite{MicnasPRB1984,KapciaAPPA2012}.
Moreover, the second-order transition temperature $T_{OD}$, which is lower than the true $T_{OD}$ of the first-order transition, was identified as the boundary of the NO phase metastability inside the CO phase region \cite{KapciaAPPA2012,KapciaJSNM2014}.

Finally, let us discuss the limit of large $|U|$, i.e., $U\rightarrow\mp\infty$.
The results for $U\rightarrow-\infty$, i.e., $k_BT_{OD}\rightarrow 2 V$ (where the {\COIPA}-{\MIP} transition is present) are in an agreement with the results for the atomic limit of the model \cite{MicnasPRB1984,KapciaPhysA2016,KapciaAPPA2012}. 
Notice that the order-disorder line exhibits local maximum for some $U<0$ and for any finite $V$. 
The local maximum moves to $U\rightarrow-\infty$ if $V\rightarrow+\infty$. 
In this limit the transition temperature decreases monotonously with increasing of $U$.
The local maximum for $U>0$ exists only for $V$ smaller than $V\approx 0.6$.
For large positive $U$ the transition temperature behaves as $k_B T_{OD}\approx 1/(2U)$ and for $U\rightarrow+\infty$ it decreases to $T_{OD}\rightarrow 0 $ (the {\COINA}-{\MIN} transition).
It is a result of an equivalence of the EFKM with the Ising model for large $U>0$ with $J\approx t^2/(2U)$  in this limit (here $t$ is explicitly given) and $k_BT_{OD}\approx 1/(2U)$ ($k_BT/J=1$ for the Ising model).

\subsubsection{Discontinuous transitions between ordered phases}
\label{sec:sub:sub:d1change}

\begin{figure}
	\includegraphics[width=\rozmiarjeden]{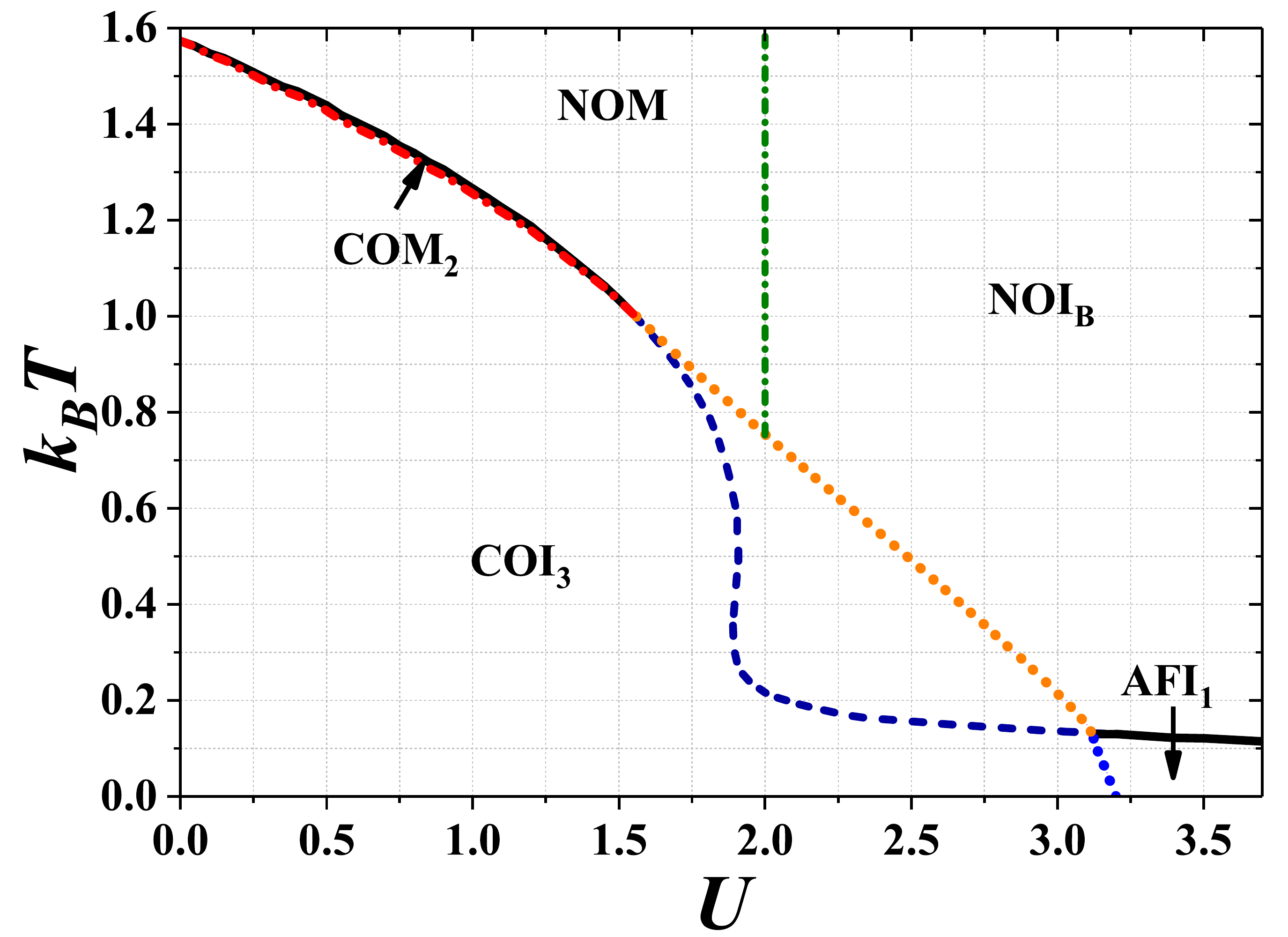}
	\caption{%
	    \label{fig:PDV16smallUpositive}
		The diagram for  $V=1.6$.
		The region of the {\COMNA} disappeared.
		Novel {\COIPC}-{\MIN} transition is present on the diagram. 
		The dashed line denotes the second-order boundary from Fig.~\ref{fig:orderdisorder}, 
		which is not a boundary between the phases with the lowest free energies.
		Other denotations as in Fig.~\ref{fig:PDV01full}.		
		}
\end{figure}

In this section we focus on  the first-order (discontinuous) transitions between
various ordered phase associated with discontinuous change of $d$.
In the model investigated in this work such transitions occur only between phases with different signs of $d_1$, i.e, only between the CO and AF phases. 
However, the criterion for the transition is equality of the free energies of both phases (cf. Sec.~\ref{sec:sub:sub:temperaturedependences}).
As it was mentioned previously, for $V=0$ the regions of the {\COIPB} and the {\COINB} are connected only  by single point at $U=0$ and $T=0$.
For $V\neq0$ three new discontinuous transitions appear on the phase diagram: {\COIPC}-{\COINB}, {\COIPC}-{\COMNA}, {\COMPB}-{\COMNA}. 
They are mentioned  in the sequence consistent with
an increase of temperature (cf. Figs.~\ref{fig:PDV01full} and \ref{fig:PDV05small}; in Fig.~\ref{fig:PDV01full} this behavior is slightly visible). 
Notice that the discussed here first-order transitions can be found only for $U_{c}<U<U_{cr}^{+}$ with an increase of temperature.
The first-order boundary on the phase diagram is a decreasing function of $U$. 
With increasing $V$ the {\COIPC}-{\COINB} line shrinks and at $V \approx 0.54$ it totally vanishes (accompanied with the disappearance of the {\COINB} region, Fig.~\ref{fig:PDV06smallUpositive}). 
The further increase of $V$ results in an appearance of {\COIPC}-{\COINA} transition.
Finally, the first-order line evolves into direct phase transitions from {\COIPC} to the nonordered phases (the {\FLI} and the {\MIN}) as described before in Sec.~\ref{sec:sub:sub:ODtrans} (Figs.~\ref{fig:PDV10smallUpositive} and \ref{fig:PDV16smallUpositive}).
Notice that the discontinuous {\COIPC}-{\COINA} boundary line is still present on the diagram and it totally vanishes only in $V\rightarrow+\infty$ limit.

\subsubsection{Continuous metal-insulator transformations}
\label{sec:sub:sub:nod1change}

On the phase diagram of the model also several continuous metal-insulator transformations between the ordered phases 
were found, namely,
\begin{itemize} 
\item[(i)] for $U<0$: {\COIPA}-{\COMPA} and {\COIPB}-{\COMPA};
\item[(ii)] for $U>0$: {\COINA}-{\COMNA}, {\COINB}-{\COMNA}, and {\COIPC}-{\COMPB}.
\end{itemize}
As one can notice, they occur only between phases with the same sign of $d_1$. 
However, the physics of the {\COIPC}-{\COMPB} transformation is different than the other ones.
Namely, it is associated with a disappearance of the energy gap between main (lower and upper) bands, whereas the rest of the transformations are connected to closing the gap between subbands inside the principal energy gap (cf. Sec.~\ref{sec:sub:sub:DOS}).

Here, we use the term ``transformation'' for the change between an insulator [with $\rho (\varepsilon_{F})=0$]
and conductor [where $\rho (\varepsilon_ {F})> 0$], because this change is not accompanied by a discontinuity of the first or second derivative of free energy, and according to the convention it cannot be classified as a phase transition of the first or second kind.

\begin{figure}
	\includegraphics[width=\rozmiarjeden]{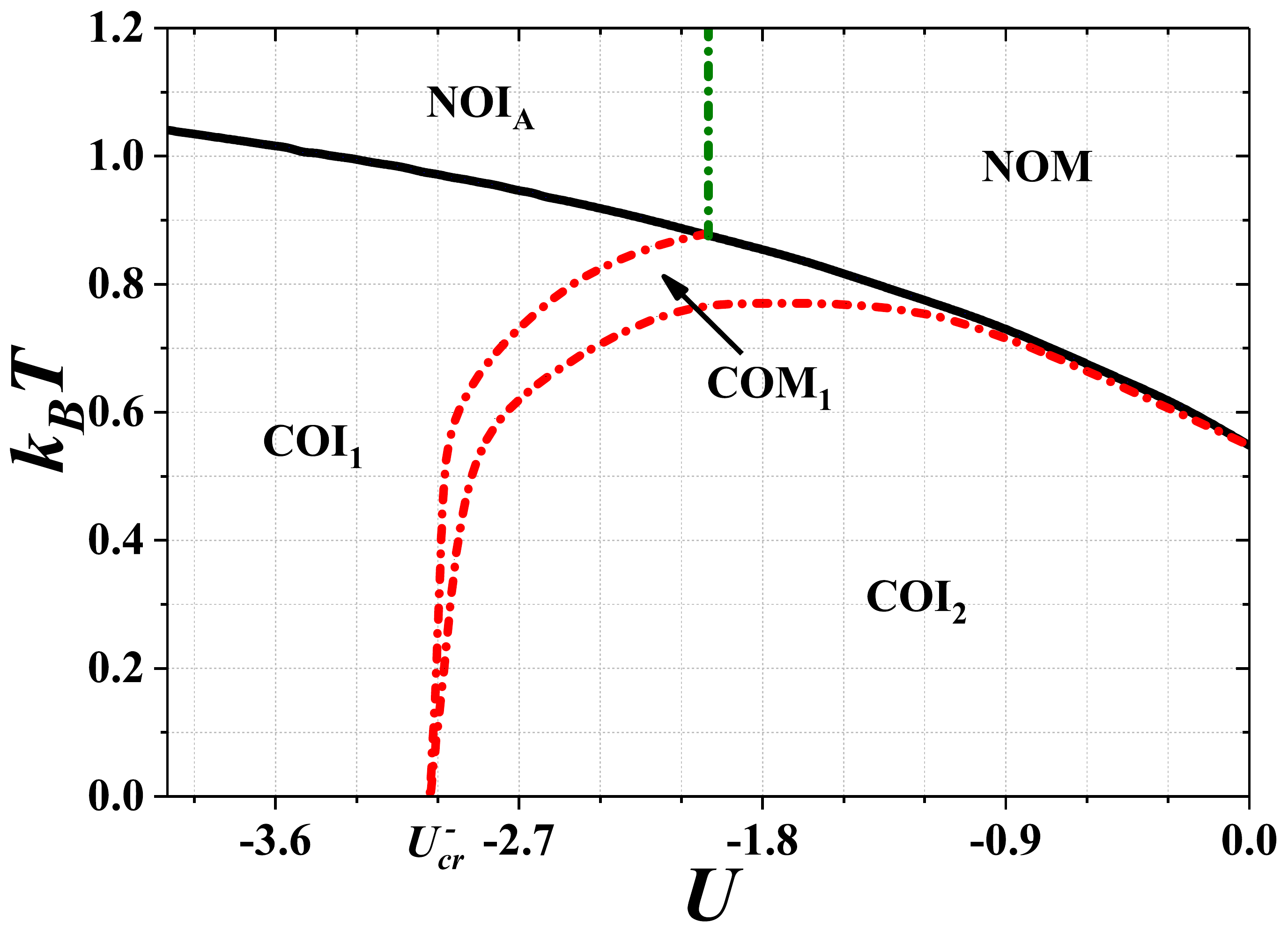}
	\caption{%
	    \label{fig:PDV06smallUnegative}
		The diagram for  $V=0.6$ and $U<0$ (cf. Fig.~\ref{fig:PDV06smallUpositive}).
		All boundaries are continuous.
		Other denotations as in Fig.~\ref{fig:PDV01full}.		
		}
\end{figure}

For $U<0$ the evolution of the boundaries with $V$ is not very complicated. 
The  region of the {\COMPA} extends from the ground state (precisely it occurs at any $T_{OD}>T>0$ for $U=U^{-}_{cr}$) due to the fact of existence of the quasi-critical point at $T=0$ and $U=U^{-}_{cr}(V)<0$ for any $V>0$ \cite{LemanskiPRB2017}.
For small $V$, the temperature of the {\COIPA}-{\COMPA} transformation decreases with $U$ from its maximal value at $U=-2$ (which is equal to $T_{OD}(U=-2)$) to zero at $U_{cr}^{-}$ (Fig.~\ref{fig:PDV01full}). 
With an increase of $V$ the {\COIPA}-{\COMPA} boundary changes  its slope at $V\approx0.268$ (in that point $U^{-}_{cr}(V\approx0.268)=-2$), and for larger $V$ the temperature of the transformation is an increasing function of $U$ (Fig.~\ref{fig:PDV06smallUnegative}).
Nevertheless, the area of the {\COMPA} separates the regions of the {\COIPA} and the {\COIPB} for any $V$.
The {\COIPA}-{\COMPA} transformation temperature is a nonmonotonous function of $U$, increasing from zero at $U_{cr}^{-}$ to its maximal value and next it decreases to the value equal $T_{OD}(U=0)$.

The situation for $U>0$ is more complex.
For any $V\neq0$ the region of the {\COMPB} phase appears for $U_{c}>U>0$ and $T>0$ (cf., e.g., Fig.~\ref{fig:PDV01full}).
The {\COIPC}-{\COMPB} boundary line decreases with $U$.
The quasi-critical point at $T=0$ exist for $V\lesssim 0.54$ \cite{LemanskiPRB2017} and thus  the {\COMNA} can occur at any $T>0$.
The {\COINB}-{\COMNA} boundary  is a nonmonotonous function of $U$.   
The region of the {\COINB} disappears at $V\approx0.54$  (cf. Figs.~\ref{fig:PDV05small} and \ref{fig:PDV06smallUpositive}) as a result of that the first-order {\COIPC}-{\COINB} transition line and the line of quasi-critical points merge at $T=0$.
The {\COINA}-{\COMNA} boundary is increasing function of $U$.
The region of the {\COMNA} moves towards higher temperatures and shrinks with increasing $V$. 
With further increase of $V$ the region of the {\COMNA} occurrence  disappears (cf. Figs.~\ref{fig:PDV10smallUpositive} and \ref{fig:PDV16smallUpositive}).

For the discussion of  the phase diagram to be complete one needs to mention also the {\COIPB}-{\COIPC} transformation.
It is a continuous one between two insulating phases with positive $d_1$.
It occurs at $U=0$ for any $V>0$ and its location is not dependent on $k_BT$ (for $V=0$ the boundary is reduced to a point at $T=0$).
It is associated with vanishing of the subband structure inside the main energy gap of the charge-ordered insulator, as it was described in Sec.~\ref{sec:sub:sub:DOS} [the lower (upper) main band and the lower (upper, respectively) subband merge together at $U=0$].
At the transformation, no discontinuities of $d$ and $d_1$, as well as $\rho(\varepsilon_F)$ and $\Delta(\varepsilon_F)$, are  found.
Similarly as for previously mentioned in this section metal-insulator transformations, at {\COIPB}-{\COIPC} boundary the derivatives of $F$ do also not exhibit standard behavior expected from second-order transitions.

Please note that all transformations discussed in this section occur between  phases with the same order and are associated with continuous changes of $d$ and $d_1$ parameters.
They are defined by the behavior of quantity $\rho(\varepsilon_F)$, but at the transformations no discontinuities $\rho(\varepsilon_F)$ and $\Delta(\varepsilon_F)$ are  found. 
Thus, they are not phase transition in the usual sense (e.g., there is no kink of free energy or entropy at the transition point) similarly to the  metal-insulator transition in the FKM~\cite{Lemanski2014}.
Nevertheless, one should underline that order-disorder transitions (both continuous and discontinuous) found in Sec.~\ref{sec:sub:sub:ODtrans} as well as discontinuous transitions between various ordered phases collected in Sec.~\ref{sec:sub:sub:d1change} are conventional phase transitions.
The details are included in Sec.~\ref{sec:sub:sub:temperaturedependences}, where dependencies of thermodynamic parameters are shown for a few exemplary boundaries discussed previously.

\subsubsection{Changes of thermodynamic quantities at the phase boundaries}
\label{sec:sub:sub:temperaturedependences}

\begin{figure}
	\includegraphics[width=\rozmiarjeden]{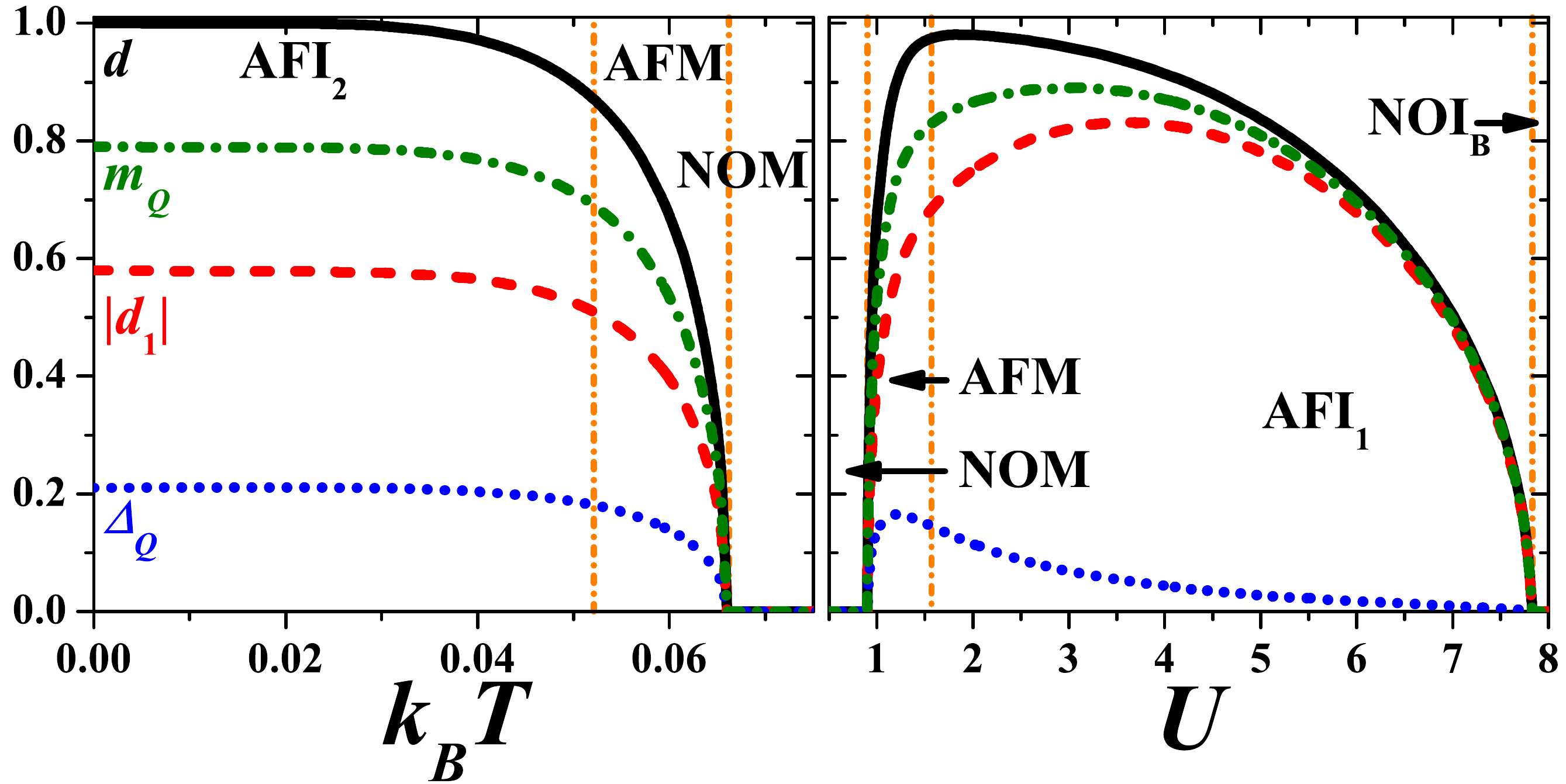}
	\caption{%
	    \label{fig:propFKM}
		Dependencies of 
		parameters $d$ (solid line), $|d_1|$ (dashed line), $\Delta_Q$ (dotted line), and $m_Q$ (dashed-dotted line) 
		for $V=0$ (the FKM) and for 
		 $U=1.0$ as a function of temperature $k_BT$ (left panel) and 
		 $k_BT=0.06$ as a function of onsite repulsion $U$ (right panel).
		Vertical dashed-dotted lines indicate the transitions.
		The locations of the {\COINB}-{\COMNA} and {\COMNA}-{\COINA} 
		boundaries are determined by vanishing of $\rho(\varepsilon_F)$, which dependence is not shown in the figure.} 
\end{figure}

Let us start this section from revisiting of the FKM [i.e., Eq.~(\ref{eq:ham}) with $V=0$].
As we wrote at the beginning of Sec.~\ref{sec:results}, in all ordered phases both charge-order and antiferromagnetism coexist.
It is not an obvious fact, because the previous works on the EFM \cite{DongenPRL1990,FreericksPRB1999,ChenPRB2003,MaskaPRB2006,HassanPRB2007,Lemanski2014}  concentrated on the analysis of the behavior of a difference of concentration of immobile particles in both sublattices, i.e.,  parameter $d$ in this work [cf. Eq.~(\ref{eq:defpard})] and parameter $d_1$ was not determined.
For $V=0$ free energy $F$ does not depend on parameter $d_1$, but it can be calculated from Eqs.~(\ref{eq:defpard1}) and (\ref{2B6}).
Parameters $d$ and $d_1$ are presented as a function of temperature for $U=1.0$ on the left panel of Fig.~\ref{fig:propFKM} (cf. also Figs. 5 and 6 of Ref.~\cite{Lemanski2014}).
This corresponds to {\COINB}-{\COMNA}-{\FLI} sequence of continuous transitions.
The right panel of Fig.~\ref{fig:propFKM} presents the parameters as a function of $U$ for fixed $k_BT=0.06$ corresponding to {\FLI}-{\COMNA}-{\COINA}-{\MIN} sequence (cf. also Fig. 7 of Ref.~\cite{Lemanski2014}).
Using relations (\ref{eq:defDeltaQmQ}), one can calculate charge polarization $\Delta_Q$ and staggered magnetization $m_Q$, which are also plotted in Fig.~\ref{fig:propFKM}.
All four discussed parameters change continuously at the phase boundaries and they vanish to zero at the order-disorder transition points.

It is clearly seen that, even for $V=0$, both $\Delta_Q$ and $m_Q$ are nonzero (what is equivalent to $d\neq|d_1|$ and $d,d_1 \neq 0$).
This implies that both charge order and antiferromagnetism exist simultaneously in the ordered phases of the FKM. 
Please also note that for $U>0$ and $V=0$ parameter $d_1<0$ and thus $\Delta_Q<m_Q$ (the antiferromagnetic order dominates). 
For $V=0$ the model exhibits a symmetry and one can obtain the results for $-U$ simply by transforming $d_1\rightarrow-d_1$, $\Delta_Q\rightarrow m_Q$, and $m_Q \rightarrow \Delta_Q$ together with changing $U\rightarrow-U$ ($d$ does not change under the transformation).
For $U<0$ parameter $d_1>0$ and thus $\Delta_Q>m_Q$ (the charge-order dominates over antiferromagnetism).

\begin{figure}
	\includegraphics[width=\rozmiarjeden]{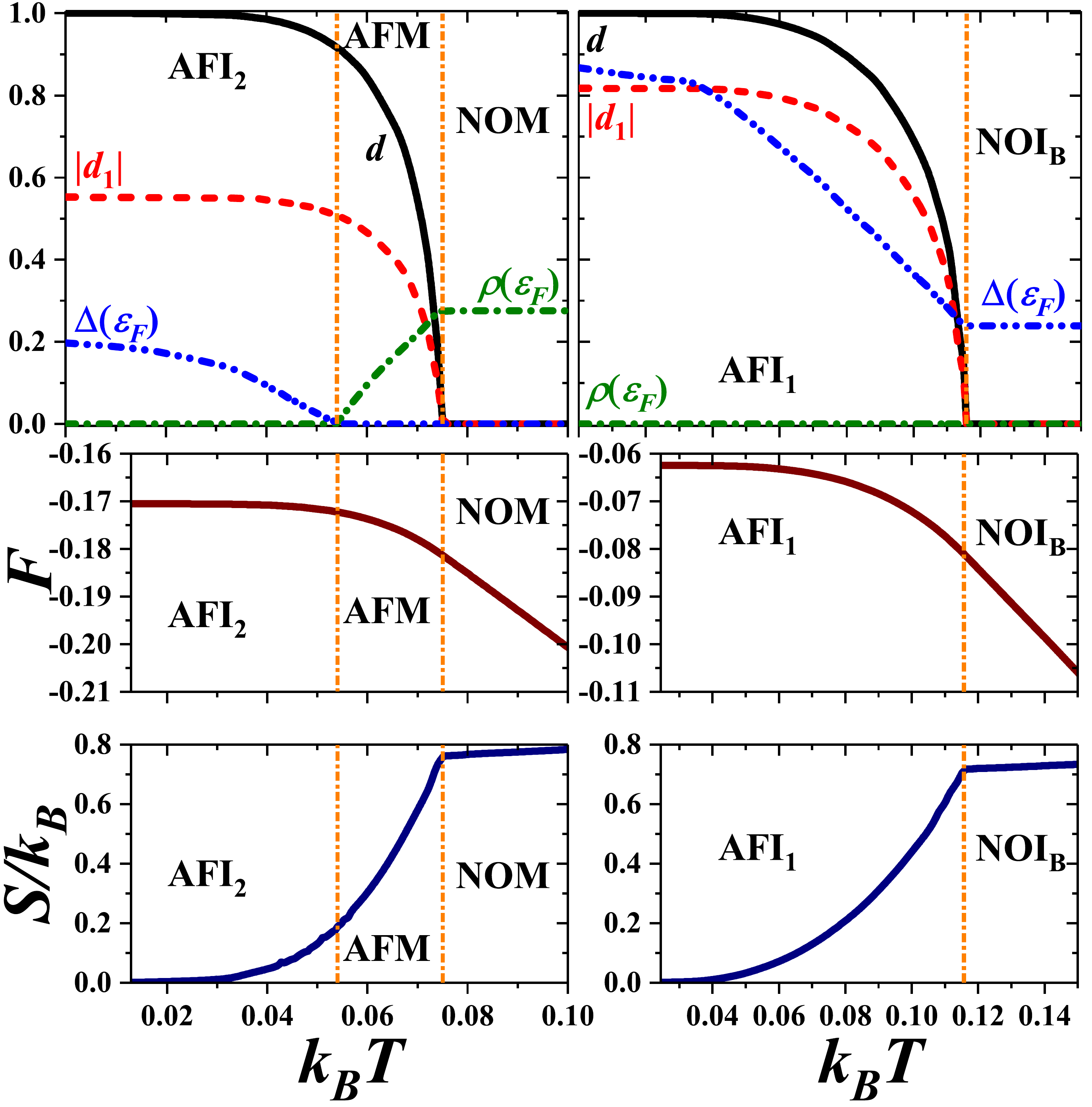}
	\caption{%
	    \label{fig:prop1}
		Top panels: temperature dependencies of parameter $d$ (solid line), 
		parameter $|d_1|$ (dashed line), 
		DOS at the Fermi level $\rho(\varepsilon_F)$ (dashed-dotted line), 
		and energy gap at the Fermi level $\Delta(\varepsilon_F)$ (dashed-dotted-dotted line).
		Lower panels: 
		temperature dependencies of free energy $F$ per site 
		and entropy $S$ per site (as indicated).
		Left panels are obtained for  $U=1.0$ ({\COINB}-{\COMNA}-{\FLI} sequence 
		of continuous transitions) whereas right panels are obtained for $U=2.5$ 
		(continuous {\COINA}-{\MIN} transition); $V=0.1$ (cf. Fig.~\ref{fig:PDV01full}). 		
		Vertical dashed-dotted lines indicate the transitions.
		Note that the {\COINB}-{\COMNA} transition 
		is not a second-order transition in the usual sense.
		}
\end{figure}

For completing our discussion on the  phase  diagram of the EFKM we present a few quantities such as parameters $d$ and $d_1$, DOS at the Fermi level $\rho(\varepsilon_F)$, and energy gap at Fermi level $\Delta(\varepsilon_F)$ as well as free energy $F$ per site and entropy $S=-\tfrac{\partial F}{\partial T}$ per site at finite temperatures for representative sets of the model parameters.

In Fig.~\ref{fig:prop1} the temperature dependencies of them are presented for $U=1.0$ and $U=2.5$ (and for $V=0.1$). 
It is clearly seen that the both order-disorder transitions: {\COMNA}-{\FLI} and {\COINA}-{\MIN} are the second-order transitions with the standard behavior of order parameter $d$, which continuously decreases with increasing temperature and goes to zero at the transition point.
The behavior of parameter $|d_1|$ is similar (notice that $d_1<0$, i.e., $\Delta_Q<m_Q$,  in the phases discussed).
It should be also noted that, obviously, both $\Delta_Q$ and $m_Q$ [as a linear combination of $d$ and $d_1$, Eq.~(\ref{eq:defDeltaQmQ})] also vanish  continuously at the transition temperature.
Gap $\Delta(\varepsilon_F)$ in the both insulating phases (i.e., {\COINB} and {\COINA}) decreases with an increase of $k_BT$.
One can distinguish two regions with different slopes, but the boundary between them cannot be undoubtedly determined.
It is a kind of smooth crossover between these different behaviors.  
At the {\COINB}-{\COMNA} transformation $\Delta(\varepsilon_F)$ goes continuously to zero, whereas at the {\COINA}-{\MIN} transition it gets the value of the gap in the {\MIN}.
$\rho(\varepsilon_F)$ increases with $k_BT$ inside the {\COMNA} from zero at the {\COINB}-{\COMNA} boundary to the value of the $\rho(\varepsilon_F)$ in the {\FLI}.
At {\FLI} it does not change with further increase of temperature.
Notice that at the {\COINB}-{\COMNA} transformation all discussed parameters change continuously, thus this boundary indeed is a continuous one
(just like another metal-insulator) 
between phases with the same signs of $d_1$, cf. Sec.~\ref{sec:sub:sub:nod1change}).

Moreover, from the bottom panels of Fig.~\ref{fig:prop1} it is clearly seen that at the transition temperatures of the both order-disorder transitions $F$ and $S$ are continuous, but the different slope of $S(T)$  in ordered and nonordered phases is visible ($S$ has a kink at the transition point). 
It is associated with the discontinuity of specific heat ($c= \tfrac{1}{T} \tfrac{\partial S}{\partial T}$) as it should be for the second-order transition with continuous change of the order parameter (a discontinuity of the second derivative of free energy).
In contrast, at the {\COINB}-{\COMNA} transformation there is no such behavior and thus one can conclude that the continuous metal-insulator transformation is not a phase transition in the usual sense [$d$ (as well as $d_1$), free energy $F$, entropy $S$, and specific heat are continuous there] (cf. also Ref.~\cite{Lemanski2014}). 

\begin{figure}
	\includegraphics[width=\rozmiarjeden]{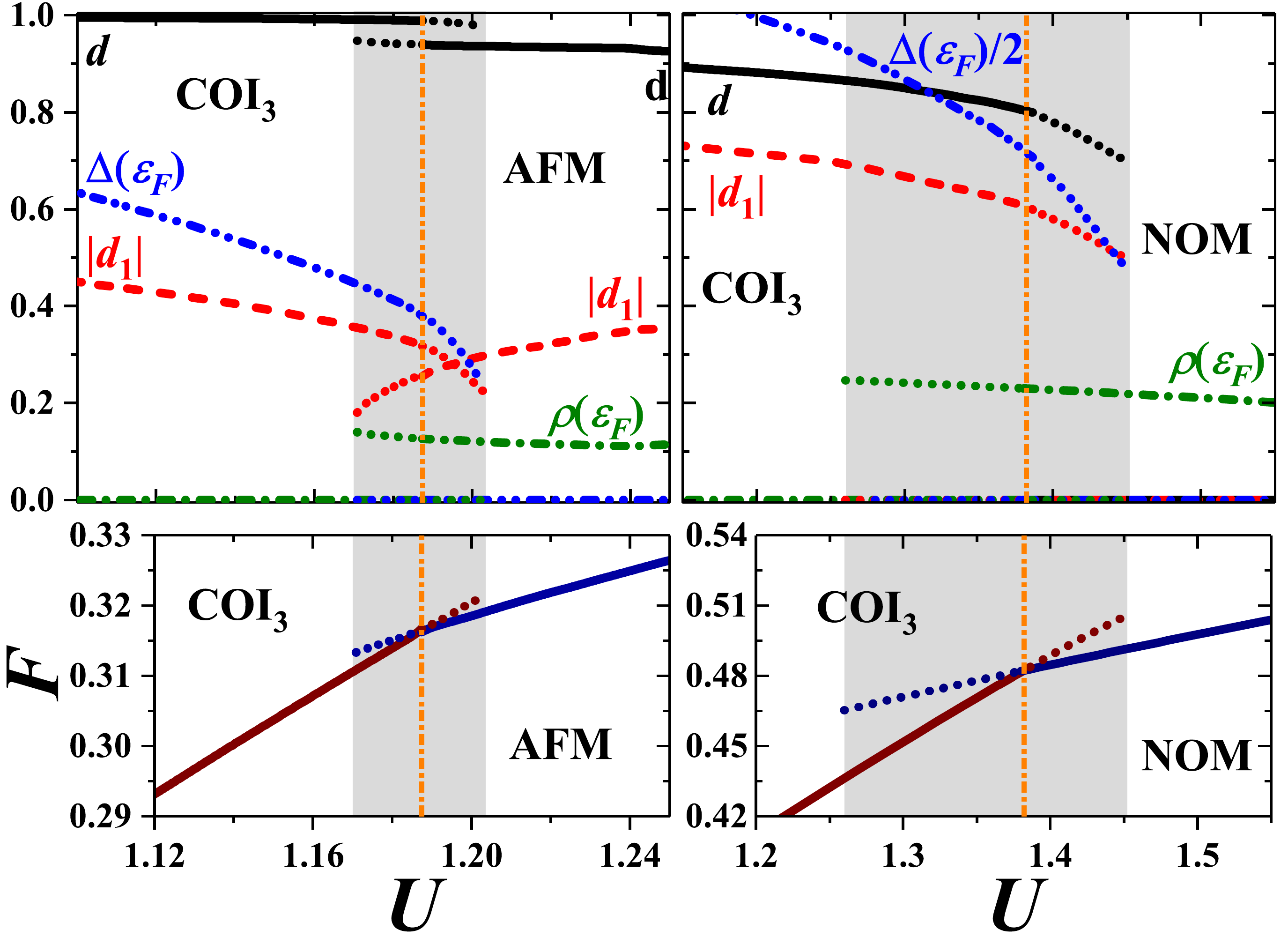}
	\caption{%
	    \label{fig:prop2}
		Top panels: $U$ dependencies of 
		parameter $d$ (solid line), 
		parameter $|d_1|$ (dashed line), 
		DOS at the Fermi level $\rho(\varepsilon_F)$ (dashed-dotted line), 
		energy gap at the Fermi level $\Delta(\varepsilon_F)$ (dashed-dotted-dotted line). 
		Bottom panels: 
		$U$ dependencies of total free energy $F$ per site (solid line).
		The dotted lines indicates the dependencies of mentioned quantities in the metastable phases.
		Left panels are obtained for  $V=0.6$ and $k_BT=0.085$ 
		(discontinuous {\COIPC}-{\COMNA}  transition, cf. Fig.~\ref{fig:PDV06smallUpositive}), 
		whereas right panels are obtained for $V=1.0$ and  $k_BT=0.4$ 
		(discontinuous {\COIPC}-{\FLI} transition, cf. Fig.~\ref{fig:PDV10smallUpositive}). 
		Vertical dashed-dotted lines indicate the transitions.
		The gray shadow indicates the coexistence regions in the neighborhood of the discontinuous phase transitions.		
		}
\end{figure}
   
Figure~\ref{fig:prop2} presents the behavior of mentioned quantities as a function of onsite interaction $U$ in the neighborhood of two discontinuous transitions: {\COIPC}-{\COMNA} and {\COIPC}-{\FLI}.
One can notice characteristic features of the first-order transitions.
Namely, in their neighborhood there is a region of a coexistence of the two phases (indicated in the figure by the gray shadow), where both solutions can coexist.
In the coexistence region one of them has a higher free energy (the metastable phase) than the other (the stable phase) (cf. the bottom panels of Fig.~\ref{fig:prop2}, where the free energies $F$ of both solutions are presented).
The value of $U$ at which energies of the both phases are the same is the transition point  (denoted by vertical dashed-dotted lines in the figure).
Moreover, in these panels also the different slope of $F(U)$ in each phase from both sides of the boundary is also visible.
It indicates that first derivative $\tfrac{\partial F}{\partial U}$ is discontinuous at the transition point as expected for a first-order transition.
These two first-order transitions are associated with a discontinuous change of all discussed quantities at the transition point, in particular, of parameters $d$ and $d_1$ (cf. the top panels of Fig.~\ref{fig:DOSfull2}).
Note also that in Fig.~\ref{fig:prop2} the dependencies of mentioned quantities in the metastable phases (i.e., in the phases, which energies are not the lowest one for given set of model parameters) are also shown (dotted lines).

The {\COIPC}-{\COMNA} transition is an example of the transition between two ordered phases, at which $d_1$ changes not only discontinuously, but it even changes its sign (i.e., $d_1>0$ in the {\COIPC} and $d_1<0$ in the {\COMNA}; $V=0.6$, $k_BT=0.085$, cf. Sec.~\ref{sec:sub:sub:d1change}). 
At the transition the discontinuous change of parameter $d$ is not so high but it is also clearly visible (note that it decreases with increasing $U$). 
Obviously, $\Delta_Q$ and $m_Q$ also exhibit discontinuous jump at the transition point.
$\Delta(\varepsilon_F)$ is decreasing functions of $U$ in the {\COIPC}, whereas $\rho(\varepsilon_F)$ slightly decreases with increasing $U$ in the {\COMNA}.

Finally, the discontinuous {\COIPC}-{\FLI} transition is an example of the order-disorder transition, which can occur for $V$ larger than $V \approx 0.7$. 
Parameters $d$ and $d_1$ as well as $\Delta(\varepsilon_F)$ decrease with increasing $U$ in the {\COIPC} and, at the transition point, they  exhibit abrupt drop to zero in the {\FLI}. 
It is worth to notice that $\rho(\varepsilon_F)$ in the {\FLI} is decreasing function of $U$ due to the enhanced correlation between electrons.

Note that a coexistence of both charge-order and antiferromagnetic order in ordered phases (for finite $U$) is a specific feature of systems, where the hopping integrals for electron species are different (cf. also Refs. \cite{DaoPRA2012,WinogradPRB2011,WinogradPRB2012,Sekania2017}).
In the case of the EFKM considered here, spin-$\downarrow$ electrons can move so they are ``more delocalized'' than the spin-$\uparrow$ electrons (ions). 
Thus, if there is an order (inhomogeneous distribution) of electrons in the system, the difference of spin-$\uparrow$ electron concentrations in the sublattices (parameter $d$) needs to be larger than those of mobile electrons with spin-$\downarrow$ (parameter $|d_1|$).
The argument is especially justified in the case of $V=0$, i.e., in the case of the FKM.
This is totally different from the extended Hubbard model with $V>0$ at $n=1$, where a phase only with charge order ($\Delta_Q\neq$ and $m_Q=0$)  or only with magnetic order ($m_Q>0$ and $\Delta_Q=$) exists, e.g., Refs.~\cite{HirschPRL1984,Lin1986,Lin2000}. 
In particular, for $U>0$ the half-filled Hubbard model ($V=0$) exhibits only the antiferromagnetic order \cite{Penn1966,GeorgesRMP1996}.

\section{Conclusions and final remarks}
\label{sec:conclusions}

In this work we investigated the extended Falicov-Kimball model [Eq.~(\ref{eq:ham})] at half-filling (i.e., $n=1$ or, equivalently, $n_\uparrow=n_\downarrow=1/2$) on the Bethe lattice within an approach which captures properly the effects of the local electron correlations.
Both onsite $U$ and intersite $V$ terms are treated in the consistent approach to give the rigorous results in the limit of large dimensions.
In this limit the intersite term reduces to the Hartree contribution.
The main achievements of the research contained in this work are as follows:
\begin{itemize}
\item[1)]{Derivation of the exact expressions (part of them are analytical) for the DOS, the energy gap at the Fermi level (for insulators) and the free energy (expressed by parameters $d$ and $d_1$)
for the extended Falicov-Kimball model in the limit of $D\rightarrow+\infty$.}
\item[2)]{Construction and analysis of the phase diagrams of the model (they appeared to be quite rich) obtained within the rigorous method for the whole range of interaction parameters $U$ and $V$ as well as temperature $T$.}
\end{itemize}

Once again, it appears that the relatively simple model of correlated electrons, when it is solved exactly, provides quite complicated phase diagrams with many ordered phases detected on them.
The differences between some of these phases are related with additional parameters, that may not be detected using approximate calculations. 
Perhaps this is why sometimes phase transitions with hidden order parameters are reported in scientific literature. On the other hand, the exact solutions permit to distinguish very precisely various ordered phases and to determine precisely of regions of their stability in the space of their model parameters.
In our case, for example, such ``hidden'' parameters that enable to distinguish some ordered insulating phases are weights of subbands of the DOS located inside the principal energy gap.

It is worth emphasizing here that a thorough examination of this system in the whole range of interaction parameters $U$ and $V$ at any $T$ was possible due to obtaining precise expressions not based on the summation over the Matsubara frequencies.
It allowed us to investigate, e.g., a nonanalyticity of the ground state and determine the quasi-critical points at $T=0$.  
Moreover, by performing integration on the real axis, the numerical results could be obtained (in practice) with any precision.
It would be not possible to attain this (in reasonable computing time) if the method of summation over Matsubara frequencies was applied due to the fact that the Green's functions vanish very slowly on the imaginary axis [$G(i\omega_n)\sim 1/\omega_n$].

Our results show that taking into account even a small Coulomb repulsive force $V$ between electrons located on neighboring sites is very important, because it causes a qualitative change in some properties of the system.
For example, if $0\leq U<2V$ (and $d_1>0$, precisely in the {\COIPC} region), then no additional subbands within finite temperatures arise inside the main energy gap.
They only arise when $U<0$ (and $d_1>0$) or $U>2V$ (and $d_1<0$). 
Whereas for $V=0$ such additional subbands arise for any $U\neq 0$.
The phase diagrams for $V>0$ at finite temperatures becomes asymmetric with the conversion of $U\rightarrow -U$ and much more complex than that with $V=0$. 
In particular, for $U>0$, the quasi-critical quantum point exists only when $V<V_{cr} \approx 0.54$, while for $U<0$ it exists for any value of $V$. 
A very interesting effect is also the existence of the discontinuous phase transition associated with a step change of $d_1$ parameter from a positive to a negative value (or vice versa).
Moreover, various order-disorder and metal-insulator transitions were found in the model.

Although our results are exact only in the limit of $D\rightarrow \infty $, they are also useful for finite dimensions.
Indeed, there is a qualitative similarity between the DOS of the cubic $D=3$ system and the DOS of the Bethe lattice in the limit $D\rightarrow \infty$ (in both these cases, the DOS close to its band edges has the square-root-type behavior).
On the other hand, some Monte Carlo calculations performed for the FKM (see, e.g., Refs.~\cite{MaskaPRB2006,ZondaSSC2009}) show that in the square $D=2$ systems subbands appear inside the principal energy gap in a similar way as it was observed for the Bethe lattice in the limit $D\rightarrow \infty$.

Let us stress here that the 
decomposition of the system into two sublattices and consideration only of such classes of solutions 
work for the half-filling case only,
when the number of both localized and itinerant particles are equal to one-half of the number of lattice sites.
This is the rigorous result, not an approximation, for any alternate lattice \cite{KennedyPhysA1986,LiebPhysA1986}.
Out of this symmetry point, e.g., in the system with a doping, one deals with either phase separation (see, e.g., Refs. \cite{FreericksPRB1999,FreericksPRB2000}) or with higher periodic or even incommensurate phases, as it was shown, for example, for 2D systems at $T=0$ in Ref.~\cite{LemanskiPRL2002}.

\begin{acknowledgments}
The authors express their sincere thanks to J. K. Freericks and M. M. Ma\'ska for useful discussions on some issues raised in this work.
K.J.K. acknowledges the support from the National Science Centre (NCN, Poland) under Grant No. UMO-2017/24/C/ST3/00276.
\end{acknowledgments}

\appendix
\section{Coefficients of the polynomial given in Eq. (\ref{eq2B3:polynomialgreen})}
\label{sect:a}

Here are the coefficients $a_0$, $a_1$, $a_2$, $a_3$, $a_4$, $a_5$ given in Eq. (\ref{eq2B3:polynomialgreen})
that are obtained from the transformation of the system of Eqs. (\ref{eq2B2:greenfuntions}).
\begin{widetext}
\begin{eqnarray}
a_0&=&-2 (4 z^2 - U^2) [8 z^3 + 4 d z^2 U - d U (-4 + U^2) - 2 z (4 + U^2)]  \\
&&+  4 [48 z^4 + 32 d z^3 U - U^2 (-4 + U^2) - 8 d z U (-2 + U^2) - 8 z^2 (6 + U^2)] (d + d_1) V \nonumber \\
&&- 16[8 z^3 + 12 d z^2 U + 2 z (-6 + U^2) -   d U (-2 + U^2)][(d + d_1) V]^2  \nonumber \\
&&+32 (-2 - 4 z^2 + 4 d z U + U^2)[(d + d_1) V]^3- 32 (-6 z + d U) [(d + d_1) V]^4 - 64[(d + d_1) V]^5, \nonumber \\
a_1&=&64 z^6 + 192 d z^3 U - 48 d z U^3 - 16 z^4 (-8 + 3 U^2) +
 U^2 (16 + 16 d^2 - U^4) + 4 z^2 (-32 - 8 U^2 + 3 U^4) \\
&&-8 [16 z^5 + 56 d z^2 U - 2 d U^3 - 8 z^3 (-4 + U^2) + z (-32 + U^4)](d + d_1) V \nonumber \\
&&+4 [-16 (2 + z^4) + 80 d z U - 8 (-1 + z^2) U^2 + 3 U^4][(d + d_1)V]^2
+64 [4 z^3 - d U + z (4 + U^2)] [(d + d_1)V]^3 \nonumber \\
&&-16 (8 + 4 z^2 + 3 U^2) [(d + d_1)V]^4-128 z [(d + d_1)V]^5 + 64  [(d + d_1)V]^6, \nonumber  \\
a_2&=&16\{-16 z^5 - 20 d z^2 U + 8 z^3 U^2 + d U (2 + U^2) - z (-4 + U^4)  \\
&&+(-4 + 16 z^4 + 24 d z U - 8 z^2 U^2 + U^4) (d + d_1) V+4 (8 z^3 - d U + 2 z U^2)[ (d + d_1) V]^2 \nonumber \\
&&-8 (4 z^2 + U^2)[(d + d_1) V]^3-16 z [(d + d_1) V]^4 + 16 [(d + d_1) V]^5\}, \nonumber\\
a_3&=&8\{48 z^4 + 24 d z U + U^4 - 16 z^2 (1 + U^2) -  8 U (d - 2 z U)(d + d_1)V \\
&&-16(-1 + 6 z^2 + U^2)[(d + d_1)V]^2 + 48 [(d + d_1) V]^4\},  \nonumber\\
a_4&=&32\{-8 z^3 - d U + 2 z (1 + U^2) - 2 (-1 + 4 z^2 + U^2)(d + d_1)  V + 8 z[(d + d_1)V]^2 + 8 [(d + d_1) V]^3\}, \\
a_5&=&16\{4 [z + (d + d_1) V]^2 - U^2\}.
\end{eqnarray}
\end{widetext}
As one can notice, the coefficients are indeed expressed explicitly by $z$, interactions $U$ and $V$, and parameters $d$ and $d_1$.
It allows us to determine the properties of the investigated system with very high precision.


\bibliography{ESHMbibliography}

\begin{thebibliography}{78}%
\makeatletter
\providecommand \@ifxundefined [1]{%
 \@ifx{#1\undefined}
}%
\providecommand \@ifnum [1]{%
 \ifnum #1\expandafter \@firstoftwo
 \else \expandafter \@secondoftwo
 \fi
}%
\providecommand \@ifx [1]{%
 \ifx #1\expandafter \@firstoftwo
 \else \expandafter \@secondoftwo
 \fi
}%
\providecommand \natexlab [1]{#1}%
\providecommand \enquote  [1]{``#1''}%
\providecommand \bibnamefont  [1]{#1}%
\providecommand \bibfnamefont [1]{#1}%
\providecommand \citenamefont [1]{#1}%
\providecommand \href@noop [0]{\@secondoftwo}%
\providecommand \href [0]{\begingroup \@sanitize@url \@href}%
\providecommand \@href[1]{\@@startlink{#1}\@@href}%
\providecommand \@@href[1]{\endgroup#1\@@endlink}%
\providecommand \@sanitize@url [0]{\catcode `\\12\catcode `\$12\catcode
  `\&12\catcode `\#12\catcode `\^12\catcode `\_12\catcode `\%12\relax}%
\providecommand \@@startlink[1]{}%
\providecommand \@@endlink[0]{}%
\providecommand \url  [0]{\begingroup\@sanitize@url \@url }%
\providecommand \@url [1]{\endgroup\@href {#1}{\urlprefix }}%
\providecommand \urlprefix  [0]{URL }%
\providecommand \Eprint [0]{\href }%
\providecommand \doibase [0]{http://dx.doi.org/}%
\providecommand \selectlanguage [0]{\@gobble}%
\providecommand \bibinfo  [0]{\@secondoftwo}%
\providecommand \bibfield  [0]{\@secondoftwo}%
\providecommand \translation [1]{[#1]}%
\providecommand \BibitemOpen [0]{}%
\providecommand \bibitemStop [0]{}%
\providecommand \bibitemNoStop [0]{.\EOS\space}%
\providecommand \EOS [0]{\spacefactor3000\relax}%
\providecommand \BibitemShut  [1]{\csname bibitem#1\endcsname}%
\let\auto@bib@innerbib\@empty
\bibitem [{\citenamefont {Micnas}\ \emph {et~al.}(1990)\citenamefont {Micnas},
  \citenamefont {Ranninger},\ and\ \citenamefont
  {Robaszkiewicz}}]{MicnasRMP1990}%
  \BibitemOpen
  \bibfield  {author} {\bibinfo {author} {\bibfnamefont {R.}~\bibnamefont
  {Micnas}}, \bibinfo {author} {\bibfnamefont {J.}~\bibnamefont {Ranninger}}, \
  and\ \bibinfo {author} {\bibfnamefont {S.}~\bibnamefont {Robaszkiewicz}},\
  }\bibfield  {title} {\enquote {\bibinfo {title} {Superconductivity in
  narrow-band systems with local nonretarded attractive interactions},}\ }\href
  {\doibase 10.1103/RevModPhys.62.113} {\bibfield  {journal} {\bibinfo
  {journal} {Rev. Mod. Phys.}\ }\textbf {\bibinfo {volume} {62}},\ \bibinfo
  {pages} {113--171} (\bibinfo {year} {1990})}\BibitemShut {NoStop}%
\bibitem [{\citenamefont {Imada}\ \emph {et~al.}(1998)\citenamefont {Imada},
  \citenamefont {Fujimori},\ and\ \citenamefont {Tokura}}]{ImadaRMP1998}%
  \BibitemOpen
  \bibfield  {author} {\bibinfo {author} {\bibfnamefont {M.}~\bibnamefont
  {Imada}}, \bibinfo {author} {\bibfnamefont {A.}~\bibnamefont {Fujimori}}, \
  and\ \bibinfo {author} {\bibfnamefont {Y.}~\bibnamefont {Tokura}},\
  }\bibfield  {title} {\enquote {\bibinfo {title} {Metal-insulator
  transitions},}\ }\href {\doibase 10.1103/RevModPhys.70.1039} {\bibfield
  {journal} {\bibinfo  {journal} {Rev. Mod. Phys.}\ }\textbf {\bibinfo {volume}
  {70}},\ \bibinfo {pages} {1039--1263} (\bibinfo {year} {1998})}\BibitemShut
  {NoStop}%
\bibitem [{\citenamefont {Yoshimi}\ \emph {et~al.}(2012)\citenamefont
  {Yoshimi}, \citenamefont {Seo}, \citenamefont {Ishibashi},\ and\
  \citenamefont {Brown}}]{YoshimiPRL2012}%
  \BibitemOpen
  \bibfield  {author} {\bibinfo {author} {\bibfnamefont {K.}~\bibnamefont
  {Yoshimi}}, \bibinfo {author} {\bibfnamefont {H.}~\bibnamefont {Seo}},
  \bibinfo {author} {\bibfnamefont {S.}~\bibnamefont {Ishibashi}}, \ and\
  \bibinfo {author} {\bibfnamefont {Stuart~E.}\ \bibnamefont {Brown}},\
  }\bibfield  {title} {\enquote {\bibinfo {title} {Tuning the magnetic
  dimensionality by charge ordering in the molecular {TMTTF} salts},}\ }\href
  {\doibase 10.1103/PhysRevLett.108.096402} {\bibfield  {journal} {\bibinfo
  {journal} {Phys. Rev. Lett.}\ }\textbf {\bibinfo {volume} {108}},\ \bibinfo
  {pages} {096402} (\bibinfo {year} {2012})}\BibitemShut {NoStop}%
\bibitem [{\citenamefont {Frandsen}\ \emph {et~al.}({2014})\citenamefont
  {Frandsen}, \citenamefont {Bozin}, \citenamefont {Hu}, \citenamefont {Zhu},
  \citenamefont {Nozaki}, \citenamefont {Kageyama}, \citenamefont {Uemura},
  \citenamefont {Yin},\ and\ \citenamefont {Billinge}}]{FrandsenNatCom2014}%
  \BibitemOpen
  \bibfield  {author} {\bibinfo {author} {\bibfnamefont {B.~A.}\ \bibnamefont
  {Frandsen}}, \bibinfo {author} {\bibfnamefont {E.~S.}\ \bibnamefont {Bozin}},
  \bibinfo {author} {\bibfnamefont {H.}~\bibnamefont {Hu}}, \bibinfo {author}
  {\bibfnamefont {Y.}~\bibnamefont {Zhu}}, \bibinfo {author} {\bibfnamefont
  {Y.}~\bibnamefont {Nozaki}}, \bibinfo {author} {\bibfnamefont
  {H.}~\bibnamefont {Kageyama}}, \bibinfo {author} {\bibfnamefont {Y.~J.}\
  \bibnamefont {Uemura}}, \bibinfo {author} {\bibfnamefont {W.-G.}\
  \bibnamefont {Yin}}, \ and\ \bibinfo {author} {\bibfnamefont {S.~J.~L.}\
  \bibnamefont {Billinge}},\ }\bibfield  {title} {\enquote {\bibinfo {title}
  {{Intra-unit-cell nematic charge order in the titanium-oxypnictide family of
  superconductors}},}\ }\href {\doibase 10.1038/ncomms6761} {\bibfield
  {journal} {\bibinfo  {journal} {{Nat. Commun.}}\ }\textbf {\bibinfo {volume}
  {{5}}},\ \bibinfo {pages} {5761} (\bibinfo {year} {{2014}})}\BibitemShut
  {NoStop}%
\bibitem [{\citenamefont {Comin}\ \emph {et~al.}({2015})\citenamefont {Comin},
  \citenamefont {Sutarto}, \citenamefont {Neto}, \citenamefont {Chauviere},
  \citenamefont {Liang}, \citenamefont {Hardy}, \citenamefont {Bonn},
  \citenamefont {He}, \citenamefont {Sawatzky},\ and\ \citenamefont
  {Damascelli}}]{CominScience2015}%
  \BibitemOpen
  \bibfield  {author} {\bibinfo {author} {\bibfnamefont {R.}~\bibnamefont
  {Comin}}, \bibinfo {author} {\bibfnamefont {R.}~\bibnamefont {Sutarto}},
  \bibinfo {author} {\bibfnamefont {E.~H. da~Silva}\ \bibnamefont {Neto}},
  \bibinfo {author} {\bibfnamefont {L.}~\bibnamefont {Chauviere}}, \bibinfo
  {author} {\bibfnamefont {R.}~\bibnamefont {Liang}}, \bibinfo {author}
  {\bibfnamefont {W.~N.}\ \bibnamefont {Hardy}}, \bibinfo {author}
  {\bibfnamefont {D.~A.}\ \bibnamefont {Bonn}}, \bibinfo {author}
  {\bibfnamefont {F.}~\bibnamefont {He}}, \bibinfo {author} {\bibfnamefont
  {G.~A.}\ \bibnamefont {Sawatzky}}, \ and\ \bibinfo {author} {\bibfnamefont
  {A.}~\bibnamefont {Damascelli}},\ }\bibfield  {title} {\enquote {\bibinfo
  {title} {{Broken translational and rotational symmetry via charge stripe
  order in underdoped {YBa$_2$Cu$_3$O$_{6+y}$}}},}\ }\href {\doibase
  10.1126/science.1258399} {\bibfield  {journal} {\bibinfo  {journal}
  {{Science}}\ }\textbf {\bibinfo {volume} {{347}}},\ \bibinfo {pages}
  {{1335--1339}} (\bibinfo {year} {{2015}})}\BibitemShut {NoStop}%
\bibitem [{\citenamefont {da~Silva~Neto}\ \emph {et~al.}({2015})\citenamefont
  {da~Silva~Neto}, \citenamefont {Comin}, \citenamefont {He}, \citenamefont
  {Sutarto}, \citenamefont {Jiang}, \citenamefont {Greene}, \citenamefont
  {Sawatzky},\ and\ \citenamefont {Damascelli}}]{NetoScience2015}%
  \BibitemOpen
  \bibfield  {author} {\bibinfo {author} {\bibfnamefont {E.~H.}\ \bibnamefont
  {da~Silva~Neto}}, \bibinfo {author} {\bibfnamefont {R.}~\bibnamefont
  {Comin}}, \bibinfo {author} {\bibfnamefont {F.}~\bibnamefont {He}}, \bibinfo
  {author} {\bibfnamefont {R.}~\bibnamefont {Sutarto}}, \bibinfo {author}
  {\bibfnamefont {Y.}~\bibnamefont {Jiang}}, \bibinfo {author} {\bibfnamefont
  {R.~L.}\ \bibnamefont {Greene}}, \bibinfo {author} {\bibfnamefont {G.~A.}\
  \bibnamefont {Sawatzky}}, \ and\ \bibinfo {author} {\bibfnamefont
  {A.}~\bibnamefont {Damascelli}},\ }\bibfield  {title} {\enquote {\bibinfo
  {title} {{Charge ordering in the electron-doped superconductor
  {Nd$_{2-x}$Ce$_x$CuO$_4$}}},}\ }\href {\doibase 10.1126/science.1256441}
  {\bibfield  {journal} {\bibinfo  {journal} {{Science}}\ }\textbf {\bibinfo
  {volume} {{347}}},\ \bibinfo {pages} {{282--285}} (\bibinfo {year}
  {{2015}})}\BibitemShut {NoStop}%
\bibitem [{\citenamefont {Cai}\ \emph {et~al.}({2016})\citenamefont {Cai},
  \citenamefont {Ruan}, \citenamefont {Peng}, \citenamefont {Ye}, \citenamefont
  {Li}, \citenamefont {Hao}, \citenamefont {Zhou}, \citenamefont {Lee},\ and\
  \citenamefont {Wang}}]{CaiNatPhys2016}%
  \BibitemOpen
  \bibfield  {author} {\bibinfo {author} {\bibfnamefont {P.}~\bibnamefont
  {Cai}}, \bibinfo {author} {\bibfnamefont {W.}~\bibnamefont {Ruan}}, \bibinfo
  {author} {\bibfnamefont {Y.}~\bibnamefont {Peng}}, \bibinfo {author}
  {\bibfnamefont {C.}~\bibnamefont {Ye}}, \bibinfo {author} {\bibfnamefont
  {X.}~\bibnamefont {Li}}, \bibinfo {author} {\bibfnamefont {Z.}~\bibnamefont
  {Hao}}, \bibinfo {author} {\bibfnamefont {X.}~\bibnamefont {Zhou}}, \bibinfo
  {author} {\bibfnamefont {D.-H.}\ \bibnamefont {Lee}}, \ and\ \bibinfo
  {author} {\bibfnamefont {Y.}~\bibnamefont {Wang}},\ }\bibfield  {title}
  {\enquote {\bibinfo {title} {{Visualizing the evolution from the {M}ott
  insulator to a charge-ordered insulator in lightly doped cuprates}},}\ }\href
  {\doibase 10.1038/NPHYS3840} {\bibfield  {journal} {\bibinfo  {journal}
  {{Nature Phys.}}\ }\textbf {\bibinfo {volume} {{12}}},\ \bibinfo {pages}
  {1047--1051} (\bibinfo {year} {{2016}})}\BibitemShut {NoStop}%
\bibitem [{\citenamefont {Hsu}\ \emph {et~al.}({2016})\citenamefont {Hsu},
  \citenamefont {Kuegel}, \citenamefont {Kemmer}, \citenamefont {Toldin},
  \citenamefont {Mauerer}, \citenamefont {Vogt}, \citenamefont {Assaad},\ and\
  \citenamefont {Bode}}]{HsuNatCom2016}%
  \BibitemOpen
  \bibfield  {author} {\bibinfo {author} {\bibfnamefont {P.-J.}\ \bibnamefont
  {Hsu}}, \bibinfo {author} {\bibfnamefont {J.}~\bibnamefont {Kuegel}},
  \bibinfo {author} {\bibfnamefont {J.}~\bibnamefont {Kemmer}}, \bibinfo
  {author} {\bibfnamefont {F.~P.}\ \bibnamefont {Toldin}}, \bibinfo {author}
  {\bibfnamefont {T.}~\bibnamefont {Mauerer}}, \bibinfo {author} {\bibfnamefont
  {M.}~\bibnamefont {Vogt}}, \bibinfo {author} {\bibfnamefont {F.}~\bibnamefont
  {Assaad}}, \ and\ \bibinfo {author} {\bibfnamefont {M.}~\bibnamefont
  {Bode}},\ }\bibfield  {title} {\enquote {\bibinfo {title} {{Coexistence of
  charge and ferromagnetic order in fcc {Fe}}},}\ }\href {\doibase
  10.1038/ncomms10949} {\bibfield  {journal} {\bibinfo  {journal} {{Nat.
  Commun.}}\ }\textbf {\bibinfo {volume} {{7}}},\ \bibinfo {pages} {{10949}}
  (\bibinfo {year} {{2016}})}\BibitemShut {NoStop}%
\bibitem [{\citenamefont {Pelc}\ \emph {et~al.}({2016})\citenamefont {Pelc},
  \citenamefont {Vuckovic}, \citenamefont {Grafe}, \citenamefont {Baek},\ and\
  \citenamefont {Pozek}}]{PelcNatCom2016}%
  \BibitemOpen
  \bibfield  {author} {\bibinfo {author} {\bibfnamefont {D.}~\bibnamefont
  {Pelc}}, \bibinfo {author} {\bibfnamefont {M.}~\bibnamefont {Vuckovic}},
  \bibinfo {author} {\bibfnamefont {H.~J.}\ \bibnamefont {Grafe}}, \bibinfo
  {author} {\bibfnamefont {S.~H.}\ \bibnamefont {Baek}}, \ and\ \bibinfo
  {author} {\bibfnamefont {M.}~\bibnamefont {Pozek}},\ }\bibfield  {title}
  {\enquote {\bibinfo {title} {{Unconventional charge order in a {Co}-doped
  high-{$T_c$} superconductor}},}\ }\href {\doibase 10.1038/ncomms12775}
  {\bibfield  {journal} {\bibinfo  {journal} {{Nat. Commun.}}\ }\textbf
  {\bibinfo {volume} {{7}}},\ \bibinfo {pages} {{12775}} (\bibinfo {year}
  {{2016}})}\BibitemShut {NoStop}%
\bibitem [{\citenamefont {Park}\ \emph {et~al.}(2017)\citenamefont {Park},
  \citenamefont {Kumar},\ and\ \citenamefont {Rabe}}]{ParkPRL2017}%
  \BibitemOpen
  \bibfield  {author} {\bibinfo {author} {\bibfnamefont {S.~Y.}\ \bibnamefont
  {Park}}, \bibinfo {author} {\bibfnamefont {A.}~\bibnamefont {Kumar}}, \ and\
  \bibinfo {author} {\bibfnamefont {K.~M.}\ \bibnamefont {Rabe}},\ }\bibfield
  {title} {\enquote {\bibinfo {title} {Charge-order-induced ferroelectricity in
  {${\mathrm{LaVO}}_{3}/{\mathrm{Sr}\mathrm{VO}}_{3}$} superlattices},}\ }\href
  {\doibase 10.1103/PhysRevLett.118.087602} {\bibfield  {journal} {\bibinfo
  {journal} {Phys. Rev. Lett.}\ }\textbf {\bibinfo {volume} {118}},\ \bibinfo
  {pages} {087602} (\bibinfo {year} {2017})}\BibitemShut {NoStop}%
\bibitem [{\citenamefont {Novello}\ \emph {et~al.}(2017)\citenamefont
  {Novello}, \citenamefont {Spera}, \citenamefont {Scarfato}, \citenamefont
  {Ubaldini}, \citenamefont {Giannini}, \citenamefont {Bowler},\ and\
  \citenamefont {Renner}}]{NovelloPRL2017}%
  \BibitemOpen
  \bibfield  {author} {\bibinfo {author} {\bibfnamefont {A.~M.}\ \bibnamefont
  {Novello}}, \bibinfo {author} {\bibfnamefont {M.}~\bibnamefont {Spera}},
  \bibinfo {author} {\bibfnamefont {A.}~\bibnamefont {Scarfato}}, \bibinfo
  {author} {\bibfnamefont {A.}~\bibnamefont {Ubaldini}}, \bibinfo {author}
  {\bibfnamefont {E.}~\bibnamefont {Giannini}}, \bibinfo {author}
  {\bibfnamefont {D.~R.}\ \bibnamefont {Bowler}}, \ and\ \bibinfo {author}
  {\bibfnamefont {Ch.}\ \bibnamefont {Renner}},\ }\bibfield  {title} {\enquote
  {\bibinfo {title} {Stripe and short range order in the charge density wave of
  {$1T\text{\ensuremath{-}}{\mathrm{Cu}}_{x}{\mathrm{TiSe}}_{2}$}},}\ }\href
  {\doibase 10.1103/PhysRevLett.118.017002} {\bibfield  {journal} {\bibinfo
  {journal} {Phys. Rev. Lett.}\ }\textbf {\bibinfo {volume} {118}},\ \bibinfo
  {pages} {017002} (\bibinfo {year} {2017})}\BibitemShut {NoStop}%
\bibitem [{\citenamefont {Nagaoka}(1966)}]{NagaokaPRB1966}%
  \BibitemOpen
  \bibfield  {author} {\bibinfo {author} {\bibfnamefont {Y.}~\bibnamefont
  {Nagaoka}},\ }\bibfield  {title} {\enquote {\bibinfo {title} {Ferromagnetism
  in a narrow, almost half-filled $s$ band},}\ }\href {\doibase
  10.1103/PhysRev.147.392} {\bibfield  {journal} {\bibinfo  {journal} {Phys.
  Rev.}\ }\textbf {\bibinfo {volume} {147}},\ \bibinfo {pages} {392--405}
  (\bibinfo {year} {1966})}\BibitemShut {NoStop}%
\bibitem [{\citenamefont {Lieb}\ and\ \citenamefont {Wu}(1968)}]{LiebPRL1968}%
  \BibitemOpen
  \bibfield  {author} {\bibinfo {author} {\bibfnamefont {E.~H.}\ \bibnamefont
  {Lieb}}\ and\ \bibinfo {author} {\bibfnamefont {F.~Y.}\ \bibnamefont {Wu}},\
  }\bibfield  {title} {\enquote {\bibinfo {title} {Absence of {Mott} transition
  in an exact solution of the short-range, one-band model in one dimension},}\
  }\href {\doibase 10.1103/PhysRevLett.20.1445} {\bibfield  {journal} {\bibinfo
   {journal} {Phys. Rev. Lett.}\ }\textbf {\bibinfo {volume} {20}},\ \bibinfo
  {pages} {1445--1448} (\bibinfo {year} {1968})}\BibitemShut {NoStop}%
\bibitem [{\citenamefont {Lieb}(1986)}]{LiebPhysA1986}%
  \BibitemOpen
  \bibfield  {author} {\bibinfo {author} {\bibfnamefont {E.~H.}\ \bibnamefont
  {Lieb}},\ }\bibfield  {title} {\enquote {\bibinfo {title} {A model for
  crystallization: {A} variation on the {Hubbard} model},}\ }\href {\doibase
  10.1016/0378-4371(86)90228-1} {\bibfield  {journal} {\bibinfo  {journal}
  {Physica A}\ }\textbf {\bibinfo {volume} {140}},\ \bibinfo {pages} {240--250}
  (\bibinfo {year} {1986})}\BibitemShut {NoStop}%
\bibitem [{\citenamefont {Georges}\ \emph {et~al.}(1996)\citenamefont
  {Georges}, \citenamefont {Kotliar}, \citenamefont {Krauth},\ and\
  \citenamefont {Rozenberg}}]{GeorgesRMP1996}%
  \BibitemOpen
  \bibfield  {author} {\bibinfo {author} {\bibfnamefont {A.}~\bibnamefont
  {Georges}}, \bibinfo {author} {\bibfnamefont {G.}~\bibnamefont {Kotliar}},
  \bibinfo {author} {\bibfnamefont {W.}~\bibnamefont {Krauth}}, \ and\ \bibinfo
  {author} {\bibfnamefont {M.~J.}\ \bibnamefont {Rozenberg}},\ }\bibfield
  {title} {\enquote {\bibinfo {title} {Dynamical mean-field theory of strongly
  correlated fermion systems and the limit of infinite dimensions},}\ }\href
  {\doibase 10.1103/RevModPhys.68.13} {\bibfield  {journal} {\bibinfo
  {journal} {Rev. Mod. Phys.}\ }\textbf {\bibinfo {volume} {68}},\ \bibinfo
  {pages} {13--125} (\bibinfo {year} {1996})}\BibitemShut {NoStop}%
\bibitem [{\citenamefont {Freericks}\ and\ \citenamefont
  {Zlati\ifmmode~\acute{c}\else \'{c}\fi{}}(2003)}]{FreericksRMP2003}%
  \BibitemOpen
  \bibfield  {author} {\bibinfo {author} {\bibfnamefont {J.~K.}\ \bibnamefont
  {Freericks}}\ and\ \bibinfo {author} {\bibfnamefont {V.}~\bibnamefont
  {Zlati\ifmmode~\acute{c}\else \'{c}\fi{}}},\ }\bibfield  {title} {\enquote
  {\bibinfo {title} {Exact dynamical mean-field theory of the {Falicov-Kimball}
  model},}\ }\href {\doibase 10.1103/RevModPhys.75.1333} {\bibfield  {journal}
  {\bibinfo  {journal} {Rev. Mod. Phys.}\ }\textbf {\bibinfo {volume} {75}},\
  \bibinfo {pages} {1333--1382} (\bibinfo {year} {2003})}\BibitemShut {NoStop}%
\bibitem [{\citenamefont {van Dongen}\ and\ \citenamefont
  {Vollhardt}(1990)}]{DongenPRL1990}%
  \BibitemOpen
  \bibfield  {author} {\bibinfo {author} {\bibfnamefont {P.~G.~J.}\
  \bibnamefont {van Dongen}}\ and\ \bibinfo {author} {\bibfnamefont
  {D.}~\bibnamefont {Vollhardt}},\ }\bibfield  {title} {\enquote {\bibinfo
  {title} {Exact mean-field hamiltonian for fermionic lattice models in high
  dimensions},}\ }\href {\doibase 10.1103/PhysRevLett.65.1663} {\bibfield
  {journal} {\bibinfo  {journal} {Phys. Rev. Lett.}\ }\textbf {\bibinfo
  {volume} {65}},\ \bibinfo {pages} {1663--1666} (\bibinfo {year}
  {1990})}\BibitemShut {NoStop}%
\bibitem [{\citenamefont {Freericks}(2006)}]{FreericksBook2006}%
  \BibitemOpen
  \bibfield  {author} {\bibinfo {author} {\bibfnamefont {J.~K.}\ \bibnamefont
  {Freericks}},\ }\href {\doibase 10.1142/9781860948824} {\emph {\bibinfo
  {title} {Transport in multilayered nanostructures. The dynamical mean-field
  theory approach}}}\ (\bibinfo  {publisher} {Imperial College Press},\
  \bibinfo {address} {London},\ \bibinfo {year} {2006})\BibitemShut {NoStop}%
\bibitem [{\citenamefont {Kotliar}\ \emph {et~al.}(2006)\citenamefont
  {Kotliar}, \citenamefont {Savrasov}, \citenamefont {Haule}, \citenamefont
  {Oudovenko}, \citenamefont {Parcollet},\ and\ \citenamefont
  {Marianetti}}]{KotliarRmp2006}%
  \BibitemOpen
  \bibfield  {author} {\bibinfo {author} {\bibfnamefont {G.}~\bibnamefont
  {Kotliar}}, \bibinfo {author} {\bibfnamefont {S.~Y.}\ \bibnamefont
  {Savrasov}}, \bibinfo {author} {\bibfnamefont {K.}~\bibnamefont {Haule}},
  \bibinfo {author} {\bibfnamefont {V.~S.}\ \bibnamefont {Oudovenko}}, \bibinfo
  {author} {\bibfnamefont {O.}~\bibnamefont {Parcollet}}, \ and\ \bibinfo
  {author} {\bibfnamefont {C.~A.}\ \bibnamefont {Marianetti}},\ }\href
  {\doibase 10.1103/RevModPhys.78.865} {\bibfield  {journal} {\bibinfo
  {journal} {Rev. Mod. Phys.}\ }\textbf {\bibinfo {volume} {78}},\ \bibinfo
  {pages} {865--951} (\bibinfo {year} {2006})}\BibitemShut {NoStop}%
\bibitem [{\citenamefont {Falicov}\ and\ \citenamefont
  {Kimball}(1969)}]{FalicovPRL1969}%
  \BibitemOpen
  \bibfield  {author} {\bibinfo {author} {\bibfnamefont {L.~M.}\ \bibnamefont
  {Falicov}}\ and\ \bibinfo {author} {\bibfnamefont {J.~C.}\ \bibnamefont
  {Kimball}},\ }\bibfield  {title} {\enquote {\bibinfo {title} {Simple model
  for semiconductor-metal transitions: {Sm${\mathrm{B}}_{6}$} and
  transition-metal oxides},}\ }\href {\doibase 10.1103/PhysRevLett.22.997}
  {\bibfield  {journal} {\bibinfo  {journal} {Phys. Rev. Lett.}\ }\textbf
  {\bibinfo {volume} {22}},\ \bibinfo {pages} {997--999} (\bibinfo {year}
  {1969})}\BibitemShut {NoStop}%
\bibitem [{\citenamefont {Brandt}\ and\ \citenamefont
  {Mielsch}(1989)}]{BrandtMielsch1989}%
  \BibitemOpen
  \bibfield  {author} {\bibinfo {author} {\bibfnamefont {U.}~\bibnamefont
  {Brandt}}\ and\ \bibinfo {author} {\bibfnamefont {C.}~\bibnamefont
  {Mielsch}},\ }\bibfield  {title} {\enquote {\bibinfo {title} {Thermodynamics
  amd correlation functions of the {Falicov-Kimball} model in large
  dimensions},}\ }\href {\doibase 10.1007/BF01321824} {\bibfield  {journal}
  {\bibinfo  {journal} {Z. Phys. B}\ }\textbf {\bibinfo {volume} {75}},\
  \bibinfo {pages} {365--370} (\bibinfo {year} {1989})}\BibitemShut {NoStop}%
\bibitem [{\citenamefont {Brandt}\ and\ \citenamefont
  {Mielsch}(1990)}]{BrandtMielsch1990}%
  \BibitemOpen
  \bibfield  {author} {\bibinfo {author} {\bibfnamefont {U.}~\bibnamefont
  {Brandt}}\ and\ \bibinfo {author} {\bibfnamefont {C.}~\bibnamefont
  {Mielsch}},\ }\bibfield  {title} {\enquote {\bibinfo {title} {Thermodynamics
  of the {Falicov-Kimball} model in large dimensions {II}},}\ }\href {\doibase
  10.1007/BF01406598} {\bibfield  {journal} {\bibinfo  {journal} {Z. Phys. B}\
  }\textbf {\bibinfo {volume} {79}},\ \bibinfo {pages} {295--299} (\bibinfo
  {year} {1990})}\BibitemShut {NoStop}%
\bibitem [{\citenamefont {Brandt}\ and\ \citenamefont
  {Mielsch}(1991)}]{BrandtMielsch1991}%
  \BibitemOpen
  \bibfield  {author} {\bibinfo {author} {\bibfnamefont {U.}~\bibnamefont
  {Brandt}}\ and\ \bibinfo {author} {\bibfnamefont {C.}~\bibnamefont
  {Mielsch}},\ }\bibfield  {title} {\enquote {\bibinfo {title} {Free energy of
  the {Falicov-Kimball} model in large dimensions},}\ }\href {\doibase
  10.1007/BF01313984} {\bibfield  {journal} {\bibinfo  {journal} {Z. Phys. B}\
  }\textbf {\bibinfo {volume} {82}},\ \bibinfo {pages} {37--41} (\bibinfo
  {year} {1991})}\BibitemShut {NoStop}%
\bibitem [{\citenamefont {Brandt}\ and\ \citenamefont
  {Urbanek}(1992)}]{BrandtUrbanek1992}%
  \BibitemOpen
  \bibfield  {author} {\bibinfo {author} {\bibfnamefont {U.}~\bibnamefont
  {Brandt}}\ and\ \bibinfo {author} {\bibfnamefont {M.~P.}\ \bibnamefont
  {Urbanek}},\ }\bibfield  {title} {\enquote {\bibinfo {title} {The f-electron
  spectrum of the spinless {Falicov-Kimball} mode in large dimensions},}\
  }\href {\doibase 10.1007/BF01318160} {\bibfield  {journal} {\bibinfo
  {journal} {Z. Phys. B}\ }\textbf {\bibinfo {volume} {89}},\ \bibinfo {pages}
  {297--303} (\bibinfo {year} {1992})}\BibitemShut {NoStop}%
\bibitem [{\citenamefont {Freericks}\ and\ \citenamefont
  {Lema\ifmmode~\acute{n}\else \'{n}\fi{}ski}(2000)}]{FreericksPRB2000}%
  \BibitemOpen
  \bibfield  {author} {\bibinfo {author} {\bibfnamefont {J.~K.}\ \bibnamefont
  {Freericks}}\ and\ \bibinfo {author} {\bibfnamefont {R.}~\bibnamefont
  {Lema\ifmmode~\acute{n}\else \'{n}\fi{}ski}},\ }\bibfield  {title} {\enquote
  {\bibinfo {title} {Segregation and charge-density-wave order in the spinless
  {Falicov-Kimball} model},}\ }\href {\doibase 10.1103/PhysRevB.61.13438}
  {\bibfield  {journal} {\bibinfo  {journal} {Phys. Rev. B}\ }\textbf {\bibinfo
  {volume} {61}},\ \bibinfo {pages} {13438--13444} (\bibinfo {year}
  {2000})}\BibitemShut {NoStop}%
\bibitem [{\citenamefont {Zlati\ifmmode~\acute{c}\else \'{c}\fi{}}\ \emph
  {et~al.}(2001)\citenamefont {Zlati\ifmmode~\acute{c}\else \'{c}\fi{}},
  \citenamefont {Freericks}, \citenamefont {Lema\ifmmode~\acute{n}\else
  \'{n}\fi{}ski},\ and\ \citenamefont
  {Czycholl}}]{ZlaticFreericksLemanskiCzycholl2001}%
  \BibitemOpen
  \bibfield  {author} {\bibinfo {author} {\bibfnamefont {V.}~\bibnamefont
  {Zlati\ifmmode~\acute{c}\else \'{c}\fi{}}}, \bibinfo {author} {\bibfnamefont
  {J.~K.}\ \bibnamefont {Freericks}}, \bibinfo {author} {\bibfnamefont
  {R.}~\bibnamefont {Lema\ifmmode~\acute{n}\else \'{n}\fi{}ski}}, \ and\
  \bibinfo {author} {\bibfnamefont {G.}~\bibnamefont {Czycholl}},\ }\bibfield
  {title} {\enquote {\bibinfo {title} {Exact solution of the multicomponent
  {Falicov-Kimball} mode in infinite dimensionsl},}\ }\href {\doibase
  10.1080/13642810110066470} {\bibfield  {journal} {\bibinfo  {journal} {Phil.
  Mag. B}\ }\textbf {\bibinfo {volume} {81}},\ \bibinfo {pages} {1443--1467}
  (\bibinfo {year} {2001})}\BibitemShut {NoStop}%
\bibitem [{\citenamefont {Chen}\ \emph {et~al.}(2003)\citenamefont {Chen},
  \citenamefont {Freericks},\ and\ \citenamefont {Jones}}]{ChenPRB2003}%
  \BibitemOpen
  \bibfield  {author} {\bibinfo {author} {\bibfnamefont {L.}~\bibnamefont
  {Chen}}, \bibinfo {author} {\bibfnamefont {J.~K.}\ \bibnamefont {Freericks}},
  \ and\ \bibinfo {author} {\bibfnamefont {B.~A.}\ \bibnamefont {Jones}},\
  }\bibfield  {title} {\enquote {\bibinfo {title} {Charge-density-wave order
  parameter of the {Falicov-Kimball} model in infinite dimensions},}\ }\href
  {\doibase 10.1103/PhysRevB.68.153102} {\bibfield  {journal} {\bibinfo
  {journal} {Phys. Rev. B}\ }\textbf {\bibinfo {volume} {68}},\ \bibinfo
  {pages} {153102} (\bibinfo {year} {2003})}\BibitemShut {NoStop}%
\bibitem [{\citenamefont {Hassan}\ and\ \citenamefont
  {Krishnamurthy}(2007)}]{HassanPRB2007}%
  \BibitemOpen
  \bibfield  {author} {\bibinfo {author} {\bibfnamefont {S.~R.}\ \bibnamefont
  {Hassan}}\ and\ \bibinfo {author} {\bibfnamefont {H.~R.}\ \bibnamefont
  {Krishnamurthy}},\ }\bibfield  {title} {\enquote {\bibinfo {title} {Spectral
  properties in the charge-density-wave phase of the half-filled
  {Falicov-Kimball} model},}\ }\href {\doibase 10.1103/PhysRevB.76.205109}
  {\bibfield  {journal} {\bibinfo  {journal} {Phys. Rev. B}\ }\textbf {\bibinfo
  {volume} {76}},\ \bibinfo {pages} {205109} (\bibinfo {year}
  {2007})}\BibitemShut {NoStop}%
\bibitem [{\citenamefont {Lema\ifmmode~\acute{n}\else \'{n}\fi{}ski}\ and\
  \citenamefont {Ziegler}(2014)}]{Lemanski2014}%
  \BibitemOpen
  \bibfield  {author} {\bibinfo {author} {\bibfnamefont {R.}~\bibnamefont
  {Lema\ifmmode~\acute{n}\else \'{n}\fi{}ski}}\ and\ \bibinfo {author}
  {\bibfnamefont {K.}~\bibnamefont {Ziegler}},\ }\bibfield  {title} {\enquote
  {\bibinfo {title} {Gapless metallic charge-density-wave phase driven by
  strong electron correlations},}\ }\href {\doibase 10.1103/PhysRevB.89.075104}
  {\bibfield  {journal} {\bibinfo  {journal} {Phys. Rev. B}\ }\textbf {\bibinfo
  {volume} {89}},\ \bibinfo {pages} {075104} (\bibinfo {year}
  {2014})}\BibitemShut {NoStop}%
\bibitem [{\citenamefont {Lema\'nski}(2016)}]{Lemanski2016}%
  \BibitemOpen
  \bibfield  {author} {\bibinfo {author} {\bibfnamefont {R.}~\bibnamefont
  {Lema\'nski}},\ }\bibfield  {title} {\enquote {\bibinfo {title} {Analysis of
  the finite-temperature phase diagram of the spinless {Falicov-Kimball}
  model},}\ }\href {\doibase 10.12693/APhysPolA.130.577} {\bibfield  {journal}
  {\bibinfo  {journal} {Acta Phys. Pol. A}\ }\textbf {\bibinfo {volume}
  {130}},\ \bibinfo {pages} {577--580} (\bibinfo {year} {2016})}\BibitemShut
  {NoStop}%
\bibitem [{\citenamefont {Haldar}\ \emph {et~al.}(2016)\citenamefont {Haldar},
  \citenamefont {Laad},\ and\ \citenamefont {Hassan}}]{haldar.laad.16}%
  \BibitemOpen
  \bibfield  {author} {\bibinfo {author} {\bibfnamefont {P.}~\bibnamefont
  {Haldar}}, \bibinfo {author} {\bibfnamefont {M.~S.}\ \bibnamefont {Laad}}, \
  and\ \bibinfo {author} {\bibfnamefont {S.~R.}\ \bibnamefont {Hassan}},\
  }\bibfield  {title} {\enquote {\bibinfo {title} {Quantum critical transport
  at a continuous metal-insulator transition},}\ }\href {\doibase
  10.1103/PhysRevB.94.081115} {\bibfield  {journal} {\bibinfo  {journal} {Phys.
  Rev. B}\ }\textbf {\bibinfo {volume} {94}},\ \bibinfo {pages} {081115(R)}
  (\bibinfo {year} {2016})}\BibitemShut {NoStop}%
\bibitem [{\citenamefont {Haldar}\ \emph
  {et~al.}(2017{\natexlab{a}})\citenamefont {Haldar}, \citenamefont {Laad},\
  and\ \citenamefont {Hassan}}]{haldar.laad.17a}%
  \BibitemOpen
  \bibfield  {author} {\bibinfo {author} {\bibfnamefont {P.}~\bibnamefont
  {Haldar}}, \bibinfo {author} {\bibfnamefont {M.~S.}\ \bibnamefont {Laad}}, \
  and\ \bibinfo {author} {\bibfnamefont {S.~R.}\ \bibnamefont {Hassan}},\
  }\bibfield  {title} {\enquote {\bibinfo {title} {Real-space cluster dynamical
  mean-field approach to the {F}alicov-{K}imball model: An alloy-analogy
  approach},}\ }\href {\doibase 10.1103/PhysRevB.95.125116} {\bibfield
  {journal} {\bibinfo  {journal} {Phys. Rev. B}\ }\textbf {\bibinfo {volume}
  {95}},\ \bibinfo {pages} {125116} (\bibinfo {year}
  {2017}{\natexlab{a}})}\BibitemShut {NoStop}%
\bibitem [{\citenamefont {Haldar}\ \emph
  {et~al.}(2017{\natexlab{b}})\citenamefont {Haldar}, \citenamefont {Laad},
  \citenamefont {Hassan}, \citenamefont {Chand},\ and\ \citenamefont
  {Raychaudhuri}}]{haldar.laad.17b}%
  \BibitemOpen
  \bibfield  {author} {\bibinfo {author} {\bibfnamefont {P.}~\bibnamefont
  {Haldar}}, \bibinfo {author} {\bibfnamefont {M.~S.}\ \bibnamefont {Laad}},
  \bibinfo {author} {\bibfnamefont {S.~R.}\ \bibnamefont {Hassan}}, \bibinfo
  {author} {\bibfnamefont {M.}~\bibnamefont {Chand}}, \ and\ \bibinfo {author}
  {\bibfnamefont {P.}~\bibnamefont {Raychaudhuri}},\ }\bibfield  {title}
  {\enquote {\bibinfo {title} {Quantum critical magnetotransport at a
  continuous metal-insulator transition},}\ }\href {\doibase
  10.1103/PhysRevB.96.155113} {\bibfield  {journal} {\bibinfo  {journal} {Phys.
  Rev. B}\ }\textbf {\bibinfo {volume} {96}},\ \bibinfo {pages} {155113}
  (\bibinfo {year} {2017}{\natexlab{b}})}\BibitemShut {NoStop}%
\bibitem [{\citenamefont {Lema\ifmmode~\acute{n}\else \'{n}\fi{}ski}\ \emph
  {et~al.}(2017)\citenamefont {Lema\ifmmode~\acute{n}\else \'{n}\fi{}ski},
  \citenamefont {Kapcia},\ and\ \citenamefont
  {Robaszkiewicz}}]{LemanskiPRB2017}%
  \BibitemOpen
  \bibfield  {author} {\bibinfo {author} {\bibfnamefont {R.}~\bibnamefont
  {Lema\ifmmode~\acute{n}\else \'{n}\fi{}ski}}, \bibinfo {author}
  {\bibfnamefont {K.~J.}\ \bibnamefont {Kapcia}}, \ and\ \bibinfo {author}
  {\bibfnamefont {S.}~\bibnamefont {Robaszkiewicz}},\ }\bibfield  {title}
  {\enquote {\bibinfo {title} {Extended {F}alicov-{K}imball model: Exact
  solution for the ground state},}\ }\href {\doibase
  10.1103/PhysRevB.96.205102} {\bibfield  {journal} {\bibinfo  {journal} {Phys.
  Rev. B}\ }\textbf {\bibinfo {volume} {96}},\ \bibinfo {pages} {205102}
  (\bibinfo {year} {2017})}\BibitemShut {NoStop}%
\bibitem [{\citenamefont {Amaricci}\ \emph {et~al.}(2010)\citenamefont
  {Amaricci}, \citenamefont {Camjayi}, \citenamefont {Haule}, \citenamefont
  {Kotliar}, \citenamefont {Tanaskovi\'c},\ and\ \citenamefont
  {Dobrosavljevi\'c}}]{AmaricciPRB2010}%
  \BibitemOpen
  \bibfield  {author} {\bibinfo {author} {\bibfnamefont {A.}~\bibnamefont
  {Amaricci}}, \bibinfo {author} {\bibfnamefont {A.}~\bibnamefont {Camjayi}},
  \bibinfo {author} {\bibfnamefont {K.}~\bibnamefont {Haule}}, \bibinfo
  {author} {\bibfnamefont {G.}~\bibnamefont {Kotliar}}, \bibinfo {author}
  {\bibfnamefont {D.}~\bibnamefont {Tanaskovi\'c}}, \ and\ \bibinfo {author}
  {\bibfnamefont {V.}~\bibnamefont {Dobrosavljevi\'c}},\ }\bibfield  {title}
  {\enquote {\bibinfo {title} {Extended {H}ubbard model: Charge ordering and
  {W}igner-{M}ott transition},}\ }\href {\doibase 10.1103/PhysRevB.82.155102}
  {\bibfield  {journal} {\bibinfo  {journal} {Phys. Rev. B}\ }\textbf {\bibinfo
  {volume} {82}},\ \bibinfo {pages} {155102} (\bibinfo {year}
  {2010})}\BibitemShut {NoStop}%
\bibitem [{\citenamefont {Kapcia}\ \emph
  {et~al.}(2017{\natexlab{a}})\citenamefont {Kapcia}, \citenamefont
  {Robaszkiewicz}, \citenamefont {Capone},\ and\ \citenamefont
  {Amaricci}}]{KapciaPRB2017}%
  \BibitemOpen
  \bibfield  {author} {\bibinfo {author} {\bibfnamefont {K.~J.}\ \bibnamefont
  {Kapcia}}, \bibinfo {author} {\bibfnamefont {S.}~\bibnamefont
  {Robaszkiewicz}}, \bibinfo {author} {\bibfnamefont {M.}~\bibnamefont
  {Capone}}, \ and\ \bibinfo {author} {\bibfnamefont {A.}~\bibnamefont
  {Amaricci}},\ }\bibfield  {title} {\enquote {\bibinfo {title} {Doping-driven
  metal-insulator transitions and charge orderings in the extended {Hubbard}
  model},}\ }\href {\doibase 10.1103/PhysRevB.95.125112} {\bibfield  {journal}
  {\bibinfo  {journal} {Phys. Rev. B}\ }\textbf {\bibinfo {volume} {95}},\
  \bibinfo {pages} {125112} (\bibinfo {year} {2017}{\natexlab{a}})}\BibitemShut
  {NoStop}%
\bibitem [{\citenamefont {Ayral}\ \emph {et~al.}(2012)\citenamefont {Ayral},
  \citenamefont {Werner},\ and\ \citenamefont {Biermann}}]{AyralPRL2012}%
  \BibitemOpen
  \bibfield  {author} {\bibinfo {author} {\bibfnamefont {T.}~\bibnamefont
  {Ayral}}, \bibinfo {author} {\bibfnamefont {P.}~\bibnamefont {Werner}}, \
  and\ \bibinfo {author} {\bibfnamefont {S.}~\bibnamefont {Biermann}},\
  }\bibfield  {title} {\enquote {\bibinfo {title} {Spectral properties of
  correlated materials: Local vertex and nonlocal two-particle correlations
  from combined $gw$ and dynamical mean field theory},}\ }\href {\doibase
  10.1103/PhysRevLett.109.226401} {\bibfield  {journal} {\bibinfo  {journal}
  {Phys. Rev. Lett.}\ }\textbf {\bibinfo {volume} {109}},\ \bibinfo {pages}
  {226401} (\bibinfo {year} {2012})}\BibitemShut {NoStop}%
\bibitem [{\citenamefont {Ayral}\ \emph {et~al.}(2013)\citenamefont {Ayral},
  \citenamefont {Biermann},\ and\ \citenamefont {Werner}}]{AyralPRB2013}%
  \BibitemOpen
  \bibfield  {author} {\bibinfo {author} {\bibfnamefont {T.}~\bibnamefont
  {Ayral}}, \bibinfo {author} {\bibfnamefont {S.}~\bibnamefont {Biermann}}, \
  and\ \bibinfo {author} {\bibfnamefont {P.}~\bibnamefont {Werner}},\
  }\bibfield  {title} {\enquote {\bibinfo {title} {Screening and nonlocal
  correlations in the extended {Hubbard} model from self-consistent combined
  {$GW$} and dynamical mean field theory},}\ }\href {\doibase
  10.1103/PhysRevB.87.125149} {\bibfield  {journal} {\bibinfo  {journal} {Phys.
  Rev. B}\ }\textbf {\bibinfo {volume} {87}},\ \bibinfo {pages} {125149}
  (\bibinfo {year} {2013})}\BibitemShut {NoStop}%
\bibitem [{\citenamefont {Huang}\ \emph {et~al.}(2014)\citenamefont {Huang},
  \citenamefont {Ayral}, \citenamefont {Biermann},\ and\ \citenamefont
  {Werner}}]{HuangPRB2014}%
  \BibitemOpen
  \bibfield  {author} {\bibinfo {author} {\bibfnamefont {L.}~\bibnamefont
  {Huang}}, \bibinfo {author} {\bibfnamefont {T.}~\bibnamefont {Ayral}},
  \bibinfo {author} {\bibfnamefont {S.}~\bibnamefont {Biermann}}, \ and\
  \bibinfo {author} {\bibfnamefont {P.}~\bibnamefont {Werner}},\ }\bibfield
  {title} {\enquote {\bibinfo {title} {Extended dynamical mean-field study of
  the {H}ubbard model with long-range interactions},}\ }\href {\doibase
  10.1103/PhysRevB.90.195114} {\bibfield  {journal} {\bibinfo  {journal} {Phys.
  Rev. B}\ }\textbf {\bibinfo {volume} {90}},\ \bibinfo {pages} {195114}
  (\bibinfo {year} {2014})}\BibitemShut {NoStop}%
\bibitem [{\citenamefont {Giovannetti}\ \emph {et~al.}(2015)\citenamefont
  {Giovannetti}, \citenamefont {Nourafkan}, \citenamefont {Kotliar},\ and\
  \citenamefont {Capone}}]{GiovannettiPRB2015}%
  \BibitemOpen
  \bibfield  {author} {\bibinfo {author} {\bibfnamefont {G.}~\bibnamefont
  {Giovannetti}}, \bibinfo {author} {\bibfnamefont {R.}~\bibnamefont
  {Nourafkan}}, \bibinfo {author} {\bibfnamefont {G.}~\bibnamefont {Kotliar}},
  \ and\ \bibinfo {author} {\bibfnamefont {M.}~\bibnamefont {Capone}},\
  }\bibfield  {title} {\enquote {\bibinfo {title} {Correlation-driven
  electronic multiferroicity in $\mathrm{TMTTF}{}_{2}\text{\ensuremath{-}}x$
  organic crystals},}\ }\href {\doibase 10.1103/PhysRevB.91.125130} {\bibfield
  {journal} {\bibinfo  {journal} {Phys. Rev. B}\ }\textbf {\bibinfo {volume}
  {91}},\ \bibinfo {pages} {125130} (\bibinfo {year} {2015})}\BibitemShut
  {NoStop}%
\bibitem [{\citenamefont {Terletska}\ \emph {et~al.}(2017)\citenamefont
  {Terletska}, \citenamefont {Chen},\ and\ \citenamefont
  {Gull}}]{TerletskaPRB2017}%
  \BibitemOpen
  \bibfield  {author} {\bibinfo {author} {\bibfnamefont {H.}~\bibnamefont
  {Terletska}}, \bibinfo {author} {\bibfnamefont {T.}~\bibnamefont {Chen}}, \
  and\ \bibinfo {author} {\bibfnamefont {E.}~\bibnamefont {Gull}},\ }\bibfield
  {title} {\enquote {\bibinfo {title} {Charge ordering and correlation effects
  in the extended {Hubbard} model},}\ }\href {\doibase
  10.1103/PhysRevB.95.115149} {\bibfield  {journal} {\bibinfo  {journal} {Phys.
  Rev. B}\ }\textbf {\bibinfo {volume} {95}},\ \bibinfo {pages} {115149}
  (\bibinfo {year} {2017})}\BibitemShut {NoStop}%
\bibitem [{\citenamefont {Ayral}\ \emph {et~al.}(2017)\citenamefont {Ayral},
  \citenamefont {Biermann}, \citenamefont {Werner},\ and\ \citenamefont
  {Boehnke}}]{AyralPRB2017}%
  \BibitemOpen
  \bibfield  {author} {\bibinfo {author} {\bibfnamefont {T.}~\bibnamefont
  {Ayral}}, \bibinfo {author} {\bibfnamefont {S.}~\bibnamefont {Biermann}},
  \bibinfo {author} {\bibfnamefont {P.}~\bibnamefont {Werner}}, \ and\ \bibinfo
  {author} {\bibfnamefont {L.}~\bibnamefont {Boehnke}},\ }\bibfield  {title}
  {\enquote {\bibinfo {title} {Influence of {Fock} exchange in combined
  many-body perturbation and dynamical mean field theory},}\ }\href {\doibase
  10.1103/PhysRevB.95.245130} {\bibfield  {journal} {\bibinfo  {journal} {Phys.
  Rev. B}\ }\textbf {\bibinfo {volume} {95}},\ \bibinfo {pages} {245130}
  (\bibinfo {year} {2017})}\BibitemShut {NoStop}%
\bibitem [{\citenamefont {Kapcia}\ \emph
  {et~al.}(2017{\natexlab{b}})\citenamefont {Kapcia}, \citenamefont
  {Bara\ifmmode~\acute{n}\else \'{n}\fi{}ski},\ and\ \citenamefont
  {Ptok}}]{KapciePRE2017}%
  \BibitemOpen
  \bibfield  {author} {\bibinfo {author} {\bibfnamefont {K.~J.}\ \bibnamefont
  {Kapcia}}, \bibinfo {author} {\bibfnamefont {J.}~\bibnamefont
  {Bara\ifmmode~\acute{n}\else \'{n}\fi{}ski}}, \ and\ \bibinfo {author}
  {\bibfnamefont {A.}~\bibnamefont {Ptok}},\ }\bibfield  {title} {\enquote
  {\bibinfo {title} {Diversity of charge orderings in correlated systems},}\
  }\href {\doibase 10.1103/PhysRevE.96.042104} {\bibfield  {journal} {\bibinfo
  {journal} {Phys. Rev. E}\ }\textbf {\bibinfo {volume} {96}},\ \bibinfo
  {pages} {042104} (\bibinfo {year} {2017}{\natexlab{b}})}\BibitemShut
  {NoStop}%
\bibitem [{\citenamefont {Terletska}\ \emph {et~al.}(2018)\citenamefont
  {Terletska}, \citenamefont {Chen}, \citenamefont {Paki},\ and\ \citenamefont
  {Gull}}]{TerletskaPRB2018}%
  \BibitemOpen
  \bibfield  {author} {\bibinfo {author} {\bibfnamefont {H.}~\bibnamefont
  {Terletska}}, \bibinfo {author} {\bibfnamefont {T.}~\bibnamefont {Chen}},
  \bibinfo {author} {\bibfnamefont {J.}~\bibnamefont {Paki}}, \ and\ \bibinfo
  {author} {\bibfnamefont {E.}~\bibnamefont {Gull}},\ }\bibfield  {title}
  {\enquote {\bibinfo {title} {Charge ordering and nonlocal correlations in the
  doped extended {H}ubbard model},}\ }\href {\doibase
  10.1103/PhysRevB.97.115117} {\bibfield  {journal} {\bibinfo  {journal} {Phys.
  Rev. B}\ }\textbf {\bibinfo {volume} {97}},\ \bibinfo {pages} {115117}
  (\bibinfo {year} {2018})}\BibitemShut {NoStop}%
\bibitem [{\citenamefont {van Dongen}(1992)}]{DongenPRB1992}%
  \BibitemOpen
  \bibfield  {author} {\bibinfo {author} {\bibfnamefont {P.~G.~J.}\
  \bibnamefont {van Dongen}},\ }\bibfield  {title} {\enquote {\bibinfo {title}
  {Exact mean-field theory of the extended simplified {Hubbard} model},}\
  }\href {\doibase 10.1103/PhysRevB.45.2267} {\bibfield  {journal} {\bibinfo
  {journal} {Phys. Rev. B}\ }\textbf {\bibinfo {volume} {45}},\ \bibinfo
  {pages} {2267--2281} (\bibinfo {year} {1992})}\BibitemShut {NoStop}%
\bibitem [{\citenamefont {van Dongen}\ and\ \citenamefont
  {Leinung}(1997)}]{DongenAP1997}%
  \BibitemOpen
  \bibfield  {author} {\bibinfo {author} {\bibfnamefont {P.~G.~J.}\
  \bibnamefont {van Dongen}}\ and\ \bibinfo {author} {\bibfnamefont
  {C.}~\bibnamefont {Leinung}},\ }\bibfield  {title} {\enquote {\bibinfo
  {title} {{Mott-Hubbard} transition in a magnetic field},}\ }\href {\doibase
  10.1002/andp.19975090104} {\bibfield  {journal} {\bibinfo  {journal} {Annalen
  der Physik}\ }\textbf {\bibinfo {volume} {509}},\ \bibinfo {pages} {45--67}
  (\bibinfo {year} {1997})}\BibitemShut {NoStop}%
\bibitem [{\citenamefont {Gajek}\ and\ \citenamefont
  {Lema\ifmmode~\acute{n}\else \'{n}\fi{}ski}(2004)}]{GajekJMMM2004}%
  \BibitemOpen
  \bibfield  {author} {\bibinfo {author} {\bibfnamefont {Z.}~\bibnamefont
  {Gajek}}\ and\ \bibinfo {author} {\bibfnamefont {R.}~\bibnamefont
  {Lema\ifmmode~\acute{n}\else \'{n}\fi{}ski}},\ }\bibfield  {title} {\enquote
  {\bibinfo {title} {Influence of the inter-ion interaction on the phase
  diagrams of the {1D} {F}alicov-{K}imball system},}\ }\href {\doibase
  10.1016/j.jmmm.2003.12.1263} {\bibfield  {journal} {\bibinfo  {journal} {J.
  Magn. Magn. Mater.}\ }\textbf {\bibinfo {volume} {272-276}},\ \bibinfo
  {pages} {e691--e692} (\bibinfo {year} {2004})}\BibitemShut {NoStop}%
\bibitem [{\citenamefont {\v{C}en\v{c}arikov\'a}\ \emph
  {et~al.}(2008)\citenamefont {\v{C}en\v{c}arikov\'a}, \citenamefont
  {Farka\v{s}ovsk\'y},\ and\ \citenamefont {\v{Z}onda}}]{Farkasovsky2008}%
  \BibitemOpen
  \bibfield  {author} {\bibinfo {author} {\bibfnamefont {H.}~\bibnamefont
  {\v{C}en\v{c}arikov\'a}}, \bibinfo {author} {\bibfnamefont {P.}~\bibnamefont
  {Farka\v{s}ovsk\'y}}, \ and\ \bibinfo {author} {\bibfnamefont
  {M.}~\bibnamefont {\v{Z}onda}},\ }\bibfield  {title} {\enquote {\bibinfo
  {title} {The influence of nonlocal coulomb interaction on ground-state
  properties of the {Falicov-Kimball} model in one and two dimensions},}\
  }\href {\doibase 10.1142/S0217979208039642} {\bibfield  {journal} {\bibinfo
  {journal} {Int. J. Mod. Phys. B}\ }\textbf {\bibinfo {volume} {22}},\
  \bibinfo {pages} {2473--2487} (\bibinfo {year} {2008})}\BibitemShut {NoStop}%
\bibitem [{\citenamefont {Hamada}\ \emph {et~al.}(2017)\citenamefont {Hamada},
  \citenamefont {Kaneko}, \citenamefont {Miyakoshi},\ and\ \citenamefont
  {Ohta}}]{HamadaJPhysSocJap2017}%
  \BibitemOpen
  \bibfield  {author} {\bibinfo {author} {\bibfnamefont {K.}~\bibnamefont
  {Hamada}}, \bibinfo {author} {\bibfnamefont {T.}~\bibnamefont {Kaneko}},
  \bibinfo {author} {\bibfnamefont {S.}~\bibnamefont {Miyakoshi}}, \ and\
  \bibinfo {author} {\bibfnamefont {Y.}~\bibnamefont {Ohta}},\ }\bibfield
  {title} {\enquote {\bibinfo {title} {Excitonic insulator state of the
  extended {Falicov-Kimball} model in the cluster dynamical impurity
  approximation},}\ }\href {\doibase 10.7566/JPSJ.86.074709} {\bibfield
  {journal} {\bibinfo  {journal} {J. Phys. Soc. Jpn.}\ }\textbf {\bibinfo
  {volume} {86}},\ \bibinfo {pages} {074709} (\bibinfo {year}
  {2017})}\BibitemShut {NoStop}%
\bibitem [{\citenamefont
  {M\"uller-Hartmann}(1989{\natexlab{a}})}]{MullerHartmannZPB1989}%
  \BibitemOpen
  \bibfield  {author} {\bibinfo {author} {\bibfnamefont {E.}~\bibnamefont
  {M\"uller-Hartmann}},\ }\bibfield  {title} {\enquote {\bibinfo {title}
  {Correlated fermions on a lattice in high dimensions},}\ }\href {\doibase
  10.1007/BF01311397} {\bibfield  {journal} {\bibinfo  {journal} {Z. Phys. B}\
  }\textbf {\bibinfo {volume} {74}},\ \bibinfo {pages} {507--512} (\bibinfo
  {year} {1989}{\natexlab{a}})}\BibitemShut {NoStop}%
\bibitem [{\citenamefont
  {M\"uller-Hartmann}(1989{\natexlab{b}})}]{mullerhartmann1989IJMPB}%
  \BibitemOpen
  \bibfield  {author} {\bibinfo {author} {\bibfnamefont {E.}~\bibnamefont
  {M\"uller-Hartmann}},\ }\bibfield  {title} {\enquote {\bibinfo {title}
  {Fermions on a lattice in high dimensions},}\ }\href {\doibase
  10.12693/APhysPolA.114.129} {\bibfield  {journal} {\bibinfo  {journal} {Int.
  J. Mod. Phys. B}\ }\textbf {\bibinfo {volume} {3}},\ \bibinfo {pages}
  {2169--2187} (\bibinfo {year} {1989}{\natexlab{b}})}\BibitemShut {NoStop}%
\bibitem [{\citenamefont {Metzner}\ and\ \citenamefont
  {Vollhardt}(1989)}]{MetznerVollhardtPRL1989}%
  \BibitemOpen
  \bibfield  {author} {\bibinfo {author} {\bibfnamefont {W.}~\bibnamefont
  {Metzner}}\ and\ \bibinfo {author} {\bibfnamefont {D.}~\bibnamefont
  {Vollhardt}},\ }\bibfield  {title} {\enquote {\bibinfo {title} {Correlated
  lattice fermions in $d=\ensuremath{\infty}$ dimensions},}\ }\href {\doibase
  10.1103/PhysRevLett.62.324} {\bibfield  {journal} {\bibinfo  {journal} {Phys.
  Rev. Lett.}\ }\textbf {\bibinfo {volume} {62}},\ \bibinfo {pages} {324--327}
  (\bibinfo {year} {1989})}\BibitemShut {NoStop}%
\bibitem [{\citenamefont {Metzner}(1989)}]{MetznerZPhysB1989}%
  \BibitemOpen
  \bibfield  {author} {\bibinfo {author} {\bibfnamefont {W.}~\bibnamefont
  {Metzner}},\ }\bibfield  {title} {\enquote {\bibinfo {title} {Variational
  theory for correlated lattice fermions in high dimensions},}\ }\href
  {\doibase 10.1007/BF01313669} {\bibfield  {journal} {\bibinfo  {journal} {Z.
  Phys. B}\ }\textbf {\bibinfo {volume} {77}},\ \bibinfo {pages} {253--266}
  (\bibinfo {year} {1989})}\BibitemShut {NoStop}%
\bibitem [{\citenamefont {Freericks}\ \emph {et~al.}(1999)\citenamefont
  {Freericks}, \citenamefont {Gruber},\ and\ \citenamefont
  {Macris}}]{FreericksPRB1999}%
  \BibitemOpen
  \bibfield  {author} {\bibinfo {author} {\bibfnamefont {J.~K.}\ \bibnamefont
  {Freericks}}, \bibinfo {author} {\bibfnamefont {Ch.}\ \bibnamefont {Gruber}},
  \ and\ \bibinfo {author} {\bibfnamefont {N.}~\bibnamefont {Macris}},\
  }\bibfield  {title} {\enquote {\bibinfo {title} {Phase separation and the
  segregation principle in the {infinite-$U$} spinless {Falicov-Kimball}
  model},}\ }\href {\doibase 10.1103/PhysRevB.60.1617} {\bibfield  {journal}
  {\bibinfo  {journal} {Phys. Rev. B}\ }\textbf {\bibinfo {volume} {60}},\
  \bibinfo {pages} {1617--1626} (\bibinfo {year} {1999})}\BibitemShut {NoStop}%
\bibitem [{\citenamefont {Ma\ifmmode~\acute{s}\else \'{s}\fi{}ka}\ and\
  \citenamefont {Czajka}(2006)}]{MaskaPRB2006}%
  \BibitemOpen
  \bibfield  {author} {\bibinfo {author} {\bibfnamefont {M.~M.}\ \bibnamefont
  {Ma\ifmmode~\acute{s}\else \'{s}\fi{}ka}}\ and\ \bibinfo {author}
  {\bibfnamefont {K.}~\bibnamefont {Czajka}},\ }\bibfield  {title} {\enquote
  {\bibinfo {title} {Thermodynamics of the two-dimensional {Falicov-Kimball}
  model: {A} classical {Monte Carlo} study},}\ }\href {\doibase
  10.1103/PhysRevB.74.035109} {\bibfield  {journal} {\bibinfo  {journal} {Phys.
  Rev. B}\ }\textbf {\bibinfo {volume} {74}},\ \bibinfo {pages} {035109}
  (\bibinfo {year} {2006})}\BibitemShut {NoStop}%
\bibitem [{\citenamefont {Hubbard}(1964)}]{HubbardPRSLA1964}%
  \BibitemOpen
  \bibfield  {author} {\bibinfo {author} {\bibfnamefont {J.}~\bibnamefont
  {Hubbard}},\ }\bibfield  {title} {\enquote {\bibinfo {title} {Electron
  correlations in narrow energy bands. {III.} {A}n improved solution},}\ }\href
  {\doibase 10.1098/rspa.1964.0190} {\bibfield  {journal} {\bibinfo  {journal}
  {Proc. R. Soc. London A}\ }\textbf {\bibinfo {volume} {281}},\ \bibinfo
  {pages} {401--419} (\bibinfo {year} {1964})}\BibitemShut {NoStop}%
\bibitem [{\citenamefont {Velick\'y}\ \emph {et~al.}(1968)\citenamefont
  {Velick\'y}, \citenamefont {Kirkpatrick},\ and\ \citenamefont
  {Ehrenreich}}]{VelickyPR1968}%
  \BibitemOpen
  \bibfield  {author} {\bibinfo {author} {\bibfnamefont {B.}~\bibnamefont
  {Velick\'y}}, \bibinfo {author} {\bibfnamefont {S.}~\bibnamefont
  {Kirkpatrick}}, \ and\ \bibinfo {author} {\bibfnamefont {H.}~\bibnamefont
  {Ehrenreich}},\ }\bibfield  {title} {\enquote {\bibinfo {title} {Single-site
  approximations in the electronic theory of simple binary alloys},}\ }\href
  {\doibase 10.1103/PhysRev.175.747} {\bibfield  {journal} {\bibinfo  {journal}
  {Phys. Rev.}\ }\textbf {\bibinfo {volume} {175}},\ \bibinfo {pages}
  {747--766} (\bibinfo {year} {1968})}\BibitemShut {NoStop}%
\bibitem [{\citenamefont {Si}\ \emph {et~al.}(1992)\citenamefont {Si},
  \citenamefont {Kotliar},\ and\ \citenamefont {Georges}}]{SiPRB1992}%
  \BibitemOpen
  \bibfield  {author} {\bibinfo {author} {\bibfnamefont {Q.}~\bibnamefont
  {Si}}, \bibinfo {author} {\bibfnamefont {G.}~\bibnamefont {Kotliar}}, \ and\
  \bibinfo {author} {\bibfnamefont {A.}~\bibnamefont {Georges}},\ }\bibfield
  {title} {\enquote {\bibinfo {title} {{F}alicov-{K}imball model and the
  breaking of {F}ermi-liquid theory in infinite dimensions},}\ }\href {\doibase
  10.1103/PhysRevB.46.1261} {\bibfield  {journal} {\bibinfo  {journal} {Phys.
  Rev. B}\ }\textbf {\bibinfo {volume} {46}},\ \bibinfo {pages} {1261--1264}
  (\bibinfo {year} {1992})}\BibitemShut {NoStop}%
\bibitem [{\citenamefont {Philipp}\ \emph {et~al.}(2017)\citenamefont
  {Philipp}, \citenamefont {Wallerberger}, \citenamefont {Gunacker},\ and\
  \citenamefont {Held}}]{PhilippEPJB2017}%
  \BibitemOpen
  \bibfield  {author} {\bibinfo {author} {\bibfnamefont {M.-T.}\ \bibnamefont
  {Philipp}}, \bibinfo {author} {\bibfnamefont {M.}~\bibnamefont
  {Wallerberger}}, \bibinfo {author} {\bibfnamefont {P.}~\bibnamefont
  {Gunacker}}, \ and\ \bibinfo {author} {\bibfnamefont {K.}~\bibnamefont
  {Held}},\ }\bibfield  {title} {\enquote {\bibinfo {title} {{M}ott-{H}ubbard
  transition in the mass-imbalanced {H}ubbard model},}\ }\href {\doibase
  10.1140/epjb/e2017-80115-7} {\bibfield  {journal} {\bibinfo  {journal} {Eur.
  Phys. J. B}\ }\textbf {\bibinfo {volume} {90}},\ \bibinfo {pages} {114}
  (\bibinfo {year} {2017})}\BibitemShut {NoStop}%
\bibitem [{\citenamefont {Shvaika}\ and\ \citenamefont
  {Freericks}(2003)}]{ShvaikaPRB2003}%
  \BibitemOpen
  \bibfield  {author} {\bibinfo {author} {\bibfnamefont {A.~M.}\ \bibnamefont
  {Shvaika}}\ and\ \bibinfo {author} {\bibfnamefont {J.~K.}\ \bibnamefont
  {Freericks}},\ }\bibfield  {title} {\enquote {\bibinfo {title} {Equivalence
  of the {Falicov}-{Kimball} and {Brandt-Mielsch} forms for the free energy of
  the infinite-dimensional {Falicov-Kimball} model},}\ }\href {\doibase
  10.1103/PhysRevB.67.153103} {\bibfield  {journal} {\bibinfo  {journal} {Phys.
  Rev. B}\ }\textbf {\bibinfo {volume} {67}},\ \bibinfo {pages} {153103}
  (\bibinfo {year} {2003})}\BibitemShut {NoStop}%
\bibitem [{\citenamefont {Mott}(1990)}]{Mott1990}%
  \BibitemOpen
  \bibfield  {author} {\bibinfo {author} {\bibfnamefont {N.~F.}\ \bibnamefont
  {Mott}},\ }\href {\doibase 10.1142/9789814360371} {\emph {\bibinfo {title}
  {Metal-Insulator Transitions}}}\ (\bibinfo  {publisher} {Taylor and Francis,
  London/Philadelphia},\ \bibinfo {year} {1990})\BibitemShut {NoStop}%
\bibitem [{\citenamefont {Montorsi}(1992)}]{Montorsi1992}%
  \BibitemOpen
  \bibinfo {editor} {\bibfnamefont {A.}~\bibnamefont {Montorsi}},\ ed.,\ \href
  {\doibase 10.1142/9789814360371} {\emph {\bibinfo {title} {The {H}ubbard
  Model. {A} Collection of Reprints}}}\ (\bibinfo  {publisher} {World
  Scientific, Singapore},\ \bibinfo {year} {1992})\BibitemShut {NoStop}%
\bibitem [{\citenamefont {Gebhard}(1997)}]{Gebhard1997}%
  \BibitemOpen
  \bibfield  {author} {\bibinfo {author} {\bibfnamefont {F.}~\bibnamefont
  {Gebhard}},\ }\href {\doibase 10.1007/3-540-14858-2} {\emph {\bibinfo {title}
  {The {M}ott Metal-Insulator Transition. {M}odels and Methods}}},\ \bibinfo
  {series} {Springer Tracts in Modern Physics}, Vol.\ \bibinfo {volume} {137}\
  (\bibinfo  {publisher} {Springer-Verlag, Berlin/Heidelberg},\ \bibinfo {year}
  {1997})\BibitemShut {NoStop}%
\bibitem [{\citenamefont {Micnas}\ \emph {et~al.}(1984)\citenamefont {Micnas},
  \citenamefont {Robaszkiewicz},\ and\ \citenamefont {Chao}}]{MicnasPRB1984}%
  \BibitemOpen
  \bibfield  {author} {\bibinfo {author} {\bibfnamefont {R.}~\bibnamefont
  {Micnas}}, \bibinfo {author} {\bibfnamefont {S.}~\bibnamefont
  {Robaszkiewicz}}, \ and\ \bibinfo {author} {\bibfnamefont {K.~A.}\
  \bibnamefont {Chao}},\ }\bibfield  {title} {\enquote {\bibinfo {title}
  {Multicritical behavior of the extended {Hubbard} model in the zero-bandwidth
  limit},}\ }\href {\doibase 10.1103/PhysRevB.29.2784} {\bibfield  {journal}
  {\bibinfo  {journal} {Phys. Rev. B}\ }\textbf {\bibinfo {volume} {29}},\
  \bibinfo {pages} {2784--2789} (\bibinfo {year} {1984})}\BibitemShut {NoStop}%
\bibitem [{\citenamefont {Kapcia}\ and\ \citenamefont
  {Robaszkiewicz}(2016)}]{KapciaPhysA2016}%
  \BibitemOpen
  \bibfield  {author} {\bibinfo {author} {\bibfnamefont {K.~J.}\ \bibnamefont
  {Kapcia}}\ and\ \bibinfo {author} {\bibfnamefont {S.}~\bibnamefont
  {Robaszkiewicz}},\ }\bibfield  {title} {\enquote {\bibinfo {title} {On the
  phase diagram of the extended {H}ubbard model with intersite density-density
  interactions in the atomic limit},}\ }\href {\doibase
  10.1016/j.physa.2016.05.056} {\bibfield  {journal} {\bibinfo  {journal}
  {Physica A}\ }\textbf {\bibinfo {volume} {461}},\ \bibinfo {pages} {487--497}
  (\bibinfo {year} {2016})}\BibitemShut {NoStop}%
\bibitem [{\citenamefont {Kapcia}\ and\ \citenamefont
  {Robaszkiewicz}(2012)}]{KapciaAPPA2012}%
  \BibitemOpen
  \bibfield  {author} {\bibinfo {author} {\bibfnamefont {K.}~\bibnamefont
  {Kapcia}}\ and\ \bibinfo {author} {\bibfnamefont {S.}~\bibnamefont
  {Robaszkiewicz}},\ }\bibfield  {title} {\enquote {\bibinfo {title} {Stable
  and metastable phases in the atomic limit of the extended {Hubbard} model
  with intersite density-density interactions},}\ }\href {\doibase
  10.12693/APhysPolA.121.1029} {\bibfield  {journal} {\bibinfo  {journal} {Acta
  Phys. Pol. A}\ }\textbf {\bibinfo {volume} {121}},\ \bibinfo {pages}
  {1029--1031} (\bibinfo {year} {2012})}\BibitemShut {NoStop}%
\bibitem [{\citenamefont {Kapcia}(2014)}]{KapciaJSNM2014}%
  \BibitemOpen
  \bibfield  {author} {\bibinfo {author} {\bibfnamefont {K.}~\bibnamefont
  {Kapcia}},\ }\bibfield  {title} {\enquote {\bibinfo {title} {Metastability
  and phase separation in a simple model of a superconductor with extremely
  short coherence length},}\ }\href {\doibase 10.1007/s10948-013-2409-8}
  {\bibfield  {journal} {\bibinfo  {journal} {J. Supercond. Nov. Magn.}\
  }\textbf {\bibinfo {volume} {27}},\ \bibinfo {pages} {913--917} (\bibinfo
  {year} {2014})}\BibitemShut {NoStop}%
\bibitem [{\citenamefont {Dao}\ \emph {et~al.}(2012)\citenamefont {Dao},
  \citenamefont {Ferrero}, \citenamefont {Cornaglia},\ and\ \citenamefont
  {Capone}}]{DaoPRA2012}%
  \BibitemOpen
  \bibfield  {author} {\bibinfo {author} {\bibfnamefont {T.-L.}\ \bibnamefont
  {Dao}}, \bibinfo {author} {\bibfnamefont {M.}~\bibnamefont {Ferrero}},
  \bibinfo {author} {\bibfnamefont {P.~S.}\ \bibnamefont {Cornaglia}}, \ and\
  \bibinfo {author} {\bibfnamefont {M.}~\bibnamefont {Capone}},\ }\bibfield
  {title} {\enquote {\bibinfo {title} {Mott transition of fermionic mixtures
  with mass imbalance in optical lattices},}\ }\href {\doibase
  10.1103/PhysRevA.85.013606} {\bibfield  {journal} {\bibinfo  {journal} {Phys.
  Rev. A}\ }\textbf {\bibinfo {volume} {85}},\ \bibinfo {pages} {013606}
  (\bibinfo {year} {2012})}\BibitemShut {NoStop}%
\bibitem [{\citenamefont {Winograd}\ \emph {et~al.}(2011)\citenamefont
  {Winograd}, \citenamefont {Chitra},\ and\ \citenamefont
  {Rozenberg}}]{WinogradPRB2011}%
  \BibitemOpen
  \bibfield  {author} {\bibinfo {author} {\bibfnamefont {E.~A.}\ \bibnamefont
  {Winograd}}, \bibinfo {author} {\bibfnamefont {R.}~\bibnamefont {Chitra}}, \
  and\ \bibinfo {author} {\bibfnamefont {M.~J.}\ \bibnamefont {Rozenberg}},\
  }\bibfield  {title} {\enquote {\bibinfo {title} {Orbital-selective crossover
  and {Mott} transitions in an asymmetric hubbard model of cold atoms in
  optical lattices},}\ }\href {\doibase 10.1103/PhysRevB.84.233102} {\bibfield
  {journal} {\bibinfo  {journal} {Phys. Rev. B}\ }\textbf {\bibinfo {volume}
  {84}},\ \bibinfo {pages} {233102} (\bibinfo {year} {2011})}\BibitemShut
  {NoStop}%
\bibitem [{\citenamefont {Winograd}\ \emph {et~al.}(2012)\citenamefont
  {Winograd}, \citenamefont {Chitra},\ and\ \citenamefont
  {Rozenberg}}]{WinogradPRB2012}%
  \BibitemOpen
  \bibfield  {author} {\bibinfo {author} {\bibfnamefont {E.~A.}\ \bibnamefont
  {Winograd}}, \bibinfo {author} {\bibfnamefont {R.}~\bibnamefont {Chitra}}, \
  and\ \bibinfo {author} {\bibfnamefont {M.~J.}\ \bibnamefont {Rozenberg}},\
  }\bibfield  {title} {\enquote {\bibinfo {title} {Phase diagram of the
  asymmetric {Hubbard} model and an entropic chromatographic method for cooling
  cold fermions in optical lattices},}\ }\href {\doibase
  10.1103/PhysRevB.86.195118} {\bibfield  {journal} {\bibinfo  {journal} {Phys.
  Rev. B}\ }\textbf {\bibinfo {volume} {86}},\ \bibinfo {pages} {195118}
  (\bibinfo {year} {2012})}\BibitemShut {NoStop}%
\bibitem [{\citenamefont {Sekania}\ \emph {et~al.}(2017)\citenamefont
  {Sekania}, \citenamefont {Baeriswyl}, \citenamefont {Jibuti},\ and\
  \citenamefont {Japaridze}}]{Sekania2017}%
  \BibitemOpen
  \bibfield  {author} {\bibinfo {author} {\bibfnamefont {M.}~\bibnamefont
  {Sekania}}, \bibinfo {author} {\bibfnamefont {D.}~\bibnamefont {Baeriswyl}},
  \bibinfo {author} {\bibfnamefont {L.}~\bibnamefont {Jibuti}}, \ and\ \bibinfo
  {author} {\bibfnamefont {G.~I.}\ \bibnamefont {Japaridze}},\ }\bibfield
  {title} {\enquote {\bibinfo {title} {Mass-imbalanced ionic {H}ubbard
  chain},}\ }\href {\doibase 10.1103/PhysRevB.96.035116} {\bibfield  {journal}
  {\bibinfo  {journal} {Phys. Rev. B}\ }\textbf {\bibinfo {volume} {96}},\
  \bibinfo {pages} {035116} (\bibinfo {year} {2017})}\BibitemShut {NoStop}%
\bibitem [{\citenamefont {Hirsch}(1984)}]{HirschPRL1984}%
  \BibitemOpen
  \bibfield  {author} {\bibinfo {author} {\bibfnamefont {J.~E.}\ \bibnamefont
  {Hirsch}},\ }\bibfield  {title} {\enquote {\bibinfo {title}
  {Charge-density-wave to spin-density-wave transition in the extended
  {H}ubbard model},}\ }\href {\doibase 10.1103/PhysRevLett.53.2327} {\bibfield
  {journal} {\bibinfo  {journal} {Phys. Rev. Lett.}\ }\textbf {\bibinfo
  {volume} {53}},\ \bibinfo {pages} {2327} (\bibinfo {year}
  {1984})}\BibitemShut {NoStop}%
\bibitem [{\citenamefont {Lin}\ and\ \citenamefont {Hirsch}(1986)}]{Lin1986}%
  \BibitemOpen
  \bibfield  {author} {\bibinfo {author} {\bibfnamefont {H.~Q.}\ \bibnamefont
  {Lin}}\ and\ \bibinfo {author} {\bibfnamefont {J.~E.}\ \bibnamefont
  {Hirsch}},\ }\bibfield  {title} {\enquote {\bibinfo {title} {Condensation
  transition in the one-dimensional extended {H}ubbard model},}\ }\href
  {\doibase 10.1103/PhysRevB.33.8155} {\bibfield  {journal} {\bibinfo
  {journal} {Phys. Rev. B}\ }\textbf {\bibinfo {volume} {33}},\ \bibinfo
  {pages} {8155--8163} (\bibinfo {year} {1986})}\BibitemShut {NoStop}%
\bibitem [{\citenamefont {Lin}\ \emph {et~al.}(2000)\citenamefont {Lin},
  \citenamefont {Campbell},\ and\ \citenamefont {Clay}}]{Lin2000}%
  \BibitemOpen
  \bibfield  {author} {\bibinfo {author} {\bibfnamefont {H.Q.}\ \bibnamefont
  {Lin}}, \bibinfo {author} {\bibfnamefont {D.K.}\ \bibnamefont {Campbell}}, \
  and\ \bibinfo {author} {\bibfnamefont {R.T.}\ \bibnamefont {Clay}},\
  }\bibfield  {title} {\enquote {\bibinfo {title} {Broken symmetries in the
  one-dimensional extended {H}ubbard model},}\ }\href@noop {} {\bibfield
  {journal} {\bibinfo  {journal} {Chin. J. Phys.}\ }\textbf {\bibinfo {volume}
  {38}},\ \bibinfo {pages} {1} (\bibinfo {year} {2000})}\BibitemShut {NoStop}%
\bibitem [{\citenamefont {Penn}(1966)}]{Penn1966}%
  \BibitemOpen
  \bibfield  {author} {\bibinfo {author} {\bibfnamefont {D.~R.}\ \bibnamefont
  {Penn}},\ }\bibfield  {title} {\enquote {\bibinfo {title} {Stability theory
  of the magnetic phases for a simple model of the transition metals},}\ }\href
  {\doibase 10.1103/PhysRev.142.350} {\bibfield  {journal} {\bibinfo  {journal}
  {Phys. Rev.}\ }\textbf {\bibinfo {volume} {142}},\ \bibinfo {pages}
  {350--365} (\bibinfo {year} {1966})}\BibitemShut {NoStop}%
\bibitem [{\citenamefont {\v{Z}onda}\ \emph {et~al.}(2009)\citenamefont
  {\v{Z}onda}, \citenamefont {Farka\v{s}ovsk\'y},\ and\ \citenamefont
  {\v{C}en\v{c}arikov\'a}}]{ZondaSSC2009}%
  \BibitemOpen
  \bibfield  {author} {\bibinfo {author} {\bibfnamefont {M.}~\bibnamefont
  {\v{Z}onda}}, \bibinfo {author} {\bibfnamefont {P.}~\bibnamefont
  {Farka\v{s}ovsk\'y}}, \ and\ \bibinfo {author} {\bibfnamefont
  {H.}~\bibnamefont {\v{C}en\v{c}arikov\'a}},\ }\bibfield  {title} {\enquote
  {\bibinfo {title} {Phase transitions in the three-dimensional
  {Falicov}-{Kimball} model},}\ }\href {\doibase 10.1016/j.ssc.2009.08.035}
  {\bibfield  {journal} {\bibinfo  {journal} {Solid State Commun.}\ }\textbf
  {\bibinfo {volume} {149}},\ \bibinfo {pages} {1997--2001} (\bibinfo {year}
  {2009})}\BibitemShut {NoStop}%
\bibitem [{\citenamefont {Kennedy}\ and\ \citenamefont
  {Lieb}(1986)}]{KennedyPhysA1986}%
  \BibitemOpen
  \bibfield  {author} {\bibinfo {author} {\bibfnamefont {T.}~\bibnamefont
  {Kennedy}}\ and\ \bibinfo {author} {\bibfnamefont {E.~H.}\ \bibnamefont
  {Lieb}},\ }\bibfield  {title} {\enquote {\bibinfo {title} {An itinerant
  electron model with crystalline or magnetic long range order},}\ }\href
  {\doibase 10.1016/0378-4371(86)90188-3} {\bibfield  {journal} {\bibinfo
  {journal} {Physica A}\ }\textbf {\bibinfo {volume} {138}},\ \bibinfo {pages}
  {320--358} (\bibinfo {year} {1986})}\BibitemShut {NoStop}%
\bibitem [{\citenamefont {Lema\ifmmode~\acute{n}\else \'{n}\fi{}ski}\ \emph
  {et~al.}(2002)\citenamefont {Lema\ifmmode~\acute{n}\else \'{n}\fi{}ski},
  \citenamefont {Freericks},\ and\ \citenamefont {Banach}}]{LemanskiPRL2002}%
  \BibitemOpen
  \bibfield  {author} {\bibinfo {author} {\bibfnamefont {R.}~\bibnamefont
  {Lema\ifmmode~\acute{n}\else \'{n}\fi{}ski}}, \bibinfo {author}
  {\bibfnamefont {J.~K.}\ \bibnamefont {Freericks}}, \ and\ \bibinfo {author}
  {\bibfnamefont {G.}~\bibnamefont {Banach}},\ }\bibfield  {title} {\enquote
  {\bibinfo {title} {Stripe phases in the two-dimensional {Falicov-Kimball}
  model},}\ }\href {\doibase 10.1103/PhysRevLett.89.196403} {\bibfield
  {journal} {\bibinfo  {journal} {Phys. Rev. Lett.}\ }\textbf {\bibinfo
  {volume} {89}},\ \bibinfo {pages} {196403} (\bibinfo {year}
  {2002})}\BibitemShut {NoStop}%
\end{thebibliography}%


\end{document}